\documentclass[reqno,11pt]{article}
\newcommand{\version}{November 11, 2005}
\usepackage{amsmath} 
\usepackage{amssymb}
\usepackage{latexsym}
\usepackage{cite}

\newcommand{\RR}{\mathbb{R}}
\newcommand{\ZZ}{\mathbb{Z}}
\newcommand{\NN}{\mathbb{N}}
\newcommand{\CC}{\mathbb{C}}

\newcommand{\Bb}{\mathbb{R}}
\newcommand{\Bi}{\mathbb{Z}}
\newcommand{\Bn}{\mathbb{N}}
\newcommand{\Bc}{\mathbb{C}}
\newenvironment{Proof}%
{\par \medskip \noindent {\em Proof.}}{\hspace*{\fill} $\square$ \par 
\medskip}
\newenvironment{Proofof}%
{\par \medskip \noindent {\em Proof}}{\hspace*{\fill} $\square$ 
\medskip}
\newtheorem{Thm}{Theorem}[section]
\newtheorem{Prop}[Thm]{Proposition}
\newtheorem{Lem}[Thm]{Lemma}

\newtheorem{Def}[Thm]{Definition}

\newenvironment{Remarks}{\par\noindent {\em Remarks. } \rm}{\par\medskip}

\newcommand{\beq}{\begin{equation}}
\newcommand{\eeq}{\end{equation}}
\renewcommand{\Re}{\text{Re }}    
\renewcommand{\Im}{\text{Im }}    
\newcommand{\lsp}{\left(\,}     
\newcommand{\rsp}{\,\right)}
\newcommand{\Lsp}{\big(\,}
\newcommand{\Rsp}{\,\big)}
\newcommand{\calA}{\mathcal{A}}          
\newcommand{\calD}{{\mathcal D}}
\newcommand{\calH}{{\mathcal H}}
\newcommand{\calL}{{\mathcal L}}
\newcommand{\calO}{{\mathcal O}}
\newcommand{\calP}{{\mathcal P}}
\newcommand{\calS}{{\mathcal S}}
\newcommand{\calN}{{\mathcal N}}
\newcommand{\Hi}{{\mathcal H}}

\newcommand{\half}{{\frac{1}{2}}} 
\newcommand{\clo}{ {\mbox{\bf --}} }

\renewcommand{\d}{{\rm d}}
\newcommand{\supp}{{\rm supp\,}}
\newcommand{\eps}{\varepsilon}

\newcommand{\unity}{{\setlength{\unitlength}{1em}
                     \begin{picture}(0.75,1)
                     \put(0,0){$1$}
                     \put(0.34,0){\line(0,1){0.65}}
                     \end{picture}
                   }}
\newcommand{\trace}{\text{\rm tr }}
\newcommand{\utilde}[1]{\underline{#1}}
\newcommand{\Lc}{H_0^+}               
\newcommand{\lc}{\xi}               
\newcommand{\Spd}{H}                
\newcommand{\spd}{e}                   
\newcommand{\spc}{{C}}              
\newcommand{\Spc}{\mathcal{C}}      
\newcommand{\Min}{\RR^4}          
\newcommand{\Hyp}{H_m^+}          
\newcommand{\HypMinus}{\dot{H}_m^+}          
\newcommand{\HypNull}{H_0^+}          
\newcommand{\HypNullMinus}{\dot{H}_0^+}          
\newcommand{\Po}{\calP_+^{\uparrow}}
\newcommand{\Poj}{\calP_+}
\newcommand{\Lor}{\calL_+^{\uparrow}}
\renewcommand{\lor}{\Lambda}     
\newcommand{\po}{g}              
\newcommand{\Boo}[2]{\lor_{#1}(#2)}         
\newcommand{\BooNull}{\lor_{W_0}}         
\newcommand{\real}{r}         
\newcommand{\WigRot}{R}       
\newcommand{\PauliLub}{\kappa}        
\newcommand{\Hirr}{\calH_1} 
\newcommand{\Uirr}{U_{1}}  
\newcommand{\Kirr}{K_{1}}          
                                  
\newcommand{\calh}{{\mathfrak h}}     

\newcommand{\strip}{{\mathcal G}}          
\newcommand{\FT}{E}    
\newcommand{\intfct}[2]{u(#1,#2)} 
\newcommand{\intfctconj}[2]{u_c({#1},#2)} 
\newcommand{\intdist}[2]{u(#1,#2)} 
\newcommand{\intdistconj}[2]{u_c({#1},#2)} 
\newcommand{\covfct}[2]{\psi({#1},#2)} 
\newcommand{\degree}{\alpha}   
\newcommand{\ReDeg}{{\degree'}}   
\newcommand{\intfctdeg}[2]{u^{\degree}(#1,#2)} 
\newcommand{\intfctdegconj}[2]{u^{\bar\degree}({#1},#2)} 
\newcommand{\covfcttwo}[2]{\psi_2({#1},#2)} 
\newcommand{\field}{\varphi}   

\newcommand{\orbit}{\Gamma}
\newcommand{\spin}{s}        
\newcommand{\RepPulBac}{\tilde{D}}

\newcommand{\intdistdeg}[2]{u^{\degree}(#1,#2)} 
\newcommand{\Spdc}{\Spd^{\rm c}}  
\newcommand{\Tub}{{\mathcal T}_+}  
\newcommand{\spdc}{\spd}  
\newcommand{\spdr}{\spd'} 
\newcommand{\spdi}{\spd''} 

\newcommand{\kre}{\real'}      
\newcommand{\sz}{k}     
\begin{document} 
\markboth{\scriptsize{MSY, \version}}{\scriptsize{MSY, \version}}
\title{String-localized Quantum Fields\\ and Modular Localization}

\author{ \hspace{-1.0cm} J.~Mund${}^{a}$, B.~Schroer${}^{b,c}$, J.~Yngvason${}^{d, e}$\\ \\
\hspace{-1.5 cm}\normalsize\it ${}^{a}$Departamento de F\'isica, ICE, Universidade Federal de Juiz de Fora,\\\hspace{-1.5 cm}\normalsize\it
36036-330 Juiz de Fora, MG, Brazil \\\hspace{-1.5 cm} 
\normalsize\it ${}^{b}$ CBPF, Rua Dr.\ Xavier Sigaud 150, 22290-180 Rio de 
Janeiro, Brazil,\\ \hspace{-1.5 cm}\normalsize\it ${}^{c}$Institut f\"ur Theoretische Physik, FU-Berlin, 
Arnimallee 14, D-14195 Berlin, Germany\\
\hspace{-1.8 cm}\normalsize\it ${}^{d}$ Erwin Schr{\"o}dinger
Institute for Mathematical Physics, Boltzmanngasse 9,
1090 Vienna, Austria\\ \hspace{-1.5 cm}${}^{e}$\normalsize\it Institut f\"ur Theoretische
Physik, Universit{\"a}t Wien, Boltzmanngasse 5,
1090 Vienna, Austria}
\date{\version }
\maketitle 
\begin{abstract}

  We study free, covariant, quantum (Bose) fields that are associated with
  irreducible representations of the Poincar\'e group and localized in
  semi-infinite strings extending to spacelike infinity. Among these
  are fields that generate the irreducible representations of mass
  zero and infinite spin that are known to be incompatible with
  point-like localized fields. For the massive representation and the
  massless representations of finite helicity, all string-localized
  free fields can be written as an integral, along the string, of
  point-localized tensor or spinor fields.  
  As a special case we discuss the string-localized vector fields
  associated with the point-like electromagnetic field and their relation to
  the axial gauge condition in the usual setting.
\end{abstract}
\tableofcontents

\section{Introduction}
In their paper \cite{BGL}, Brunetti, Guido and Longo (BGL) established a
general connection between positive energy representations of the
Poincar\'e group and localization properties of state vectors in the
Hilbert space of the representation.  This {\it modular localization} is not
associated with any position operators, which are known to be
problematic in the relativistic case, but rather with the Lorentz
boosts corresponding to wedge-like regions in Minkowski space and the
PCT operator.  Using these concepts, the authors of \cite{BGL} show
that every irreducible, positive energy representation of the
Poincar\'e group admits dense sets of vectors that are naturally
localized in space-like cones in Minkowski space with arbitrarily
small opening angles.

It is well known that in the case of the irreducible representations
of finite spin or helicity this localization can be sharpened to
double cone localization, by making use of the concrete realization of
the representation in the one particle space of a covariant Wightman
field.  The BGL concept, however, applies also to the Wigner
representations of zero mass and infinite spin, where a localization
in the sense of point-like fields is not possible \cite{Yng70}.  This
latter result excludes compact localization in the sense of Wightman
fields, even with infinitely many components, and applies also to the
special constructions in \cite{IM, Abbott, Hirata}. We note in passing that these 
representations have recently found applications 
in work on 'tensionless strings' in String Theory \cite{Mourad,Savvidy}. 

The localization spaces for space-like cones of \cite{BGL} are
abstractly defined in terms of intersections of wedge-localized spaces
without concrete formulas for their generation.  In a previous letter
\cite{MSY1} we showed that the spaces for the zero mass and infinite spin
representations can be explicitly described in terms of
{\it string-localized fields}.  The strings can be depicted as the cores of
the space-like cones of \cite{BGL}. More precisely, the fields
considered in \cite{MSY1} are operator valued distributions $\field(x,e)$
where $x$ is a point in Minkowski space and
$e$ is in the manifold of space-like
directions
\begin{equation} \label{eqSpd}
\Spd :=\{ e\in\RR^d:\, e\cdot e=-1\}.
\end{equation}
The localization region of $\field(x,e)$ is the space-like string (or 
ray) $x+\RR_0^+e$ in the
sense that if
the strings
$x_1+\RR_0^{+}e'_1$  and $x_2+\RR_0^{+} e_2$ are
space-like separated  for all $e'_1$ in some open neighborhood of
$e_1$,\footnote{\label{StringSep} That is,  $x_1+\RR_0^{+}e_1$ and 
$x_2+\RR_0^{+} e_2$ are space-like separated {\em and} 
$e_1$ and $e_2$ are space-like separated, c.f.~Lemma~\ref{StringWedge}.}   then
\begin{equation} \label{eqFieldLoc} 
[\field(x_1,e_1),\field(x_2,e_2)]=0.
\end{equation}
The field  transforms in a covariant way under a 
unitary representation $U$ of the Poincar\'e group $\Po$
according  to\footnote{We denote elements of $\Po$ by pairs $(a,\lor)$
with $a\in\RR^d$ and $\lor$ in the Lorentz group $\Lor$.}
\begin{align}  \label{eqFieldCov}  
U(a,\lor)\;\field(x,\spd)\; U(a,\lor)^{-1}=
\field(\lor x+a,\lor\spd)\,,\quad(a,\lor)\in\Po.
\end{align}
Thus, the space-like direction, $e$, substitutes for the usual 
the Lorentz index. The restriction of $U$ to the translation group is assumed to satisfy
the spectrum condition, i.e., the joint spectrum of its generators 
is a subset of the forward light cone. 
These properties essentially define what is meant by a
`string-localized field' in this paper. Further restrictions but also some
generalizations will be
introduced later.

The purpose of the present paper is twofold.  On the one hand we will
supply the mathematical details of the construction in \cite{MSY1}. 
On the other hand, we put this construction into a wider context by
exploring the relation between the modular localization and string-like 
localization in quantum field theory (QFT).  Our considerations
are restricted to free fields but we expect our findings also to be of
relevance in more general situations in particle physics.  In
particular it is our desire to find a path to a (possibly
perturbative) construction of massive interacting string-localized
objects whose existence and general properties are predicted on
structural grounds in the setting of algebraic quantum field theory
(AQFT) \cite{BF}.  We hope that our analysis of string-localized free
fields will turn out to be a useful step into that direction.\\

The quest for the understanding of string-like extended objects is
almost as old as the Lagrangian quantization approach to point-like
quantum field theory and it is appropriate to start by recalling some
of its history.  The idea that string-localized objects are useful
dates back to the early days of pre-renormalization QED when P.~Jordan 
\cite{Jordan1} proposed to use exponential line integrals over
electromagnetic vector potentials in order to arrive at gauge
invariant composites involving matter fields.  His completely
algebraic proof of the Dirac monopole quantization was a nice application of
string-like objects in QED that
unfortunately has remained largely unknown up to this date.

In more recent (post-renormalization) times Mandelstam
\cite{Mandelstam} and Wilson \cite{Wilson} made extensive use of
expressions involving finitely- or infinitely-extended integrals over
local gauge fields.  A more recent rigorous treatment of the
perturbative aspects of such objects can be found in \cite{St}. 
Jordan \cite[and earlier papers quoted therein]{Jordan2} in his series of 
publications under the somewhat misleading title
``neutrino theory of light'' was also the first to use such nonlocal
expressions in space-time dimension $d=1+1$ for what we now call
bosonization/fermionization, apparently not quite aware that this
trick is limited to $d=1+1$. Later this formalism was used for several purposes.
In \cite{Klaiber} exponential line integrals served to obtain an
improved treatment of the Thirring model, and in \cite{Streater-Wild} as
well as in \cite{Froehlich} it was used to illustrate the abstract
setting of the Doplicher-Haag-Roberts theory of superselection
sectors.

The first systematic structural analysis for semi-infinite strings in
massive QFT is due to Buchholz and Fredenhagen (B-F) \cite{BF}.  In
this case the string-like objects are massive charge-carrying fields
within the setting of AQFT whose localization core is
a semi-infinite space-like string and whose particle and symmetry
aspects are the same as for point-like interpolating fields.  
The B-F strings are thus dynamical objects, i.e. their
string--localization is due to interaction.

There is an important case where strings appear naturally without
interaction, namely the massive representations of the Poincar\'e
group in space-time dimension $d=1+2$ with non-integer (or
non-half-integer) spin.  The occurrence of braid group statistics in
this case was first explored in \cite{LM} and a realization of anyons
(particles with abelian braid group statistics) along the line of a
Aharonov-Bohm effect was proposed in \cite{Wilczek}.  A description of
the general case of plektonic statistics, the string-like nature of
the associated operators as well as their scattering theory appeared
in \cite{FG,FRSII,FGR}.  The first construction of string-localized
anyon one-particle states is due to one of the present authors
\cite{M02a} who in a previous paper \cite{M} also showed that a
mass-shell description of the associated fields is not possible.  A
relativistic field theory of anyons, even in the absence of genuine interactions,
does not yet exist.

The strings of String Theory have little relation to string
localization in the sense of the present paper.  This is not
surprising in view of the different history and motivation.  Whereas
string-localized fields are objects which fit naturally into the
conceptual framework of QFT, String Theory is an attempt to transcend
QFT and whose main contemporary motivation is the incorporation of all
interactions (including gravity) into a scheme which at least on a
perturbative level remains ultraviolet-finite.  The word "string"
refers in this case to its historical connections with quantized
Nambu-Goto string Lagrangians which, however, do not lead to
string-localized quantum fields \cite{Dimock, Erler-Gross}.  Since
some of the details behind these differences are quite interesting we
will return to this issue in a separate section at the end of this
paper. \\

After this digression on the history of string-like objects in QFT we come
back to the contents of the present paper.  Starting from an
irreducible representation of the Poincar\'e group in $d=3$ or
$d=4$ space-time dimensions our aim is to find the most general
string-localized field that generates this representation when applied
to the vacuum and is moreover free in the sense that it is completely
determined by the two-point function.  Our main findings are as
follows:
\begin{itemize}
    \item Such fields exist for all irreducible representations where 
    the representation of the `little group' is either faithful, or trivial. 
    In $d=4$ this applies to all 
    massive representations, the massless scalar
  (helicity zero) representations  and the massless infinite spin 
    representations. In $d=3$ this holds  for all massive 
    representations of integer spin, the massless scalar
 representation  and the massless infinite spin 
    representations\footnote{
    For $d=3$, `infinite spin' is really a 
    misnomer because the representation of the little group is 
    one-dimensional in this case. See Section 6.2.}.
     \item For the massive and the scalar massless 
    representations all string localized fields can be written as a 
    line integral over point-like fields. This is not possible 
    for the infinite spin representations and the corresponding string 
    localized fields are thus truly elementary.
   \item  For the massless representations in $d=4$ with finite, nonzero 
   helicity, string fields can be defined if the previous definition 
   is modified and a  tensor (or spinor) index is added to the field in 
   addition to the space-like direction $e$. In particular, photons 
   can be described by a string localized field with a 4-vector index 
   in addition to $e$. The requirement that this field is a 
   vector potential for the (point-localized) electromagnetic 
tensor field fixes it 
   uniquely and leads naturally to the axial gauge condition.
   \item String localization improves the short distance behavior of 
   propagators in such a way that the singularities do not get worse with 
   increasing spin.
   \end{itemize}
   The third point above is related to the well known fact that the
possibilities to intertwine the Wigner canonical representations with
covariant spinorial representations is more restricted for massless
finite helicity representations than for the massive ones.  The group
theoretical reason lies in the different stabilizer groups (`little
groups') for time-like and light-like vectors.  What matters is the
restriction to the little group of the representation of the Lorentz
group occurring in the covariant transformation law.  This restriction
must contain the canonical representation considered as a
subrepresentation and this requirement excludes in the massless case
certain covariant transformation laws.  The best known case is that of
free photons which, in a Hilbert space with positive definite metric,
can be described by a point-like field strength tensor but not by a
point-like vector potential that would have a better short distance
behavior than the field strength.  In Section \ref{Photons} we shall
discuss this case further and in particular show how the photon can be
described by a `vector string' $A_{\mu }(x,e)$ which in addition to
Lorentz transformations of $x$ and $e$ that determine the localization
suffers a matrix transformation of the `internal' vector index $\mu$. 
This vector string satisfies the so-called `axial
gauge condition' in conventional quantum electrodynamics, but in 
the latter case $e$ is not
considered as specifying a string direction and therefore is kept
unchanged under Lorentz transformations.  This is achieved at the
prize of an (abelian) gauge transformation.  
 We also overcome the singularity of the axial gauge at $e\cdot p=0$ (contributing to its
unpopularity) by treating the potential as a distribution in $e$.

We hope to return to this interesting alternative viewpoint to the
gauge theoretical setting in a separate work.  We will also refrain
here from investigating possible links between the string localized
vector potentials and the Jordan-Mandelstam-Wilson string-like objects
and for the  massless representations we shall limit explicit 
constructions to helicity 0 and 1.  Since in
most of the present work we will be dealing with scalar string
localized fields in the sense of Eqs.\ \eqref{eqFieldLoc}-\eqref{eqFieldCov}
we will usually omit the pre-fix `scalar'.  Without loss of generality
we may restrict our considerations to self-conjugate (hermitian,
Majorana) Bose fields.  The extension to half-integer spins and
Fermi fields does not bring in fundamentally new aspects and will not
be treated for reasons of space.

The organization of the subsequent sections is as follows:
In the next section we discuss the concept of
modular localization, emphasizing its difference to
the Newton-Wigner localization of particle states.
The third section presents the key concept for the
modular localization of  positive energy representations of
the Poincar\'e group, namely the interwiners between the Wigner
canonical form of the representation and the covariant string localized 
form, and discusses their uniqueness.
The string-localized fields are represented in terms of these 
intertwiners and creation and annihilation operators for the Wigner 
particle states in the basic formula \eqref{eqField} in Theorem 
\ref{StringField}. 
In Section \ref{massive_bosons} this formula  is specialized to the massive 
representations by calculating the intertwiner. Here it is 
also shown that all string-localized fields for these representations 
can be written as line integrals over point-fields.
Section 5
contains the discussion of string-localized vector fields for 
photons, while Section 6 is concerned with the massless, `infinite spin' 
representations in $d=4$ and $d=3$ where 
point-localized fields do not exist. In Section 7 we return briefly to the 
comparison of string localized fields and the strings of String Theory.
The final section 8 contains a resum\'e of the main results and an
outlook. In order not to burden the main text with too many technical 
details we present the proofs of several lemmas in the Appendix.

\section{Modular Localization}

Localization and causality are central concepts which have accompanied
relativistic quantum field theory right from its beginning through all stages
of its development. Since these properties first appeared in the quantum
setting as a result of quantizing classical fields, it was natural to assume
that the classical relativistic notions of locality and causality continue to
apply in the quantum realm. However the conceptual difference between
observables and states, which in QFT becomes more accentuated by the
omnipresence of vacuum polarization, required a more careful adaptation of
these concepts.

Historically the first step towards an intrinsic formulation of
relativistic quantum physics independent of any classical analogies
was undertaken by Wigner in 1939 when he identified relativistic
particle states with irreducible positive energy representations of
the Poincar\'{e} group.  These representations come with two
 notions of localization: the Newton-Wigner (NW)
localization \cite{NW} that was formulated some years afterward,
and the more recent modular localization
\cite{BGL,M02a,FS02}.

The NW localization is the result of the adaptation of Born's quantum
mechanical probability density for particle positions to Wigner's relativistic
representation theoretical setting.  Newton and Wigner define, in the
single particle space, a position operator whose spectral projectors
are supposed to measure the probability of detecting a (single)
particle in different space-time regions. States localized in disjoint space regions at 
{\it fixed time} in some given frame of reference
are orthogonal. This localization 
incorporates macro-causality and the cluster property, and is
perfectly well-suited for scattering theory.  
On the other hand it is
not consistent with relativistic covariance and 
causality,  except in an approximate sense for distances of the order of the
Compton
wave length or smaller.  In fact, it is by now well understood that 
{\it any} notion of localization 
that requires the set of states
localized in a space-time region ${\mathcal O}$ to be  
orthogonal
to the states localized in the causal
complement ${\mathcal O'}$ is incompatible with translational
covariance and positivity of the energy \cite{PerezWilde, Malament}. 

A localization concept for quantum systems compatible with
relativistic covariance and 
causality is contained in the formalism of local quantum
field theory.  This notion of localization refers not to positions of
particles, but to local measurements of observables and to charge
creation.  The algebra $\mathcal A$ of observables in quantum field
theory has a natural net structure which assigns to each space time
region $\mathcal O$ a sub-algebra $\mathcal A(\mathcal O)\subset
\mathcal A$.  Typically, the algebra $\mathcal A(\mathcal O)$ is
generated by smeared field operators $\Phi(f)$ (or their neutral currents in case the fields  are charged) with test functions $f$
supported in $\mathcal O$.  A key point is that the net structure of
the observables allows a {\it local comparison} of states: Two states
are locally equal in a region $\mathcal O$ if and only if the
expectation values of all operators in $\mathcal A(\mathcal O)$ are
the same in both states.  Local deviations from any state, in
particular the vacuum state, can be measured in this manner, and
states that are indistinguishable from the vacuum in the causal
complement of some region (`strictly localized states' \cite{Licht}) can
be defined.

Due to the unavoidable correlations in the vacuum state in
relativistic quantum theory (the Reeh-Schlieder 
property\cite{Reeh-Schlieder}), the
space $\Hi(\mathcal O)$ obtained by applying the operators in $\mathcal
A(\mathcal O)$ to the vacuum is, for any open region $\mathcal O$,
dense in the Hilbert space and thus far from being orthogonal to
$\Hi(\mathcal O')$.  This somewhat counterintuitive fact is inseparably
linked with a structural difference between the local algebras and the
algebras encountered in non-relativistic quantum mechanics or the global
algebra of a quantum field, associated with the entire Minkowski space-time.  Whereas the
latter has minimal projections (corresponding to optimal
observations), the local algebras are type III in the terminology
of Murray and von Neumann.  Some physical consequences of this
difference are reviewed in \cite{TypeIII}.  The Reeh-Schlieder property also
implies that the expectation value of a projection operator localized in a
bounded region can not be interpreted as the probability of detecting
a single particle in that region since it is necessarily nonzero in
the vacuum state.  This is not surprising because {\it
strict} localization requires arbitrarily high energies which in a
relativistic theory may be accompanied by the creation of particles. A 
direct comparison with NW localization can be made in the case of free 
fields which are well defined as operator valued distributions in the 
space variables at a fixed time.
The one-particle states that are NW localized in a given space 
region at a fixed time are not the same as the states obtained by
applying field operators smeared with test functions supported in this 
region to the vacuum. The difference lies in the non-local energy factor $({\bf 
p^2}+m^2)^{1/2}$ linking the non-covariant NW states with the states 
defined in terms of the covariant field operators.

Causality in relativistic quantum field theory is mathematically
expressed through {\it local commutativity}, i.e., mutual
commutativity of the algebras $\mathcal A(\mathcal O)$ and $\mathcal
A(\mathcal O')$.  There is an intimate connection of this property
with the possibility of preparing states that exhibit no mutual
correlations for a given pair of causally disjoint regions.  In fact,
in the recent paper \cite{BuchholzSummers2004} Buchholz and Summers
show that local commutativity is a necessary condition for the
existence of such uncorrelated states.  Conversely, in combination
with some further properties ({\it split property} \cite{DL}, existence of scaling
limits), that are physically plausible and have been verified in
models, local commutativity leads to a very satisfactory picture of
statistical independence and local preparabilty of states in
relativistic quantum field theory.  We refer to \cite{Summers, Wern} for
thorough discussions of these matters and \cite{TypeIII} for a brief review.

Consequent application of the above mentioned concepts avoids the
defects of the NW localization and resolves spurious problems rooted
in assumptions that are in conflict with basic principles of
relativistic quantum physics.  An example is the apparent difficulty
\cite{Hegerfeldt} with Fermi's famous gedankenexperiment \cite{Fermi} which he
proposed in order to show that the velocity of light is the limiting
propagation velocity in quantum electrodynamics.  An argument which
takes into account the progress on the conceptual issues of causal
localization and in mathematical rigor since the times of Fermi and
confirms his conclusion can be found in \cite{Bu-Yng}, see also
\cite{TypeIII}.

{\it Modular localization} of single particle states is a concept that
is intrinsically defined within the representation theory of the
Poincar\'e group but draws its motivation from 
local quantum field theory.  The space $\Hi(\mathcal O)$ obtained by
applying the operators of a local algebra $\mathcal A\mathcal(\mathcal
O)$ to the vacuum vector $\Omega$ can be regarded as the domain of the
Tomita involution $S_{\mathcal O}$ that maps $A\Omega$ to
$A^*\Omega$.  In the special case where $\mathcal O$ is a space-like
wedge, the Tomita involution has a geometrical interpretation
according to the Theorem of Bisognano and Wichmann \cite{BiWi}: It is
determined by the PCT operator combined with a rotation and the
generator of the Lorentz boosts associated with the wedge.  It has
been realized in recent years by Brunetti, Guido and Longo~\cite{BGL}
and by B.~Schroer~\cite{FS02} that by appealing to
this interpretation of the Tomita involution for wedges and using the
spatial counterpart of Tomita-Takesaki theory \cite{RvD} it is possible to
partially invert the above procedure of passing from local algebras 
to localized states.  Namely, there is a natural localization
structure on the representation space for any positive energy
representation of the proper Poincar\'e group which upon second 
quantization gives rise to a local net of operator algebras on the Fock space over
the representation Hilbert space.

In the context of Wigner's description
of elementary relativistic systems 
the starting point is an irreducible 
representation $U_1$ of the Poincar\'e group on a Hilbert space $\Hi_1$ 
that after second
quantization becomes the single-particle subspace of the Hilbert 
space (Fock-space) $\Hi$ 
of the field. (We emphasize, however,  that the construction works for 
arbitrary positive energy representations, not only irreducible ones.)
The construction then
proceeds according to the following 3 steps \cite{BGL,M02a,FS02}. 
To maintain simplicity we limit our presentation to the
bosonic situation and refer to \cite{M02a} and \cite{FS02} for the general
treatment. 
\paragraph{Step 1.}
Fix a reference wedge region, e.g. 
\begin{equation}\label{eqW0}
W_0=\left\{  x\in\mathbb{R}^{d};x^{d-1}>\left|  x^{0}\right|  \right\},  
\end{equation}
and consider the one-parameter group $\Boo{W_0}{\cdot}$ of Lorentz boosts 
which leave $W_0$ invariant, and the 
reflection $j_{W_0}$ across the edge of the wedge. More specifically, 
$\Boo{W_0}{t}$ acts as 
$$
\left( \begin{matrix} \cosh(t) & \sinh(t) \\ \sinh(t) & \cosh(t) 
\end{matrix}\right) 
$$ 
and $j_{W_0}$ acts as the reflection on the coordinates $x^0$ and $x^{d-1}$, 
leaving the other coordinates unchanged.   
Then use the Wigner representation $U_1(\cdot)$
of the boosts and the reflection\footnote{In certain cases an irreducible
representation of ${\mathcal P}_{+}^\uparrow$ has to be doubled in order to 
accommodate the anti-unitary (since
time is inverted) reflection. This is always the case with zero mass finite
helicity representations and more generally if particles are not
self-conjugate.},  to define  
\begin{align}    \label{eqModOp}
\Delta_{W_0}^{it}& :=U_1(\Lambda_{W_0}(-2\pi t)),\; J_{W_0} :=U_1(j_{W_0})\\
S_{W_0} &  :=J_{W_0}\Delta_{W_0}^{\frac{1}{2}}. \label{eqTom} 
\end{align}
The operator $\Delta_{W_0}^{\frac{1}{2}}$ is  unbounded (in general), closed and 
positive, $J_{W_0}$ is an anti-linear involution commuting with 
$\Delta_{W_0}^{it}$, and 
$S_{W_0}$ is anti-linear and closed with $S_{W_0}^{2}\subset1$.
These properties characterizes $S_{W_0}$ as a {\it Tomita
involution}\footnote{Operators with this property are the corner stones of the
Tomita-Takesaki modular theory of operator algebras. Here they arise in the
spatial Rieffel van Daele setting \cite{RvD} of modular theory from a realization
of the geometric Bisognano-Wichmann situation within the Wigner representation
theory.} which is uniquely determined by its eigenspace 
to the eigenvalue +1, i.e., 
\begin{equation}  \label{eqKW}
    K(W_0) :=\left\{  \psi\in \hbox{domain of $\Delta_{W_0}^{\frac{1}{2}}$} ,\,S_{W_0}\psi=\psi\right\}.
    \end{equation}
 This is a closed, real linear subspace of $\Hi_1$ satisfying   
\begin{equation}
\overline{K(W_0)+iK(W_0)}  =\Hi_1,\,\,K(W_0)\cap 
iK(W_0)=0,\label{standard}\end{equation}
and
\begin{equation}
J_{W_{0}}K(W_0)=K(W_{0}')=K(W_0)^{\bot}\label{duality}%
\end{equation}
where $\bot$ refers to orthogonality in the sense of the symplectic
form ${\rm Im}(\cdot,\cdot)$ on $\Hi_{1}$.  Eq.  \eqref{standard} means that the
complex subspace, spanned by $K(W_0)$ together with the eigenspace
$iK(W_{0})$ of $S_{W_0}$ to eigenvalue $-1$ is dense\footnote{The
complex subspace $K(W_0)+iK(W_0)$ is closed in the graph norm
associated with the Tomita operator $S_{W_0}$.  Its denseness in the
Wigner norm is a one-particle version of the Reeh-Schlieder theorem.} in 
$\Hi_1$. 
This property and the absence of nontrivial vectors in the
intersection of the two real spaces means that $K(W_{0})$ is a
real {\it standard subspace}
in the sense of \cite{RvD}.  Conversely, a real standard subspace
$K$ determines uniquely a Tomita involution $S$ (generally not related
to group representation theory) with domain $K+iK$, defined by

$S(\psi+i\varphi):=\psi-i\varphi$ for $\psi,\varphi\in K$. Its polar 
decomposition then leads to an anti-unitary involution $J$ with $JK=K^\bot$ 
and a unitary group $\Delta^{it}$ leaving $K$ invariant.
The application of Poincar\'{e} transformations to the reference space
$K(W_0)$ generates a family of wedge spaces
$K(W)=U_1(a,\Lambda)K(W_0)$ if $W=(a,\Lambda)W_0$, with corresponding
Tomita involutions $S_{W}$.  
(The definition is consistent because every Poincar\'e transformation which 
leaves $W_0$ invariant commutes with $\Boo{W_0}{t}$ and $j_{W_0}$, cf.~\cite{BGL}.) 
There is an equivalent view on the construction of his family, which we introduce 
here for later reference. Namely, one associates to a wedge $W=gW_0$, $g\in\Po$, 
the boosts 
$\Boo{W}{t}:= g \,\Boo{W_0}{t}\,g^{-1}$ and reflection $j_W:= g\, j_{W_0} \, g^{-1}$. 
Then the operators $\Delta_W$, $J_W$ and $S_W$ are defined as in eqs.~\eqref{eqModOp} and \eqref{eqTom} with $W_0$ replaced by $W$. Note that 
in particular, by \eqref{duality}, 
\begin{equation}K(W')=K(W)^\bot.\label{duality1}\end{equation}
    
    The above scheme applies also to 
ray representations of the Poincar\'e group corresponding the 
half-integral spin 
 \cite{FS02} but \eqref{duality} generalizes to
 \begin{equation}K(W)^{\bot} =ZK(W^{\prime})\label{twist}\end{equation}
where the ``twist'' operator $Z$ satisfies $Z^2=-1$. An
interesting situation arises if the spin $s$ is not half-integer, as it happens in
$d=1+2$ dimensions for anyons \cite{M02a}. In that case the spin-statistics factor 
$Z^{2}=e^{i2\pi s}$ is not a real number.

\paragraph{Step 2.} 
A sharpening of the  localization is obtained by intersecting 
the localization spaces for wedges, defining for any causally closed 
region $\mathcal O$ contained in some wedge
\begin{equation}
K(\mathcal{O}):=\cap_{W\supset\mathcal{O}}K(W).\label{int}%
\end{equation}
The crucial question is whether these spaces are standard. 
According to an important
theorem of Brunetti, Guido and Longo \cite{BGL} standardness holds, for all irreducible positive
energy representations of the proper Poincar\'e group,
if $\mathcal{O}$ is a 
{\it space-like cone}, i.e. a set of the form
\begin{equation}\spc=a+\cup_{\lambda\geq0}\lambda D\end{equation} where 
$a$ (the apex of the cone) is a point in Minkowski space and $D$ 
is a double cone,
space-like separated from the origin. The double cone regions $D$ are 
conveniently envisaged as intersections
of a forward light cone with a backward cone whose apex is inside the 
forward cone.

The resulting family $\spc\rightarrow { K}(\spc)$ of closed real
subspaces  of $\calH_1$, indexed by the set $\Spc$ of space-like cones $\spc$, 
has the 
following properties: 
\\
1. {\em Isotony:} 
If  $\spc_1 \subset\spc_2$, then 
\[ { K}(\spc_1)\subset{ K}(\spc_2) \;.  \]
2. {\em Locality:} 
If $\spc_1$ is causally separated from $\spc_2$, i.e., $(x-y)^2<0$ 
for $x\in C_{1}$, $y\in C_{2}$, then 
\begin{align}  \label{eqLoc} 
{K}(\spc_1)\:\subset\: { K}(\spc_2)^\bot\;. 
\end{align}
3. {\em Poincar\'e covariance:} 
For all $\spc$ and $g\in\mathcal P^\uparrow_{+}$ 
\begin{align*}
 U_1(\po)\:{ K}(\spc)& \:= \: {K}(\po\spc) \;.
  \end{align*}
4. 
{\em Standardness:}  ${K}(\spc)$ is standard for all
space-like cones $\spc$. 
\\ 
It is a remarkable fact that these properties, plus Haag Duality~\eqref{duality1} 
for space-like cones $C$, uniquely characterize the family $C\to K(C)$ constructed 
above in the massive case. 
This follows from the algebraic Bisognano-Wichmann theorem~\cite{M01a}, whose proof 
uses precisely the above properties.
\\
In case of the finite spin/helicity
representations standardness also holds if $\mathcal O$ is an
(arbitrary small) double cone.

The constructive clout of modular localization is revealed in two 
applications, the first of which is the basis of our construction of 
string-localized fields. 
\paragraph{Application 1: Construction of interaction-free algebraic nets 
\cite{BGL, FS02}.}

Given a family of real subspaces ${}K(\mathcal{O})\subset \Hi_1$ as 
defined by \eqref{int}, 
one can apply the
CCR (Weyl) respectively CAR second quantization functor to obtain a covariant
$\mathcal O$-indexed net of von Neumann algebras $\mathcal{A}(\mathcal{O})$ 
acting on the Fock space $\Hi=\mathcal F(\Hi_1)$ built 
over $\Hi_1$. For
integer spin/helicity values~\cite{FS02} the
modular localization in Wigner space  implies the identification of 
the symplectic
complement with the complement in the sense of relativistic
causality, i.e. $K(\mathcal{O})^{\perp}=K(\mathcal{O}^{\prime})$ (spatial Haag
duality). The Weyl functor takes the spatial version of Haag duality
into its algebraic counterpart. One proceeds as follows: For each Wigner wave
function $\psi\in \Hi_1$ the associated (unitary) Weyl operator is defined as%
\begin{equation}
{\rm Weyl}(\psi):=\exp i\left\{  a^{\ast}(\psi)+a(\psi)\right\}  ,\text{ }%
{\rm Weyl}(\psi)\in {\mathcal B}(\Hi)
\end{equation}
where $a^{\ast}(\psi)$ and $a(\psi)$ are the usual creation and 
annihilation operators on Fock space.  We then define the von Neumann algebra corresponding to the localization
region $\mathcal{O}$ in terms of the operator algebra generated by the image
of the localized subspace $K(\mathcal{O})$%
\begin{align}
\mathcal{A(O)}  &  := 
\left\{ {\rm Weyl}(\psi)|\,\,\psi\in K(\mathcal{O})\right\}  ^{\prime\prime
}.\nonumber
\end{align}
(By the von Neumann double commutant theorem, our generated operator algebra is weakly closed by 
definition.)  The functorial 
relation between real subspaces and von Neumann algebras preserves the causal
localization structure and commutes with the improvement of localization
through intersections ($\cap$) according to $K(\mathcal{O})=\cap
_{W\supset\mathcal{O}}K(W),\,\mathcal{A(O)=}\cap_{W\supset\mathcal{O}%
}\mathcal{A}(W)$ as expressed in the commuting diagram  \cite{Schroer
AOP}
\begin{equation}%
\begin{array}
[c]{ccc}%
\big\{K(W)\big\}_{W} & \longrightarrow & 
\big\{\mathcal{A}(W)\big\}_{W}\\
\downarrow{ \cap} &  & \downarrow{ \cap}\\
K(\mathcal{O})_{{}} & \longrightarrow & \mathcal{A(O)}%
\end{array}
\label{square}%
\end{equation}
where the vertical arrows denote the tightening of localization by
intersection and the horizontal denote the action of the Weyl functor.
The case of half-integer spin representations is analogous \cite{M01a,FS02}. 
The only significant difference is the mismatch between the causal 
and symplectic complements that is taken care of by the twist 
operator $Z$, cf.\eqref{twist}. 

It is important to note that while the spaces $K(W)$ for wedges are
uniquely determined in the {\it one-particle} space by the
representation of the Poincar\'e group alone, also in the presence of
interaction \cite{M02a}, this is in general no longer so for the space
${\mathcal K}(W)=\mathcal A(W)^{\text{sa}}\Omega$ 
generated by the wedge algebra in the {\it  whole}
Hilbert space.  In fact, the Tomita involution associated with 
${\mathcal K}(W)$
involves, besides the Lorentz boosts and rotations, the PCT operator
and hence the scattering matrix.  The PCT operators for the in- and
out-fields of an interacting field differ, despite the fact that both transform
with respect to the same representation of the orthochronous, proper
Poincar\'e group and their PCT operators coincide on the
one-particle space.

The scheme of passing from particle- to field- localization described 
above works in particular for
Wigner's infinite spin representations; one only must be aware that in this
case one cannot achieve a better localization than that in space-like cones
since the infinitely many degrees of freedom coming from the faithful
representation of the little group do not allow a compact localization. The
generating fields in this case are operator-valued distributions supported on
semi-infinite strings (the cores of space-like cones) and their construction
and derivation of their properties constitutes the main content of the present work.

A different mechanism which leads to string localization is that of 
$d=1+2$
``anyons'' i.e., Wigner representations with anomalous spin which activates the
universal, instead of the standard two-fold, 
covering group of the Lorentz group. 
Such a generalization leads to a
spin-statistics situation characterized by a complex modification of the
spatial Haag duality $K(\mathcal{O})^{\perp}=ZK(\mathcal{O}^{\prime}%
),\,Z^{2}=e^{2\pi is}$. This requires string-localization, but contrary to the
previous case the passing from the spatial to the algebraic setting cannot be
done in a functorial way 
even if no genuine physical
interaction is present \cite{M}. Since the methods of construction of
localized operator algebras are significantly different from the present ones,
this matter will be pursued in a separate work.
%
\paragraph{Application 2: Partial results on constructive aspects of
modular localization in presence of interaction \cite{SchroerAOP, BBS}.}

In presence of interactions there do not exist any compactly localized
operators which create a one-particle state without a vacuum
polarization admixture when acting on the vacuum (`polarization free
generators' (PFG)).  It comes therefore as a pleasant surprise that
the first line of the commuting square \eqref{square} remains intact
in the following sense: modular theory secures the existence of
wedge-localized PFG which are unbounded operators affiliated to the
algebra $\mathcal{A}(W)$ \cite{BBS}$.$ In physical-intuitive terms:
wedge localization is the best compromise between field states and
Wigner particle states.

Wedge localized PFG that  are operator-valued distributions on a
translation-invariant dense domain (`tempered PFG') are in
more than two space-time dimensions only compatible with
trivial scattering, but in $d=1+1$ they lead
precisely to the Zamolodchikov-Faddev (Z-F) algebra setting for
factorizing models \cite{BBS,SchroerAOP}.  This observation
brings a wealth of new insights: (1) It attributes a space-time
interpretation to the hitherto rather abstract auxiliary Z-F algebra
(which extends the creation/annihilation operators of free theories
without affecting their ``on-shell nature'').  (2) It decouples the
bootstrap-formfactor program for factorizing models from the
quantization of classically integrable systems (the necessity to find
a complete system of infinitely many conserved anomaly-free currents)
and replaces the recipes of that program by derivations of its rules
from first principles of general QFT using modular theory.  (3) It
strengthens the idea that there is nothing intrinsic about the
ultraviolet problems of the standard approach; they are simply the
unavoidable price to pay if one enters QFT via the classical parallelism 
referred to as quantization (which worked so well for passing from mechanics to
quantum mechanics).  Whereas intrinsic formulations of QFT which avoid
singular generators have been known for several decades, it is only
the recent progress of modular localization which is opening an avenue
for new constructions.  The ideal situation would be to be able to
construct QFT in analogy to  what has been done in factorizing models~\cite{crossing}, 
namely in terms of a two-step process in which the first step
consists in constructing generators of wedge-localized algebras and
the second step in tightening localization by intersecting wedge
algebras as it was already successfully achieved for the factorizing
models.  Only such an approach is capable of revealing the true
frontiers of QFT beyond those generated by the use of singular field
coordinatizations.

Our study of string-localized fields in this paper is a less ambitious
step in this direction: Free string-localized fields $\varphi(x,e)$
are less singular than point-like fields $\varphi(x)$ because
intuitively speaking they transfer part of the quantum fluctuations to
fluctuation in the space-like string direction $e$ so that the
power-counting allows more possibilities. As a matter of fact the
short-distance behavior of free string-localized fields does not
become worse with increasing spin and there is no clash any more
between quantum physics and the technical necessity to use
(point-like) vector-potentials for photons since the physical photon
space supports stringlike-localized vector potentials. It is expected
that their use in suitably defined interactions will lead to a more
complete understanding why in QFT the renormalizability requirement
alone determines a unique interaction for vector particles~\cite{DS}
and the role of the gauge formalism is at best to facilitate its
construction. This is different from classical field theory where
there are many possible interactions involving classical vector
potentials and one needs the gauge principle in order to select the
Maxwellian one.
\\

An interesting but difficult question is whether modular localization has
directly verifiable observable consequences.  Clearly, the states in $K(\mathcal O)$ 
are in general not strictly localized in $\mathcal O$ in the sense of \cite{Licht}, 
i.e.\ giving the same expectation values as the vacuum state for observables localized 
outside the region $\calO$.  (The strictly localized states for some region do not form a linear space.) 
However, they should practically look like the vacuum state for observables localized 
outside $\calO$. We can substantiate this quantitatively in the case of a free field. 
In this case, let $\phi$ be a single particle vector in $K(\calO)+i K(\calO)$. Then the 
deviation of the corresponding state $\omega=(\phi,\cdot \,\phi)$ from the vacuum state 
$\omega_0$ is dominated by the vacuum fluctuations for observables localized outside $\calO$. 
More precisely, there is a constant $c>0$ such that 
\begin{equation} \label{eqVacFluc} 
|\omega(A)-\omega_0(A)| \leq c \,(\Delta A)_{\omega_0} 
\end{equation}
for all self-adjoint $A\in\calA(\calO')$.  Here, $(\Delta A)_{\omega_0}$ denotes the vacuum 
fluctuation, $(\Delta A)_{\omega_0}^2 = \omega_0(A^2)-\omega_0(A)^2$. (This can be shown by the methods used 
in~\cite[Lemma 4.1]{BrBu94})
The important point is that the physical significance (in contrast to the mathematical definition) 
of the modular localized subspaces is {\em not} intrinsic to 
the single particle theory or the representation of the Poincar\'e group, but relates to the 
local observables of an underlying quantum field theory. This also holds outside the realm of free fields. \\

As a last point in this section we return to the comparison of the
modular, or field theoretical, localization in relativistic QFT and
localization in terms of projection operators as in non-relativistic
many-body quantum mechanics.  Characteristic for the latter is that
the algebra of observables can be written as the tensor product of two
type I factors, corresponding respectively to observables localized (at
a fixed time) in a spatial region and in its complement\footnote{In
Fock space, these are the algebras generated by creation and
annihilation operators for wave functions with support in complementary
regions in ${\mathbb R}^{d-1}$.}.  In relativistic quantum field
theory the local algebras ${\mathcal A}({\mathcal O})$ are type III
(physically, this can ultimately be attributed to vacuum fluctuations)
and can therefore not be regarded as factors in a tensor product.  On
the other hand, the split property \cite{DL} mentioned earlier
goes a long way towards recovering the quantum mechanical picture.  By
this property (which is a consequence of very reasonable bounds on
phase space degrees of freedom \cite{BW, BDF, Bu-Yng_nucl}) local
observable algebras separated by a positive security distance can be
regarded as sub-algebras of commuting type I factors.  Pictorially,
each type I algebra can be thought of as the algebra of a sharply
localized core region, augmented by a ``halo'' in the security region
where the localization is ``fuzzy''.  The minimal projectors in the
type I algebra can be regarded as QFT analogs of localizing projectors
but there is no concept corresponding to an $x$-space probability
density \`a la Born.  Another difference is that the vacuum restricted
to the split type I factor is not a pure state but rather a thermal
equilibrium state at temperature $(2\pi)^{-1}$ with respect to a
\lq modular Hamiltonian\rq\ which is determined by the canonical split
construction \cite{DL}.

The factorization in transverse direction to light-like regions and the
split property have recently played a crucial role in the QFT
formulation of "hologaphic projection" which repairs the loose ends of
the old "light-cone quantization" and converts some of the underlying
ideas into valuable constructive instruments of rigorous local quantum
physics~\cite{crossing}. All algebraic modular localization results
mentioned in this paper (including the split property) have a spatial
counterpart in the Wigner representation theoretical setting and can
(for free fields) 
be obtained by a functorial construction from the latter.

\section{General Construction and Uniqueness of the Fields}
 \label{intertwiner}
We want to sketch the construction of string--localized
fields, and discuss the question to what extent they are 
fixed by our assumptions. Exploiting the fact that free fields are
fixed by the single particle states which they create (by a generalization~\cite{St} 
of the Jost-Schroer theorem~\cite{SW}), this is reduced to the construction of certain intertwiners 
and the question of their uniqueness. 

To set the stage, let us recall the irreducible unitary positive-energy  
representations $\Uirr$  of the Poincar\'e group  $\Po$ in $d$-dimensional 
Minkowski space, $d=3,4$. Namely, $\Uirr$ is determined by the mass value $m\geq0$ and an 
irreducible unitary representation $D$ of a certain subgroup $G$ of
$\Po$,  the so--called little group. It is the stabilizer subgroup in
$\Lor$ of a fixed vector $\bar p$ in the mass shell for $m\geq0$, 
$$
\Hyp:=\{ p \in\RR^d:\; p\cdot p=m^2,\,p^0>0 \}.   
$$
If $m>0$, then $G$ is the rotation subgroup, and if $m=0$ then 
$G$ is isomorphic to the euclidean group in $d-2$ dimensions. 
(If in this case $D$ is faithful, the resulting representation $\Uirr$ is a
so--called ``infinite spin'' representation. If $D$ is not faithful  
but non-trivial, we shall speak of a ``helicity representation''.)  
The representation $\Uirr$ is induced by $D$ as follows. 
The representation space is $\Hirr:=L^2(\Hyp,d\mu;\calh)$, where $d\mu$ be the
Lorentz invariant measure on $\Hyp$ and $\calh$ is the representation space of $D$. 
On this Hilbert space, $\Uirr$ acts according to 
\begin{equation} \label{eqUirr} 
\big(\Uirr(a,\lor)\psi\big) (p) = e^{ia\cdot p}\,D(\WigRot(\lor,p))\,\psi(\lor^{-1}p)\,.
\end{equation}
Here $\WigRot(\lor,p)\in G$ is the so--called Wigner rotation, defined by 
\begin{equation}\label{eqWigRot} 
 R(\lor,p) :=  B_p^{-1}\,\lor\;B_{\lor^{-1}p}, 
\end{equation} 
where for almost all $p\in\Hyp$,  $B_p$ is a Lorentz transformation which maps $\bar{p}$ to $p$. We will denote the set of $p$ for which $B_p$ is defined by $\HypMinus$. 

Each of the considered representations extends to a representation of the
proper Poincar\'e group $\Poj$ as follows.\footnote{$\Uirr(\Poj$) acts in
the same Hilbert space as  $\Uirr(\Po$), 
except for $m=0$ and {\em finite} helicity, where the Hilbert space
has to be doubled: One has to take the
direct sum of representations for helicity $n$ and $-n$, and $\Uirr(j_0)$,
defined in Eq.~\eqref{eqUj}, also flips $n\leftrightarrow -n$.}  
Let $j_0$ be the reflection at the edge of the standard wedge
$W_0$, cf.~\eqref{eqW0}. Choosing $\bar p$ invariant under $-j_0$, the
adjoint action of $j_0$ leaves $G$ invariant,   
and $D$ extends to a representation of the subgroup of $\calL_+$ generated by $G$
and $j_0$ by an anti--unitary involution $D(j_0)$. 
(Explicit expressions for these representers will be given in the relevant cases later on.)  
One now defines an anti-unitary involution $\Uirr(j_0)$ by 
\begin{align} \label{eqUj} 
 \big(\Uirr(j_0)\psi\big) (p) :=  D(j_0) \, \psi(-j_0p).
\end{align}
If the family  $B_p$, $p\in\HypMinus$, is chosen so that 
\begin{equation}\label{eqjBpj} 
 j_0B_pj_0=B_{-j_0p},
\end{equation} 
then one checks that $\Uirr(j_0)$ extends $\Uirr$ to an (anti-) unitary  representation of $\Poj$ (which is generated by $\Po$ and $j_0$). 

For later reference, we fix some notations concerning the manifold $\Spd$ of space-like directions, cf.\ \eqref{eqSpd}. 
The Poincar\'e group acts on $\Spd$ by letting the translations act trivially, i.e.\  
\begin{align} 
g \,e&:= \Lambda e &  \text{ if } g&=(a,\Lambda)\in\Po, \label{eqPoSpd}\\
j \,e&:= \Lambda\,j_0\,\Lambda^{-1}e & \text{ if }  j&=(a,\Lambda)\,j_0\,(a,\Lambda)^{-1}\in \calP^\downarrow_+. 
\label{eqjSpd} 
\end{align}
Similarly, one gets wedge regions $W_H$ in $\Spd$ arising from Minkowski wedges $W$ as follows. 
For $W = (a,\Lambda)W_0$, we define 
\begin{equation} \label{eqWH}
W_H:= \Lambda W_0 \,\cap \Spd. 
\end{equation} 
($\Lambda W_0$ is the wedge which arises from $W$ by translation and contains the origin in its
edge.) 

\subsection{The Concept of Intertwiners; Uniqueness} 
The concrete formula \eqref{eqUirr} for $\Uirr$ 
(the realization of the representation in a ``Wigner base'') is not
suitable as it stands for the construction of covariant, local fields
by second quantization. The problem is twofold: The transformation
matrix $D(R(\lambda,p))$ depends on $p$ (except in the scalar case),
which leads to a nonlocal transformation law in $x$ upon Fourier
transformation, and the Wigner rotation factors have singularities 
which cause problems  with
local commutativity.  In the standard setting of point-localized
fields this difficulty is overcome by replacing the ``Wigner
bases'' by  ``covariant bases''. 
This is achieved 
with the help of so--called intertwiner functions which intertwine the representer of 
the Wigner rotation factor $R(\Lambda,p)$ with a representer of $\Lambda$ 
and lead to the well-known formulas for the point-like free fields (\cite{Wei-book}, see also our Section~\ref{StringVsPoint}).  
Here, we consider a new solution, which in contrast to the mentioned one 
works also for the massless infinite spin representations 
{which remained outside the covariant spinorial formalism}. Namely,  
our string-localized fields will be constructed with the help of intertwiner 
functions $\intfct{\spd}{\cdot}$ which depend on the points $e$ in the set $\Spd$ 
of space-like directions, and absorb the Wigner rotation factor $R(\Lambda,p)$, trading 
it with a transformation $e\to \Lambda e$.  

We now define these intertwiner functions in detail.  
Let $\Spdc$ be the complexification of $\Spd$, 
\begin{align} \label{eqSpdc}
\Spdc  & :=  \{ \spdc \in\CC^d,\; \spd\cdot \spd = -1  \},   
\end{align} 
where the dot denotes bilinear extension of the Minkowski metric to $\CC^d$, 
\begin{align} \label{eqScalProdC}
\spd\cdot \spd := \spdr\cdot\spdr-\spdi\cdot\spdi + 2i\,\spdr\cdot\spdi
\quad \text{ if}\quad \spdc=\spdr+i\spdi. 
\end{align} 
Let further $\Tub$ be the tuboid consisting of all $\spdc=\spdr+i\spdi\in\Spdc$ such that
$\spdi$ is in the interior of the forward light cone (in $\RR^d$). 
We will consider 
subsets $\Theta$ of $\Tub$ of the form 
\begin{equation} \label{eqTheta}
\Theta= \Spdc\cap \big(\Omega_1+i\RR^+\Omega_2\big),
\end{equation} 
where $\Omega_1$ and $\Omega_2$ are compact subsets of $\RR^d$ and
$\Omega_2$ is contained in the  forward light cone.
Note that $\Theta$ is bounded due to the condition $e''\cdot e''\leq1$ for $e=e'+ie''\in\Tub$, and its compact closure is given by $\Theta\cup (\Omega_1\cap\Spd)$.
\begin{Def}[Intertwiners] \label{DefIntFct}
A function $u:\Tub\times \HypMinus\to \calh$ is called an
intertwiner function for $D$ if it satisfies the following conditions. 
Firstly, it has the ``intertwiner property'' 
\begin{align} \label{equCov}
D(\WigRot(\lor,p))\,  u(\lor^{-1}e,\lor^{-1}p) =  u( e,p) 
\end{align} 
for $(e,p)\in\Tub\times \HypMinus$ and $\lor\in\Lor$. 
Secondly,  for almost all $p$ the  function $e \mapsto  u(e,p)$
is analytic on the tuboid $\Tub$. Finally, the following bound is satisfied. 
There is a constant $N\in\Bn_0$ and a function $M$ on $\HypMinus$ which is locally $L^2$ w.r.t.\ $d\mu$ 
and polynomially bounded,  
and for every $\Theta\subset\Tub$ of the form indicated in Eq.~\eqref{eqTheta}, there is  
a constant $c=c_\Theta$ such that for all $e\in\Theta$ holds 
\begin{equation} \label{eqModGrowth}
\|u(e,p)\| \leq c \, M(p) \, |e''|^{-N}. 
\end{equation}
Here, $|e''|$ denotes any norm in $\RR^{d}$ {and the norm of u refers to the little Hilbert space $\calh$}.  
\end{Def}
\begin{Remarks}
1. If the growth order $N$ of $e\rightarrow u(e,p)$ in
\eqref{eqModGrowth} is zero, then the function 
$e\mapsto u(e,p)$ has a unique extension to the real boundary $\Spd$ 
as a weakly continuous $\calh$-valued function  
which we denote by the  same symbol.

2. Given $u$, we define the {\it conjugate intertwiner} 
\begin{align}  
 \intfctconj{e}{p} := D(j_0) \, u(j_0e,-j_0p). \label{eqv} 
\end{align}
It transforms as in Eq.~\eqref{equCov} and satisfies the
bounds~\eqref{eqModGrowth}, and is anti-analytic in 
$-\Tub$. It is noteworthy that we find ``self-conjugate'' intertwiners in all cases. 
\end{Remarks}
The bound \eqref{eqModGrowth} is chosen so that for fixed $p$ the function
$\spdc\mapsto u({\spdc},{p})$ is  of moderate growth near the ``real boundary'' $\Spd$ in
the sense of \cite{BrosMos} and therefore admits a distributional boundary
value in $\calD'(\Spd)$. In particular, for every $h\in\calD(\Spd)$, the 
(weak) integral 
\begin{align} \label{eqIntDist} 
u({h},{p}):= \int d\sigma(e) \,h(e) \, u({e},{p}), 
\end{align}
where $\sigma$ is the Lorentz invariant measure on $\Spd$, 
can be defined by letting the argument $\spd$ of $u({\spd},{p})$
approach $\Spd$ from $\Spdc$ inside the tuboid $\Tub$ after the integration,
cf.~\cite[Thm.\ A.2]{BrosMos}. 

These smeared intertwiner functions give rise to a family of single particle vectors 
which behave covariantly and are modular--localized in 
``truncated space-like cones''(for $N>0$),  or in  ``space-like half-cylinders'' (for $N=0$). By a {\em truncated space-like cone}, we mean a region in Minkowski space of the form 
$\calO+\RR_0^+\Omega$, where $\calO$ and $\Omega$\footnote{\label{HDoKe}$\Omega$ must be small enough, namely contained in some wedge.} are bounded subsets of $\RR^{d}$ and $\Spd$.    By a {\em space-like half cylinder} we mean a region in Minkowski space of the form 
$\calO+\RR_0^+ e$, with $e\in\Spd$. 
These single particle vectors are constructed as follows. 
For $f\in \calS(\RR^{d})$ and $h\in\calD(\Spd)$, we define 
$\psi({f},{h})$ and $\psi_c({f},{h})$ as 
\begin{align} 
\psi({f},{h})(p) &:= \FT{f}(p) \, u(h,p) , \label{eqCovFct'}\\
\psi_c({f},{h})(p) &:= \FT{f}(p) \, \intfctconj{h}{p}, \label{eqCovFctConj'} 
\end{align} 
$p\in\Hyp$, where $\FT{f}$ is the restriction to the mass shell of the Fourier transform of $f$.  
If the growth order $N$ of $e\mapsto u(e,p)$ in \eqref{eqModGrowth} is
zero, we define $\psi(f,e)$ and $\psi_c(f,e)$ analogously without
the smearing with $h(\spd)$. 
The bound~\eqref{eqModGrowth} makes sure that the $\calh$-valued functions $\psi(f,h)$ etc.\ on $\Hyp$ 
are in $L^2$, hence in $\Hirr$. 
The intertwiner property \eqref{equCov} implies that these single particle vectors behave covariantly under the Poincar\'e group. 
Most importantly, the analyticity of $u$, together with the bound \eqref{eqModGrowth}, implies that 
$\psi(f,h)$ is modular--localized in the truncated space-like cone $\supp f + \RR_0^+ \supp h$.  
The idea behind this assertion is as follows. Let $\strip$ denote the strip 
\begin{align}   \label{eqStrip}
\strip := \RR+i(0,\pi) 
\end{align} 
and $\strip^\clo$  its closure, $\strip^\clo:=\RR+i[0,\pi]$. 
It is known that for $e\in W_H$, the map $z\mapsto
\Boo{W}{z}e$ is analytic and has values in $\Tub$ for $z\in\strip$, cf.~Eq.~\eqref{eqBooW}.\footnote{$W_H$ has  been defined in \eqref{eqWH} and $\Boo{W}{z}e$ refers to the action of $\Po$ on $\Spd$ defined in \eqref{eqPoSpd}.} 
Hence for $e\in W_H$, the function  
$z\mapsto\intfct{\Boo{W}{z}e}{p}$ is  analytic on $\strip$. 
The bound~\eqref{eqModGrowth} ensures that the same holds for $\intdist{\Boo{W}{z}_*h}{p}$ if $\supp h\subset W_H$, 
and that the latter is polynomially bounded for large $p$. This implies that $\psi(f,h)$ is in the domain of the operator $S_W$ whenever $W$ contains $\supp f$ and $W_H$ (or its closure) contains $\supp h$ or, equivalently, whenever $W$ contains the truncated space-like cone $\supp f + \RR_0^+ \supp h$. 
%
The details are spelled out in Appendix~\ref{Proofs}. We get the following result. 
\begin{Prop}[Properties of single particle state vectors] \label{Cov}
Let $u(e,p)$ be an intertwiner function as in
Definition~\ref{DefIntFct}, and $\psi(f,h)$, $\psi_c(f,h)$  be defined as in
\eqref{eqCovFct'}, \eqref{eqCovFctConj'}. 

0) For $f\in\calS(\RR^d)$ and $h\in\calD(\Spd)$, the state vector 
$\psi({f},{h})$ is in $\Hirr=L^2(\Hyp,\d\mu)\otimes \calh$. 
Furthermore, a single particle version of the Reeh-Schlieder theorem holds: 
Let $\calO$ and ${\mathcal U}$ be arbitrary open sets in $\RR^d$ and $\Spd$, respectively. Then the linear span of $\psi(f,h)$, $\supp f\subset \calO$, $\supp h\subset {\mathcal U}$, is dense in the single particle space. 

i) The family transforms covariantly under $\Uirr$:  
\begin{align}  \label{eqPsiCov} 
\Uirr(g) \psi({f},{h})&=\psi(g_*f,g_*h) \,, \quad g\in\Po ,\\
\Uirr(j) \psi({f},{h})&=\psi_c(j_*\bar{f},j_*\bar{h}),  \quad j\in
P^{\downarrow}_+,   \label{eqPsiCovj} 
\end{align} 
where  $g_*$ denotes the push--forward, 
$(g_*f)(x)=f(g^{-1}x)$.\footnote{Here, $ge$ and $je$, for $e\in\Spd$,  is meant as in equations~\eqref{eqPoSpd} and \eqref{eqjSpd}.}   
The same holds for $\psi_c$. 

ii) 
Let $f$ be a smooth function on $\RR^d$ with support in a double cone 
$\calO$, and let $h$ be a smooth function on $\Spd$ with support in a 
compact set $\Omega$.\footnote{$\Omega$ must be small enough, cf.\ footnote~\ref{HDoKe}.} 
Then $\psi({f},{h})$ is localized in the truncated space-like cone  
$\calO+\RR_0^+\Omega$. More precisely, if $W$ contains $\calO+\RR_0^+\Omega$, then it is in the domain of $S_W$, and 
\begin{align} 
S_W \psi({f},{h}) &=\psi_c(\bar{f},\bar{h}) \,.\label{eqSPsi} 
\end{align}

iii) If the growth order $N$ of $e\mapsto u(e,p)$ in \eqref{eqModGrowth} is
zero, then 0) and i) hold with $h$ replaced by $e$ and $g_*h$ replaced by $ge$. 
Moreover, if $\supp f\in\calO$ then 
$\psi({f},{\spd})$ is localized in the space-like half cylinder $\calO+\RR_0^+e$, and 
$S_W \psi({f},{\spd}) =\psi_c(\bar{f},e)$ whenever $W$ contains $\calO+\RR_0^+e$. 
\end{Prop} 
Note that $ii)$ is equivalent to 
\begin{equation}  \label{eqPsifhLoc} 
\psi({f},{h})+\psi_c(\bar{f},\bar{h}) \,\in\, \Kirr(\calO+\RR_0^+\Omega).
\end{equation}
For the self-conjugate intertwiners which we find in the subsequent sections, $\psi({f},{h})$ is in 
$\Kirr(\calO+\RR_0^+\Omega)$ for real valued $f$, $h$. 
The proof of the proposition is given in Appendix~\ref{Proofs}. 

Second quantization then yields a string-localized free quantum field. We also assert the converse, namely that every string-localized free quantum field arises this way, and we also discuss to what extent the corresponding intertwiner functions are unique. 
Let $a^*(\psi)$ and $a(\psi)$,  $\psi\in\Hirr$, denote the
creation and annihilation operators in the bosonic Fock space over
$\Hirr$. We shall write symbolically\footnote{Note that $\psi(p)\in \calh$, and ``$\cdot$'' stands for the contraction over the indices of a basis of $\calh$} 
$$
a^*(\psi)=: \int_{\Hyp} d\mu(p)\,\psi(p)\cdot a^*(p)\quad\text{and}\quad
a(\psi)=: \int_{\Hyp} d\mu(p)\,\overline{\psi(p)}\cdot a(p). 
$$ 
Our results are summarized in the following theorem, which is valid for all bosonic 
particle types except those corresponding to the helicity representations. 
%
\begin{Thm}[Existence and uniqueness of string 
fields]  \label{StringField}
Let $\Uirr$ be any irreducible positive energy representation of the 
Poincar\'e group with faithful or trivial representation 
of the little group. 

i) Let $u$ be an intertwiner function for $D$, and let $u_c$ be defined as above.  
Then the field $\field(x,e)$ defined by 
\begin{align} \label{eqField} 
\field(x,\spd) := \int_{\Hyp}\d\mu(p)\left\{  \;
e^{ip\cdot x}  \; u(e,p) \cdot a^*(p) 
+ e^{-ip\cdot x} \; \overline{\intfctconj{e}{p}} \cdot a(p)\;  \right\}   
\end{align}
satisfies our requirements \eqref{eqFieldLoc} and \eqref{eqFieldCov}. 
It further satisfies the Reeh-Schlieder and Bisognano-Wichmann
properties (see below).  Moreover, if the growth order $N$ of
$e\mapsto u(e,p)$ in \eqref{eqModGrowth} is zero, then the field is a
{\em function} in $e$, and the commutativity~\eqref{eqFieldLoc}
already holds if $x_1+\RR_0^+e_1$ is space-like separated from
$x_2+\RR_0^+e_2$.\footnote{By Lemma~\ref{StringWedge}, this is
compatible with $e_1,e_2$ light-like separated or identical, in
contrast to the general case, c.f.~Footnote~\ref{StringSep}.  In
particular, this case covers the scenario envisaged by Steinmann in
\cite{St}, where all string directions $e$ coincide.}
 
ii) A non--trivial intertwiner  function $u$ with these properties
exists for all mentioned representations. It is unique up to multiplication with a
function of $e\cdot p$, which is meromorphic in the upper half plane. 
(That is to say, if $\hat u$ is another intertwiner  function, then
for almost all $e\in\Tub$ and almost all $p$, 
\begin{align}  
  u(e,p) = F(e\cdot p) \, \hat u(e,p), \label{equhatu}  
\end{align} 
where $F$ is a numerical function, meromorphic on the complex upper half 
plane.) 

iii) Conversely, let $\field$ be a string-localized field, in the sense of \eqref{eqFieldLoc} 
and \eqref{eqFieldCov}, which is free in the sense that it creates only single particle vectors 
from the vacuum, and which satisfies in addition the 
Bisognano-Wichmann property\footnote{In the massive case, the Bisognano-Wichmann 
property is not an extra assumption, cf.\cite{BiWi,M01a}.}. Then it is 
of the form \eqref{eqField} up to unitary equivalence, 
with $u$ as in Definition~\ref{DefIntFct} and with $u_c$ as in~\eqref{eqv}. 
Further, it satisfies a ``PCT theorem'', namely, it is covariant under reflections: 
\begin{equation} \label{eqPCT}
 U(j) \field(x,e) U(j)^{-1} = \field(jx,je)^*,  \quad j\in\Poj^{\downarrow}.  
\end{equation}
\end{Thm} 
By the {\it Reeh-Schlieder property} we mean that products of the fields already generate 
a dense set from the vacuum when smeared within arbitrary, fixed, open sets 
$\calO\in\RR^d$ and $U\in\Spd$.  
The {\it Bisognano-Wichmann property} means that the modular group $\Delta_W^{it}$ 
and modular conjugation $J_W$ of the algebra associated to a wedge $W$ 
coincides with the representers of the boosts $\Boo{W}{t}$ and reflection 
$j_W$ associated to $W$, respectively. 
\begin{Remarks}
1. Clearly, one can construct new intertwiner functions from given ones via Eq.~\eqref{equhatu}, with $F$ analytic on the upper half plane, polynomially bounded at infinity and of moderate growth near the reals. 
The corresponding fields all belong to the same Borchers
class, i.e., they are relatively string-localized with respect to each other.

2. The intertwiner functions appearing in the theorem are more precisely speaking ``scalar'' string-localized interwiners, i.e.\ functions $u(p,e)$ which have no explicit vector (tensor) index. We show in Section~\ref{Photons} 
that for massless helicity representations with helicity $n\neq0$ no such intertwiners exist. 
This and an analogous situation for higher helicity mass zero representations is the reason for our exception of these representations from the theorem. If one wants to describe photons (corresponding to the direct sum of the 
representations with helicity $\pm 1$) in terms of vector potentials instead of field strengths then one has to admit a vector index. 
Similarly, string-localized intertwiners for fermions need a spinor index. 
\end{Remarks}
A special case, for which we give examples in Section~\ref{massive_bosons}, 
is a string-field $\field(x,e)$ which transforms as in~\eqref{eqFieldCov}, 
but is {\em point-like} localized, i.e.\ the space-like commutation 
property~\eqref{eqFieldLoc} depends only on $x$ and not on $e$. 
This corresponds to the following analyticity property of its intertwiner. 
\begin{Prop}[Point-localized fields]  \label{PointField}
A string-localized field, in the sense of Eq.~\eqref{eqFieldCov},
is {\em point-like} localized if, and only if, the function $e\mapsto u(e, p)$ 
is analytic on the entire complexified $\Spdc$ and the bound~\eqref{eqModGrowth}, 
with growth order $N=0$, holds for all compact subsets of $\Spdc$. 
\end{Prop}

The rest of this subsection is devoted to the proof of the proposition 
and of the structural part 
of the theorem, namely $i)$, $iii)$ and the uniqueness part of $ii)$. 
The existence claim of $ii)$ will be proved by explicit construction: 
We sketch the construction of intertwiners in the next subsection, and the 
proof that they have the required properties will be given in each case 
separately, in Sections \ref{massive_bosons} through \ref{massless}.  
\begin{Proofof} {\em of Theorem~\ref{StringField}}. 
$i)$ follows from Proposition~\ref{Cov} via second quantization. The proof of the Bisognano-Wichmann property is contained in~\cite[proof of Thm.\ I.3.2]{LRT}, the argument being as follows.  
By construction, the second quantization of the Tomita involution of $K(W)$ coincides with the Tomita involution for the closure of the real space $\calA(W)^{\text{sa}}\Omega$. But the latter is just the Tomita operator associated with the von Neumann algebra $\calA(W)$ and $\Omega$. Since second quantization preserves the polar decomposition, this proves the Bisognano-Wichmann property. 

To show the uniqueness statement of $ii)$, let $u$
and $\hat u$ be functions with the stated intertwiner and analyticity
properties. 
Considering equation~\eqref{equCov} with $\lor=B_p$ and using
$\WigRot(B_p,p)=1$, one finds that $u$ is of the form 
\begin{align}  \label{equu0}
u(e,p)=u_0(B_p^{-1}e),  \quad\text{ where }\quad  u_0(e):= u(e,\bar p). 
\end{align} 
Then $u_0$ transforms under the little group $G$ according to 
\begin{align}  \label{equ0Cov}
D(\lor)\, u_0(e) = u_0(\lor e),  \quad \lor\in G. 
\end{align}
(Note for later reference that, conversely, $u(e,p)$ is fixed by $u_0(e)$ through the left equation in \eqref{equu0}, and $u_0$ transforming as in~\eqref{equ0Cov} implies that $u(e,p)$ satisfies the intertwiner relation~\eqref{equCov}.)  
Let now $u_0$ and $\hat u_0$ be solutions to this equation. We first show that in $d=4$, they are linearly dependent (which in $d=3$ is a tautology since there the little Hilbert space is one--dimensional). 
This can be seen as follows. Eq.~\eqref{equ0Cov} implies that for fixed $e$ the vector $u_0(e)$ must be invariant under the restriction of $D$ to the stability subgroup, in $G$, of $e$. But this condition, as we show in Lemma~\ref{OrbitsStab}, fixes the $u_0(e)\in\calh$ up to a factor if $e$ is in the ``real boundary'' $\Spd$ of $\Tub$, i.e.\ $u_0(e)$ and $\hat u_0(e)$ are linearly dependent for $e\in\Spd$. By the edge-of-the-wedge theorem for tuboids~\cite[Thm.\ A\,3]{BrosMos}, this implies that $u_0(e)$ and $\hat u_0(e)$ are linearly dependent for all $e\in\Tub$. Hence in $d=4$, as well as in $d=3$, we have 
\begin{align} \label{equhatu0'}
u_0(e) = f(e)\,\hat u_0(e),\quad f(e)\in\CC,  
\end{align}
for all $e\in\Tub\setminus\calN$, where $\calN$ denotes the set (of measure zero) where $\hat u_0$ vanishes. On this domain, the function $f$ must be analytic. Further, Eq.~\eqref{equ0Cov} implies that 
$f$ is invariant under $G$. For $e\in\Tub$, let now $G^c_e$ denote the set of complex Lorentz transformations which leave $\bar p$ invariant, map $e$ into $\Tub$, and are path-connected with the unit. Analyticity of $f$ on $\Tub$ then implies that for $e\in\Tub$ the function $\Lambda\mapsto f(\Lambda e)$ is analytic on $G^c_e,$  and invariance of $f$ under $G$ implies that this function is constant not only on $G$, but on $G^c_e$. 
But we show in Lemma~\ref{OrbitsStab} that for (almost -- cf.\ below) each pair $e$, $\hat e\in\Tub$ satisfying $\bar p\cdot \hat e = \bar p\cdot e$, there is some $\Lambda\in G^c_e$ such that $\hat e=\Lambda e$. 
Therefore, $f$ is constant on all $e\in\Tub$ with $\bar p\cdot e=$ constant. (In fact, in the massive case the 
complexification of $G$ does {\em not} act transitively on  the set $\bar p\cdot e=im$, 
namely here one has to exclude the point $e=(i,0,0,0)$, cf.\ Lemma~\ref{OrbitsStab}. 
But by continuity, 
the conclusion also holds for the set $\bar p\cdot e=im$.)  
It follows that $f$ can be written as $f(e)=F(\bar p\cdot e)$, 
which implies Eq.~\eqref{equhatu} of the 
theorem  by virtue of~\eqref{equu0}.   
It also follows that the function $F$ is analytic on the upper half plane 
except, possibly, for those points $\bar p\cdot e$ with $e\in\calN$, i.e.\ $\hat u_0(e)=0$. 
But these are isolated points in $\CC$, which can be seen as follows. If 
$e\in\calN$, then by the covariance condition~\eqref{equ0Cov} $\hat u_0$ 
vanishes on the entire orbit of $e$ under $G$, and by analyticity it vanishes 
on the orbit of $e$ under $G^c_{e}$. Again by Lemma~\ref{OrbitsStab}, this implies 
that $\hat u_0$ vanishes on the entire hyper-surface $\bar p\cdot e=$constant. 
Suppose now that the set of $\bar p\cdot e$, $e\in\calN$, has an accumulation point. 
Then $\hat u_0$ vanishes on all corresponding hyper-surfaces and hence vanishes altogether.  
Hence all points $\bar p\cdot e$, $e\in\calN$, are isolated. Further, analyticity of 
$u_0$ implies that $F$ has no  essential singularity on any of these points. It follows that 
$F$ is a meromorphic function.      

Ad $iii)$ By a version of the  Jost--Schroer theorem~\cite{St}, a string--localized Wightman
field $\field(x,e)$ creating only single particle states from the vacuum 
is,  up to unitary equivalence,
of the form $a^*(\field(x,e)\Omega) + a(\field(x,e)^*\Omega)$. Here $a^*$ and
$a$ are the creation and annihilation operators acting on the second
quantization of the single particle space. 
Covariance \eqref{eqFieldCov} under translations implies that $\field(x,e)\Omega$
and $\field(x,e)^*\Omega$ are of the form\footnote{The subsequent formulas are to be understood in the sense of distributions.} 
\begin{align} 
\field(x,e)\Omega \,(p) =:  e^{ix\cdot p} \, u(e,p), 
\quad \field(x,e)^*\Omega \,(p) =:e^{ix\cdot p}  \, {u_c(e,p)},
\end{align} 
where $u$ and $u_c$ are $\calh$--valued distributions. 
Hence our fields are, up to equivalence, of the form~\eqref{eqField}. 
The  covariance property \eqref{eqFieldCov} then implies that $u$ must
satisfy the intertwining property \eqref{equCov}. 

We now show that $u$ must also have the analyticity
property. To this end, let $W$ be a wedge and 
let $L_W$ denote the generator of the unitary one-parameter
group representing the boosts $\Boo{W}{t}$. 
The Bisognano-Wichmann property implies that if $x+\RR_0^+e\in W$, then
firstly $\field(x,e)\Omega$ is in the domain of the unbounded operator
$\exp(-\pi L_W)$, and secondly   
\begin{equation} \label{eqF*SF'} 
 \field(x,e)^*\Omega = U(j_W) \,\exp(-\pi L_W)\,\field(x,e)\Omega.   
\end{equation}
But the first assertion implies that the $\calH$-valued function $t\mapsto U(\lor_W(t))\field(x,e)\Omega$ 
has an analytic extension into the strip $\RR+i(0,\pi)$, weakly continuous at
the boundary. (The value at  $t=i\pi$ then coincides with $\exp(-\pi L_W)\field(x,e)\Omega$.) 
It follows that for almost all $p$ the  function 
\begin{align} \label{equAna'}
 t \mapsto  u(\lor_W(t)_*h,p) 
\end{align}
is the  boundary value of an analytic function in the strip
$\strip:=\RR+i(0,\pi)$ if $h$ is a test function on $\Spd$ with support 
contained in $W_H$. 
(We have for a moment restored the distribution notation in order 
not to miss the point.) 
One concludes from the above that $u$ is the boundary value, in the sense 
indicated after Eq.~\eqref{eqIntDist}, of an analytic function $e\mapsto u(e,p)$ 
on the tuboid $\Tub$, of moderate growth near the real ``boundary'' $\Spd$.    
(To this end, one represents the distribution $u$ as a first order 
derivative (in the sense of distributions) of a 
continuous function on $\Spd$, and recalls that the set of $\lor_W(t)e$, 
with $t\in\strip$ and $W$, $e\in W$ varying, exhausts the entire tuboid, cf.~Lemma~\ref{LWte}.) 
Since $U(\Boo{W}{t}) \field(x,e)\Omega$ must be in $\Hirr$ for all $t\in\strip$, $e\in W$, this 
also shows the bound~\eqref{eqModGrowth}. 

To show that $u$ and $u_c$ are related as in Eq.~\eqref{eqv}, we consider first
$e\in W_0$. Then equation~\eqref{eqF*SF'} implies  
that $u_c(e,p)$ coincides with $D(j_0)u(\lor_0(t)e,-j_0p)$ at ${t=i\pi}$. 
Using that $\lor_0(i\pi)=j_0$, this implies equation~\eqref{eqv} for
$e\in W_0$. 
Let now $e$ be an arbitrary point in $\Spd$. Then $e\in\lor W_0$ for some
$\lor$. Using the intertwining property \eqref{equCov} of $u$ and the 
identity \eqref{eqjBpj}, one finds that equation~\eqref{eqv} also holds for such $e$. 

As to the PCT theorem, we now have the two identities: 
\begin{align} \label{eqF*SF} 
\big( U(j_0) \field(x,e)\Omega\big) (p) &= D(j_0) \,  e^{ix\cdot (-j_0p)} \,
u(e,-j_0p),  \\
 \big( \field(j_0x,j_0e)^*\Omega\big) (p) &= e^{ij_0x\cdot p} \,u_c(j_0e,p). 
\end{align}
By Eq.~\eqref{eqv} and antilinearity of $D(j_0)$, the right hand sides
coincide.
We therefore get equation~\eqref{eqPCT} for $j=j_0$ by the
Reeh--Schlieder property, and for all $j\in\Poj^\downarrow$ by covariance~\eqref{eqFieldCov}. 
\end{Proofof}
\\
We finally prove the proposition on point-localized fields. 
\begin{Proofof} {\em of Proposition~\ref{PointField}.} 
Suppose $u$ is analytic on the entire complexified $\Spdc$ and
satisfies the mentioned bound.  Then the function $t\mapsto
u(\Boo{W}{t}e)$ is analytic in the strip $\strip$ whether or not $e$
is contained in $W$.  The proof of Proposition~\ref{Cov} reveals that
then the single particle vectors $\psi(f,h)$, $\supp f\subset \calO$,
are localized in $\calO$ in the modular sense.  This proves the ``if'' part via second
quantization.  Conversely, suppose $\field(f,h)$ is localized, in the
sense of commutators, at the support of $f$, independently of $\supp
h$.  Then the reasoning of the above proof, ad $iii)$, shows that the
function in Eq.~\eqref{equAna'} is analytic in the strip $\strip$,
independently of $\supp h$.  But the set of $\lor_W(t)e$, with $W$,
$e\in\Spd$ and $t\in\strip$ arbitrary, exhausts the entire $\Spdc$,
cf.~Lemma~\ref{LWte}.  As above, one concludes that $u(e,p)$ is
analytic on $\Spdc$.  The bound~\eqref{eqModGrowth} also follows as in
the above proof, with $N=0$ since $u(e,p)$ is analytic.
\end{Proofof}

\subsection{Construction of the Intertwiners: General Recipe.} 
\label{Construction}
We now describe the idea of the construction of intertwiner functions $u$ with the properties
required in  Theorem~\ref{StringField} for irreducible positive energy representations 
of the Poincar\'e group with a faithful (or scalar) representation $D$ of the little group $G$, i.e.\ 
for massive bosons and the massless infinite spin particles. 
The proofs of the relevant properties will be 
given in Sections~\ref{massive_bosons} (massive case) and \ref{massless} 
(massless infinite spin case). 

We will exploit the fact that 
all of these irreducible unitary representations $D$ of $G$ occur 
in the decomposition of the
pullback representation acting naturally on functions on suitable $G$--orbits. 
Namely,  let $\orbit$ be the $G$--orbit defined by  
\begin{align} \label{eqOrbit}
\orbit :=\{ q\in\HypNull:  \, q\cdot \bar p=1\}.   
\end{align}
(Recall that $\bar{p}$ is the base point in $\Hyp$ whose stabilizer group is $G$.) 
It turns out that $\orbit$ is isometric to the sphere 
$S^{d-2}$ for $m>0$, and to $\RR^{d-2}$ for $m=0$, cf.~Lemma~\ref{Orbit}. Since every isomorphism of $\orbit$ extends, by linearity, to a Lorentz transformation which leaves $\bar p$ invariant, it follows that 
$G$ is precisely the isometry group of $\orbit$. Thus, the isometry 
$\orbit\cong S^{d-2}$ or $\RR^{d-2}$ establishes the isomorphism 
$G\cong SO(d-1)$ or $E(d-2)$, respectively for $m>0$ or $m=0$. 
Let now $d\nu$ denote the $G$--invariant measure on $\orbit$, 
and let $\RepPulBac$ be the  unitary representation of $G$ acting on $L^2(\orbit,d\nu)$ as 
\begin{align} \label{eqRepPulBac}
\big(\RepPulBac(R)v\big)(q)&:=v(R^{-1}q),\quad R\in G. 
\end{align}
It is well--known that $\RepPulBac$ decomposes into the 
direct sum of all irreducible representations of $G\cong SO(d-1)$ for $m>0$ 
and into a direct integral of all faithful irreducible representations of
$G\cong E(d-2)$ for $m=0$. 
Hence, for any faithful representation $D$ of $G$ 
there exists a partial isometry $V$ from $L^2(\orbit,\d\nu)$ 
into $\calh$ which intertwines the representations $\RepPulBac$ and $D$: 
\begin{align} \label{eqInter} 
D(R)\,V = V\,\RepPulBac(R),\quad R\in G. 
\end{align}
(In the case $m=0$, $V$ is a generalized partial isometry defined only on a dense 
set in $L^2(\orbit,\d\nu)$.) 

We now solve the intertwiner equation~\eqref{equCov} by projecting a corresponding 
$L^2(\orbit,d\nu)$-valued solution $\tilde u(e,p)$ onto $\calh$. Namely, let $F$
be a suitable numerical function and define 
\begin{align} \label{equF}
\tilde u(e,p)\,(q) &:= F(q\cdot B_p^{-1}e),\\
u(e,p) & := V\, \tilde u(e,p). \label{equExp}
\end{align}
Then $\tilde{u}$ solves the analogue of \eqref{equCov} with $D$ replaced 
by $\RepPulBac$, and 
$u$ solves Eq.~\eqref{equCov} by construction.   Now note that the imaginary part 
of $q\cdot e$ is strictly positive if $q\in\HypNull$, $e\in\Tub$. 
Hence the analyticity property can be satisfied if 
$F$ has an analytic extension into the upper complex half plane.  
It turns out that a proper choice for $F$ is 
\begin{equation} \label{eqPower}
F(w):=w^\degree, 
\end{equation} 
for suitable
$\alpha\in\CC$. 
In case $\degree\not\in\Bi$, the power $w^\degree$ is understood 
via the branch of the logarithm on $\CC\setminus \RR^-_0$ with 
$\ln 1=0$, and by continuous extension from the upper half plane if 
$w\in\RR^-$, i.e.,   
$\lim_{\eps\rightarrow 0+}(w+i\eps)^{\degree}$. 
The intertwiner $u$ obtained this way will be denoted $u^\degree$ in the sequel. 
In general, the function $q\mapsto (q\cdot e)^\degree$ will be in $L^2(\orbit,d\nu)$ 
only after smearing with a test function $h\in\calD(\Spd)$, $\int d\sigma(e) \,h(e)\,(q\cdot e)^\degree$. 
The representation $\RepPulBac$ extends naturally to the Lorentz group on the (dense) set of functions 
of this form via push-forward. 

In fact, Bros et al.\ have shown~\cite{BrosMos} that this 
representation is unitary in $L^2(\orbit,d\nu)$ if the real part of $\degree$ is 
$-(d-2)/2$, and that in this case it is equivalent to the irreducible principal 
series representation with value $\degree(\degree+d-2)=-|\degree|^2$ of the Casimir operator. 
This choice is also distinguished by the fact that the resulting intertwiner, and hence the associated free field, satisfies the Klein Gordon equation in the variable $e\in\Spd$, with mass $|\degree|^2$, cf.~\cite{BrosMos}. 
%
The connection of our approach with the work of Bros et al. (which we sketch in the appendix) has been elaborated in~\cite{MundSaclay}. 

\section{Massive Bosons} \label{massive_bosons}
We construct intertwiners for massive bosons, arriving at explicit expressions for the intertwiners and the ensuing two-point functions. 
We obtain fields with genuine string-like (in contrast to point-like) localization, and clarify their relation to point-like localized fields.  
We also show that they have better UV behavior than the point-like 
localized usual free  fields. 

Although intertwiners can be easily constructed from the known point-localized 
intertwiners, cf.~Subsection~\ref{StringVsPoint} or Lemma~\ref{us}, we shall construct them along the lines of the last subsection. Our main motivation is that these intertwiners have the additional interesting feature that they satisfy  the Klein Gordon equation in the variable $e\in\Spd$, as already mentioned. 
Let us first recall the irreducible representations ${D}$
of the little group ${G}$ and of $j_0$ in the massive case. 
For $m>0$, $G$ is isomorphic to $SO(d-1)$. Recall that the irreducible representations
of $SO(2)$ are labeled by $s\in\ZZ$ and act in $\CC$ as
$R(\omega)\mapsto e^{is\omega}$, 
and that the irreducible representations of $SO(3)$ are labeled by
$s\in\NN_0$ and act in 
$\CC^{2s+1}\cong \text{span}\{Y_{s,\sz},\sz=-s,\ldots,s\}\subset L^2(S^2)$, 
with $Y_{s,\sz}$ the spherical harmonics,  according to 
\begin{align}  \label{eqDsR'} 
(D(R) Y_{s,\sz}) (n) &:= Y_{s,\sz}(R^{-1}n). 
\end{align} 
for $n\in\mathbb R^3$, $\Vert n\Vert=1$. Since $\bar p=(m,0,0,0)$ is invariant under $-j_0$, $G$ is invariant under the
adjoint action of $j_0$, and the subgroup of $\Lor$ generated by $G$
and $j_0$ is a semi-direct product.  
The above representations $D$ of $G$ extend to this group 
via an anti--unitary involution $D(j_0)$. Namely, in the case $d=4$, $D(j_0)$ is the operator defined by
anti--linear extension of 
\begin{equation} \label{eqDjm}
D(j_0)Y_{s,\sz} := (-1)^\sz Y_{s,-\sz}. 
\end{equation} 
In $d=3$, $D(j_0)$ is just complex conjugation. (We show in Lemma~\ref{Dj} that $D(j_0)$ indeed satisfies the representation properties $D(j_0)D(\Lambda)D(j_0)=D(j_0\Lambda j_0)$, $\Lambda\in G$.)

\subsection{Intertwiners.} 
We now specify the general construction of Section~\ref{Construction}, arriving at completely explicit expressions, cf.\ Proposition~\ref{um>0}. 
For $m>0$, and base-point $\bar p:=(m,0,0,0)$ or $(m,0,0)$ in $\Hyp$, the set $\orbit$ of all $q\in\HypNull$ with $q\cdot \bar p=1$, cf.~\eqref{eqOrbit}, is 
isometric to the sphere $S^{d-2}$ via the parametrization of $\orbit$ given by 
\begin{align} \label{eqqtheta} 
q(\theta)&:=(1,\cos\theta,\sin\theta)/m, & d&=3,\\
q(n)&:=(1,n_1,n_2,n_3)/m, \quad |n|^2=1, & d&=4. \label{eqqn}
\end{align}
The isomorphism $q$ from $S^{d-2}$ onto $\orbit$ identifies the action of $G$ in $\orbit$ with the action of $SO(d-1)$ in $S^{d-2}$, and $-j_0$ acts as $\theta\mapsto \pi-\theta$ or $(n_1,n_2,n_3)\mapsto (-n_1,-n_2,n_3)$, respectively. 
The isometric intertwiners 
$V=V_{s}$ from the representation $\RepPulBac$, cf.~\eqref{eqRepPulBac}, to 
the irreducible representation for spin $s$ come out as 
\begin{align} 
 V_{s}v &:= \int_{S^1}d\theta e^{is\theta} v(q(\theta)),& d=3, \label{eqIntm>0} \\
 (V_{s}v)_\sz &:= (Y_{s,\sz},v)=\int_{S^2}d\sigma(n) \overline{Y_{s,\sz}(n)}
v(q(n)), & d=4.   \label{eqIntm>0'} 
\end{align} 
Here, $v\in L^2(\orbit,d\nu)$, $q$ is the parametrization of $\orbit$ defined in 
\eqref{eqqtheta} and \eqref{eqqn}, and $d\sigma(n)$ denotes the rotation invariant 
measure on the sphere. 
Thus, our construction \eqref{equF}, \eqref{equExp} and \eqref{eqPower} leads to the following
intertwiners: 
\begin{align}
\intfctdeg{e}{p} & = e^{-i\pi\degree/2}\, 
                  \int_{S^1} d\theta  e^{is\theta} \,(B_pq(\theta)\cdot e)^\degree, & d=3 , \label{equ0m>0d=3}\\
\intfctdeg{e}{p}_\sz & = e^{-i\pi\degree/2}\,
                  \int_{S^2} d\sigma(n) \, \overline{Y_{s,\sz}(n)}\,(B_pq(n)\cdot e)^\degree,  & d=4.
                  \label{equ0m>0d=4} 
\end{align}
(We have introduced a factor $e^{-i\pi\degree/2}$ so that $(u^\degree)_c=u^{\bar\degree}$. This follows from a calculation analogous to \eqref{eqIntJ'} below.) 
Let us discuss the three-dimensional case in more detail. 
For $e$ in the real boundary $\Spd$, the integrand in \eqref{equ0m>0d=3} turns out two have two distinct zeroes of order $1$ as a function of $\theta$. The corresponding pole, for  real $\spd\in\Spd$, of the integrand is therefore integrable iff $\Re\degree>-1$. Hence, for this range of $\degree$ we expect $u^\degree$ to be an intertwiner function with growth order zero, thus leading to localization in space-like half cylinders, and to fields which do not have to be smeared in $\spd$. This is indeed the case: 
\begin{Prop}[Intertwiners of growth order 0]  \label{IntMassive3d}
Consider the three-dimens\-ion\-al case and let $\Re\degree>-1$.  Then
$\intfctdeg{e}{p}$ is an intertwiner function in the sense of
Definition~\ref{DefIntFct}, with growth order $N=0$ in
\eqref{eqModGrowth}.  More specifically, it is bounded, uniformly in
$e\in\Tub$ and $p\in\Hyp$.
\end{Prop}
In $d=4$, we also expect that $\intfctdeg{e}{p}$ is an intertwiner function with growth order zero for $\Re \degree >-1$, since $q(n)\cdot e$ also has a zero of order one as a function of $\cos \theta$. (This follows from Eq.~\eqref{eqqne} below if rotational covariance~\eqref{equ0Cov} is used to put $e_1=0=e_2$.)  
\begin{Proof}
{}For $e\in\Tub$ the imaginary part of the integrand
$B_p^{-1}q(\theta)\cdot e$ is strictly positive.  This allows one to
find, for $e$ in any given compact subset of $\Tub$, a dominating
function for the integrand.  Therefore the analyticity in $e$ of the
integrand implies that $u^\degree$ is analytic.  To prove the uniform
boundedness, we denote $e_\pm:=e_1\pm ie_2$ and calculate
\begin{align}
m\, q(\theta)\cdot e & = e_0 -\half\big(e_+e^{-i\theta}+e_-e^{i\theta}\big)  
       = -\half e_- e^{-i\theta}(e^{i\theta}-z_+)(e^{i\theta}-z_-)      \label{eqxie}      \\
& = -ie^{-i\theta}\big( (e^{i\theta}-z_+)^{-1}-(e^{i\theta}-z_-)^{-1}\big)^{-1} ,  \label{eqxie'}
\end{align}
where $z_\pm:= e_-^{-1}(e_0\pm i)$ are the zeroes of the polynomial
$z^2-2(e_0/e_-)z  + e_+/e_-$.  Therefore $u_0^\degree$ (corresponding to $u^\degree$ as in Eq.~\eqref{equu0}) satisfies 
$$
|u_0^\degree(e)|\leq c\,\int d\theta \big( |e^{i\theta}-z_+|^{\ReDeg}+|e^{i\theta}-z_-|^{\ReDeg}\big),  
$$
where $\ReDeg:=\Re \degree$. (We have used that for $w\in\RR+i\RR_0^+$, $|w^\degree| \leq c |w|^{\ReDeg}$,
where $c=\max\{1,e^{-\pi\Im\degree}\}$.)  
By rotational invariance, we may assume $z_\pm\in [0,\infty)$. Then $|e^{i\theta}-z_\pm|\geq |\sin\theta|\geq \frac{2}{\pi}|\theta|$ in the interval $\theta\in(-\pi/2,\pi/2)$, and $|e^{i\theta}-z_\pm|\geq 1$ in its  complement in $S^1$. Hence the integral has a bound independent of $z_\pm$, hence of $e$. 
This proves the claim. 
\end{Proof}
We now consider the case $\degree = n\in\NN_0$. Since the intertwiners $u^n$ are analytic in all of $\Spdc$, they lead to point-localization according to Proposition~\ref{PointField}.   
In the sequel, will refer to the 1-1 correspondence between $u(e,p)$ and $u_0(e)$ given in Eq.~\eqref{equu0}. 
It is clear from eqs.~\eqref{equ0m>0d=3}, \eqref{equ0m>0d=4} that 
$u^n_0$ is an $n$-linear form on $\CC^d$ and can therefore be written as 
\begin{equation} \label{eqIntStringPoint}
u^n_0(e)_\sz = \sum_{\mu_1,\ldots,\mu_n} u^{\mu_1 \ldots\mu_n}_\sz\,e_{\mu_1}\cdots e_{\mu_n}.  
\end{equation} 
By the covariance condition~\eqref{equ0Cov}, the matrices $u^{\mu_1 \ldots\mu_n}_\sz$ build up an intertwiner from  the natural representation of $SO(d-2)$ on the symmetric $n$-tensors to the irreducible representation with spin $s$. 
It is then clear (and also follows from eqs.~\eqref{eqxie}, \eqref{eqqne} below) that $u^n$ vanishes unless $n\geq |s|$, $s$ being the spin of the particle. Thus, the simplest point-like localized cases is $\degree=|s|$. We now exhibit explicit expressions for this case in $3$ and in $4$ dimensions.  
\begin{Lem}   \label{us} 
The intertwiner $u^s$ is given as $u^s(e,p)=u_0^s(B_p^{-1}e)$, with $u_0^s$ as follows. 
\\
In $3$ dimensions, $u^{|s|}_0$ is given, up to a real factor,  
by 
\begin{align}  \label{eqIntFct0Exp}
u_0^{|s|}(e)&= i^{|s|} 
\, \times 
\begin{cases}
(e_1 + ie_2)^{s} &\text{ if } s \geq 0, \\ 
(e_1 - ie_2)^{|s|} & \text{ if } s < 0. 
\end{cases}
\end{align}
In $4$ dimensions, $u^{s}_0$ is given, up to a real factor, by 
\begin{align} 
u_0^s(e)_\sz 
&= i^{s}\sqrt{\frac{(s+\sz)!}{(s-\sz)!}}\,
\big\{(e_1+ie_2)\partial_{e_3}-(\partial_{e_1}+i\partial_{e_2})e_3\big\}^{s-\sz}\,(e_1-ie_2)^s . \label{equ0m}
\end{align}
For real $e\in\Spd$, it coincides with 
\begin{equation} 
u_0^s(e)_\sz = (-i)^{s}\,(1+e_0^2)^{s/2} \,  \overline{Y_{s,\sz}(n(e))},  \label{equ0m'} \\  
\end{equation} 
where $Y_{s,\sz}$ are the spherical harmonics, and 
$n(e):= (1+e_0^2)^{-1/2}$$(e_1,e_2,e_3)\in S^2$.  
\end{Lem}
Note that Eq.~\eqref{equ0m} exhibits analyticity of the intertwiner on 
the whole of $\Spdc$, while Eq.~\eqref{equ0m'} exhibits its covariance~\eqref{equ0Cov} under rotations. 
\begin{Proof}
The 3-dimensional case follows straightforwardly from 
$$
\big(q(\theta)\cdot e\big)^{|\spin|}  = 
(-2m)^{-|s|} \,\big((e_1+ie_2)^{|s|}e^{-i{|s|}\theta} + (e_1-ie_2)^{|s|} e^{i{|s|}\theta}\big)+
\sum_{\nu=-|s|+1}^{{|s|}-1}c_\nu e^{-i\nu\theta},    
$$ 
which is a consequence of equation~\eqref{eqxie}.  
To prove the $4$--dimensional case, define $\hat u^s_0(e)$ for real $e=(e_0,e_1,e_2,e_3)\in\Spd$ by the r.h.s.\ of Eq.~\eqref{equ0m'}. Recalling that $Y_{s,\sz}$ is the restriction of a polynomial to the sphere, homogeneous of degree $s$, it is clear that multiplying $Y_{s,\sz}(n(e))$ with the factor $(1+e_0^2)^{s/2}$ amounts to restricting the same polynomial to the sphere $\Spd\cap\{e_0=\text{const}\}$. This implies that $\hat u_0^s$ coincides with the 
r.h.s.\ of Eq.~\eqref{equ0m}.  
It remains to show that  $\hat u^s$ coincides with $u^s$, as defined in \eqref{equ0m>0d=4}, up to a real factor. To this end, one first checks that $\overline{Y_{s,\sz}(n(e))}$, and hence $\hat u_0^s(e)$, is a solution to \eqref{equ0Cov}. Hence, in view of the uniqueness property, it suffices to show that the  $s$-components coincide up to a real factor. 
To this end, we write $n\in S^2$ as $n=(\sin\theta\cos\phi,\sin\theta\sin\phi,\cos\theta)$, and  have  
\begin{align}  \label{eqqne}
m\; q(n)\cdot e &= e_0
-\half\sin\theta\big((e_1+ie_2)e^{-i\phi}+(e_1-ie_2)e^{i\phi}\big) -
\cos\theta e_3. 
\end{align}
This implies  
$$
(q(n)\cdot e)^\spin= (-2m)^{-\spin}\,(\sin\theta)^{\spin}e^{i\spin\phi}\,(e_1-ie_2)^\spin
\,+\,\sum_{\sz=-\spin}^{\spin-1}c_\sz(\theta)e^{i\sz\phi}.  
$$
Using $Y_{s,s}(\theta,\phi)=c_s \,(\sin\theta)^s\,e^{i s\phi}$, this
yields $u_0^{\spin}(e)_s= e^{-i\pi \spin/2}\,c\,(e_1-ie_2)^s$, 
But this coincides with $\hat u_0^s(e)_s$ up to a real factor. This completes the proof.  
\end{Proof}

{}From the uniqueness statement $(ii)$ in Theorem~\ref{StringField}, cf.\
equation~\eqref{equhatu}, we then have the form of the most general intertwiner function:  
\begin{Prop}[The  general form of massive intertwiners]  \label{um>0}
Let $F$ be an analytic function on the upper half plane which is polynomially bounded at infinity and has moderate growth near the reals. Then
\begin{equation}
u(e,p):=F(e\cdot p) \, u^{|s|}(e,p), \label{equFhatu}
\end{equation}
where $u^{|s|}$ is given by Eq.~\eqref{eqIntFct0Exp} or \eqref{equ0m},
is an intertwiner function, in the sense of
Definition~\ref{DefIntFct}, for mass $m>0$ and spin $s$.

Conversely, every such intertwiner function is of this form. 
\end{Prop}
\begin{Proof} 
Since the proof of the first statement is straightforward, we
only show the ``converse'' statement.  In view of the uniqueness
assertion in Theorem~\ref{StringField}, it only remains to prove the properties
of $F$ apart from being a meromorphic function on the upper half plane. 
We first show that it must be analytic.  To this end, let us determine
the zeroes of $u^s$ in the four-dimensional
case with $s>0$.  Firstly, $u^s(e)_s=0$ implies, by Eq.~\eqref{equ0m}
for $\sz=s$, that $e_1=ie_2$.  Consider then the $0$-component of
$u^s(e)$.  By Eq.~\eqref{equ0m}, $u^s(e)_{\sz=0}$ is a sum with one term
proportional to $(e_3)^s$, while all other summands contain a factor
$(e_1-ie_2)^n$, $0<n\leq s$.  Now these terms vanish due to
$e_1=ie_2$, and therefore $u^s(e)_0=0$ implies that $e_3=0$.  On the
other hand, $e_1=ie_2$ and $e_3=0$ obviously imply $u^s(e)=0$.  It
follows that $u_0^s(e)=0$ if and only if $e$ is of the form
$(e_0,ie_2, e_2,0)$.  Such $e$ is in $\Spdc$ if and only if $e_0=\pm
i$, and in $\Tub$ if and only if $e_0=+i$ and $|e_2|^2<1$.  In
particular, for all zeroes in $\Tub$ holds $\bar p\cdot e=im$.  Hence
the only possible pole of $F$ in the upper half plane is at $im$.  But
there are points $e\in\Tub$ with $\bar p\cdot e=im$ and $u^s(e)\neq0$,
for example $(i,-ie_2,e_2,0)$.  Hence $F$ may not have a pole at $im$,
and must therefore be analytic on the upper half plane.  The same
conclusion holds, of course, if $s=0$, and a similar consideration
holds in the three-dimensional case.

To show the boundedness condition on $F$, note that $u_0^s$ is a
polynomial in $e$, and that $B_p$ is a polynomial in $p$.  
Therefore $u^s(e,p)\equiv u_0^s(B_p^{-1}e)$ does not fall off for large $p$, hence the bound~\eqref{eqModGrowth} on $u$ implies a similar bound for $F(e\cdot p)$, and it follows that $F$ must be polynomially bounded. 
Similarly, one concludes that $F$ must have moderate growth near the reals. 
\end{Proof}
\begin{Remarks} 1.  The intertwiner leads to point-like localized
fields if and only if $F$ is entire, that is, analytic on the complex
plane, cf.\ Proposition~\ref{PointField}.  Note that the
boundedness condition then implies that $F$ is a polynomial.  This
complies well with the fact that the (mass shell restriction of the)
momentum space two-point function of a compactly--localized observable
is an entire function of $p$ on the complex mass
hyperboloid~\cite{DHRII} (which coincides with $\Spdc$ up to a scaling
factor $m$), and in fact a polynomial in the case of a Wightman field~\cite{Strocchi}.

2. By a calculation analogous to \eqref{eqIntJ'}, one finds that the intertwiners $u^s$ coincide with their ``conjugate'' intertwiners $(u^s)_c$, as defined in \eqref{eqv}. Hence for $u$ as in the proposition, we have 
$$ 
u_c(e,p)=\overline{F(-e\cdot p)}\, u(e,p).
$$ 

3. For spin $1$ in $4$ dimensions, we get an explicit formula for the two-point function of the field corresponding (as in \eqref{eqField}) to $u$: Namely, from Eq.~\eqref{equ0m} we have (up to an overall factor) 
$u_0^1(e)_{\pm 1} = \mp i(e_1\mp ie_2)$ and $u_0^1(e)_{0} = i \sqrt{2} \, e_3$.
The above remark then yields   
\begin{align}
\lsp \Omega,\field(x,e)\field(x',e')\Omega\rsp & =
\int d\mu(p) e^{ip\cdot(x'-x)} 
F(-e\cdot p) F(e'\cdot p)\big\{(e\cdot p)(e'\cdot p) -e\cdot e'\big\}. \label{eq2PointSpin1}
\end{align}
\end{Remarks} 
\subsection{String-localized Fields from Point--Fields} \label{StringVsPoint}

In order to obtain a good vantage point for the issue of \textit{point-like
fields versus proper strings} it is necessary to remind the reader of the
basic results of Wigner's particle-based representation theoretical approach
to interaction-free fields and their associated algebras. In case of a massive
particle there are intertwiners $v(p)$ which connect the $(m,s)$ irreducible 
one-particle Wigner representation with wave functions (and their associated quantum fields) transforming under 
certain finite-dimensional (non-unitary) representations $D'$ of the Lorentz group. More precisely, $v(p)$ is a linear map from the representation space of $D'$ onto the little Hilbert space $\CC^{2s+1}$ , satisfying     
\begin{align}
D(R(\Lambda,p))\,v(\Lambda^{-1}p) = v(p)\, D'(\Lambda), \label{eqPointIntertwiner} 
\end{align}
where $D=D^{(s)}$ denotes the spin $s$ representation of $SO(3)$, as before. The associated quantum field then transforms covariantly under $D'$\cite{Wei-book}:  
\begin{align}
& U(a,\Lambda)\,\Phi_r(x)\,U(a,\Lambda)^{\ast}=\Phi_{r'}(a+\Lambda x)\, D'(\Lambda)_{r'r} \label{field}\\
& \Phi_r(x)=( 2\pi)^{-\frac{3}{2}} \int d\mu(p) \sum_{\sz=-s}^{s}
           \big\{e^{ipx} v(p)_{\sz,r} a^{\ast}(p,\sz)+e^{-ipx}v_c(p)_{\sz,r} b(p,\sz)\big\}.\nonumber 
\end{align}
Here, $v_c(p):= D(i\sigma_2)\circ v(p)$ is the conjugate intertwiner\footnote{The usual notation $u(p)$ would lead to confusion with our notation $u(e,p)$.}, and $a,b$ are the Wigner annihilation operators as before (the
self-conjugate situation $b=a$ being always a special case).   
In fact, it is well-known~\cite{Wei-book,Joos}  that for given $(m,s)$ there is a countably infinite number of intertwiners and corresponding covariant fields. Namely, for any two half-integers $A,\dot{B}$ satisfying the restriction 
\begin{equation}\label{AB}
\left|  A-\dot{B}\right|  \leq s\leq A+\dot{B}%
\end{equation}
there is an intertwiner from the $(m,s)$ representation to the representation $D':=D^{(A,\dot{B})}$, the  representation in the space of $2A$ undotted and $2\dot{B}$ dotted symmetrized spinors. 
For given $(m,s)$, the  infinitely many different associated fields (and their derivatives)  
form the linear part of the \textit{Borchers equivalence class} of point-like fields\footnote{The full class is formed by the (Wick-ordered) composites of these fields.}. 

For the comparison with string-localized fields it is helpful to emphasize the following points.

\begin{itemize}
\item The intertwiners above are determined by the
covariant transformation law for the field $\Phi_r$, but we would be
lead to the same family of distribution valued intertwiners if modular
localization was required instead. In other words, covariance in the
sense of the (classical) tensor/spinor calculus is in this case
equivalent to the quantum requirement of modular localization. This
is, of course, the reason why historically it was possible for Pascual
Jordan to kick-start quantum field theory by (Lagrangian) field
quantization without having to wait for Wigner's more intrinsic
approach that does not rely on classical concepts.  The equivalence is
lost, however, if one leaves the realm of point localization, in
particular when no such localization is possible as for Wigner's
infinite spin representation \cite{Yng70}. In fact, the covariant
field equations for these representations found by Wigner \cite{Wig48}
have no localization properties. The argument $x$ of his
covariant wave function on which the Poincar\'{e} group acts
covariantly does not admit the interpretation as a point of
localization, not even as the end point of a string, if vanishing of
quantum mechanical commutators is taken as a criterion for
localization. On the other hand, the fields we construct in Section
\ref{massless} are both covariant and string localized.

\item  Only some of the fields in the infinite family indexed by the pairs $,\dot{B}$ satisfying \eqref{AB} permit a description in
terms of Lagrangian quantization i.e. are associated to an action principle
(canonical quantization, functional integration). For the lowest spins up to
$s=4$ these Lagrangian quantization descriptions of the Wigner approach to
massive particles have been explicitly computed (Dirac, Duffin-Kemmer.
Rarita-Schwinger for $s=\frac{1}{2},1,\frac{3}{2}$) \cite{Chang}. As emphasized
by Weinberg \cite{Wei-article} and formalized in the Epstein-Glaser approach
\cite{Ep-Gl}, one does not need a Lagrangian (but only an interaction
polynomial) in order to set up causal perturbation theory\footnote{The reader
should be aware that although Weinberg's book contains the broadest exposition
of the Wigner representation theory, the underlying philosophy (of lending
support to Lagrangian quantization) is very different from that in his
previous articles \cite{Wei-article} on higher spin fields (and certainly also
different from the spirit of the present article).}, a fact which has been
confirmed in subsequent work on renormalization theory in the mathematical
physics setting.
\end{itemize}

The absence of a Lagrangian description for string-localized fields is less
surprising after one becomes aware that the existence of a Lagrangian (and the
formulation of an action for the purpose of a functional integral formulation
of QFT) is the exception rather then the rule even in the point-like case;
although for each  $(m,s)$ there is a lagrangian
realization, most fields in the admissible family (\ref{field}) are not
``Euler-Lagrangian''. Suppose that the space of covariant wave functions
$H_{(m,s)}^{[  A,\dot{B}]  }$ is the solution space of an
Euler-Lagrange equation and let $H_{(m,s)}^{[  A^{\prime},\dot{B}%
^{\prime}]  _{{}}}$ with $A^{\prime}\geq A,\,\,\dot{B}^{\prime}\geq
\dot{B}$ be another covariant description of the same Wigner representation.
Then the distribution-valued wave function of the larger description are
related to the Euler-Lagrangian wave functions by a rectangular matrix
whose entries involve derivatives $\partial$. Consider for example the
spinless case $H_{(m,0)}^{[ 0,0] }$ and $H_{(m,0)}^{[ A,\dot{A}] }$. The
components of the multi-component $[A,\dot{A}]$  description involve
derivatives. Such wave functions are not of Euler-Lagrange type since the mass-shell condition as well as the nontrivial Poincar\'{e} transformation property and the identification of the Lorentz
indices as coming through the derivatives cannot be obtained in an
Euler-Lagrange setting.\\

An apparently pedestrian method to construct genuine string-localized fields 
(which in fact turns out to be the most general one, cf.\ below), is to smear a point-like 
field over a semi-infinite space-like line
\begin{align} \label{eqPointString}
\Phi(x,e)   =\int_{0}^{\infty} dt \,f(t)\sum_{r}\Phi_{r}(x+te)w(e)_r,  
\end{align}
where $f(t)$ is supported in the interval $[0,\infty)$ and $w(e)$ is 
a tensor formed from $e$ which 
is Lorentz invariant in the sense that 
\begin{align}   \label{eqw}
 D^{\prime}(\Lambda)w(\Lambda^{-1}e)=w(e),\quad \Lambda\in\Lor. 
\end{align}
One easily verifies that $\Phi(x,e)$ is string-localized in the sense of Eq.~\eqref{eqFieldLoc} and satisfies the string-covariance condition~\eqref{eqFieldCov}. 
In fact it is not difficult to see that in agreement with Theorem~\ref{StringField}, $\Phi(x,e)$ is of the form~\eqref{eqField}, with intertwiner given by 
\begin{align}
u(p,e)  &  = \tilde{f}(e\cdot p)\, u_{\text{point}}(p,e) \label{equFu}\\
u_{\text{point}}(p,e)_\sz  &  =\sum_{r}v(p)_{\sz,r} w(e)_{r} ,\quad \sz=-s,\ldots,s, \label{equw}  
\end{align}
with $\tilde{f}$, the Fourier transform of $f$, being analytic in the upper half-plane
but not in the whole plane (in which case one falls back to point-like localization%
\footnote{This follows from the first remark after Proposition~\ref{um>0}, which asserts that $\tilde{f}$ is a polynomial, hence $f$ has support in a point.}). 

It turns out that also the converse holds: 
\begin{Thm}[Massive, free string fields are 
integrals over point fields] \label{StringPoint}
In the massive case {\em every} string-localized free field can be written  as in Eq.~\eqref{eqPointString}, i.e.\  as an integral, along the string, of a point-localized tensor field. 
\end{Thm} 
\begin{Proof}
First note that $u^s(e,p)$, as given in Lemma~\ref{us}, is of the same form as $u_{\text{point}}$, cf.~\eqref{equw}. Namely, Eq.~\eqref{eqIntStringPoint} implies that 
$$
u^s(e,p)= v^s\, D'(B_p^{-1}) \,w(e)
$$
with $w(e)$ the $s$-fold symmetric tensor power of $e$, $D'$ the natural representation of the Lorentz group on the symmetric $s$-tensors, and $v^s$ the intertwiner from $D'|SO(d-2)$ to the irreducible representation with spin $s$ furnished by the matrix $u_\sz^{\mu_1\ldots \mu_s}$ of Eq.~\eqref{eqIntStringPoint}.  
Now $v^s(p):= v^s\circ D'(B_p^{-1})$ satisfies the intertwiner relation~\eqref{eqPointIntertwiner} and defines a particular point-localized field for spin $s$.  
Let now $u$ be the  intertwiner corresponding, according to Theorem~\ref{StringField}, to the given string-localized field $\field(e,x)$. Then Proposition~\ref{um>0} implies that 
\begin{equation*} 
u(e,p) = F(e\cdot p) u^s(e,p)= F(e\cdot p)\sum_{r=(\mu_1,\ldots,\mu_s)} v^s(p)_r \, w(e)_r, 
\end{equation*}
where $F$ is analytic in the upper
half plane. But this implies that $\field(e,x)$
is indeed of the form \eqref{eqPointString}, with ${f}$ being the
inverse Fourier transform of the boundary value of $F$ at $\RR$. The properties 
of $F$ asserted by Proposition~\ref{um>0}, namely polynomial boundedness at 
infinity and moderate growth near $\RR$, then imply that $f$ has support in 
the non-negative reals~\cite[Thm.\ IX.16]{ReSi}.  
\end{Proof}
It is interesting to go through the details in the example $s=1$. Here we take 
$D'(\lor):=\lor$ acting in $\CC^4$, $w(e):=e$, and intertwiner function $v(p):=V\circ D'(B_p^{-1})$ with $V:\CC^4\to \CC^3$  given as  $(Ve)_{\pm1}:=e_1\mp i e_2$,
$(Ve)_0:=\sqrt{2} e_3$. The resulting $u_{\text{point}}(e,p)$ coincides, again in agreement with our uniqueness statement, with  $u^1(e,p):=u_0^1(B_p^{-1}e)$ from equation~\eqref{equ0m} up to a
factor. 

\subsection{UV--Behavior} 
We show that the distributional character of our free fields is, in the massive case,  less singular 
than that of the usual point--like free fields, even more so in
the direction of the localization string. 
This fact should lead to a larger class of admissible interactions in a perturbative approach, as compared to taking the standard point-like localized free fields as
starting point.  

To this end, we determine the large $p$ behavior of the intertwiner
function $\intfctdeg{e}{p}$.  We already know that in $3$ dimensions
it is bounded in $p$, cf.\ Proposition~\ref{IntMassive3d}.  We now
show that the same holds in $d=4$, and it even falls off in the
direction of $e$.  This is a considerable improvement to the
point-localized usual free field for spin $s$, whose intertwiner
function goes at least like $|p|^{s}$.  We consider both the 3- and
the 4-dimensional case.
\begin{Prop}[Spin-independent bounds] \label{UVm>0d=4}
i) Let $\intfctdeg{e}{p}$ be the 4-d intertwiner function defined in Eq.~\eqref{eqIntm>0'}, 
with $\Re \degree =: \ReDeg>-1$.  Then there is a constant $c>0$ such that for all $e\in\Spd$ and $p\in\Hyp$ the following estimate holds: 
\begin{align} \label{equps}
  \|u^\degree(p,e)\|^2  \leq c \, \big(m^2 + (e\cdot p)^2\big)^\ReDeg. 
\end{align}
ii) The 3-d intertwiner function  $\intfctdeg{e}{p}$ satisfies the same estimate 
\eqref{equps} as the 4-d version if $\ReDeg>-1/2$.
\end{Prop} 
\begin{Proof} 
Ad $i)$
By the covariance equation~\eqref{equ0Cov}, it suffices to consider  
$u_0^\degree(e)$ with $e$ of the form $e=(e_0,0,0,e_3)$, $e_0^2-e_3^2=-1$. 
Then Eq.~\eqref{eqqne} implies that  
$$
m\; q(n)\cdot e= e_0-\cos\theta e_3 \quad \text{ for }n=(\sin\theta\sin\phi,\sin\theta\cos\phi,\cos\theta), 
$$
and for $\sz=-s,\ldots,s$ we have 
$$
|\big(u_0^\degree(e)\big)_\sz|\leq c'\,\int\d\cos\theta\d\phi \,
|e_0-\cos\theta e_3|^\ReDeg \leq c \,|e_3|^\ReDeg
$$
if $\ReDeg>-1$. Using $e_3^2=1+e_0^2$ and $\intfctdeg{e}{p}=u_0^\degree(B_p^{-1}e)$ and 
$(B_p^{-1}e)_0=e\cdot p/m$, this yields the claim. 

Ad $ii)$ 
We write the integrand $q(\theta)\cdot e$ as in Eq.~\eqref{eqxie'} and note that $(e^{i\theta}-z_\pm)^\alpha$ is {square} integrable if $\ReDeg>-1/2$. Then the Cauchy-Schwarz inequality implies 
\begin{align} \label{equ0Est}
|u_0^\degree(e)|^2\leq c\, |e_1-ie_2|^{2\ReDeg} = c\,(1+e_0^2)^{\ReDeg},
\end{align}
where $c:=1/4 (\int d\theta |e^{i\theta}-1|^{2\ReDeg})^2$. 
This proves the claim. 
\end{Proof}

A similar result can be achieved for a general 
intertwiner of the form as in Proposition~\ref{um>0}, $u(e,p)=F(e\cdot p)$ $u^s(e,p)$, 
with $u_0^s(e)$ as in equation~\eqref{eqIntFct0Exp} or \eqref{equ0m'} (for $d=3$ or $4$, respectively):  
Namely, the latter equations imply that 
$\|u_0^s(e)\|^2 = c\, (1+e_0^2)^s$. With $(B_p^{-1}e)_0=e\cdot p/m$, 
this proves the following 
\begin{Prop}[Norm of intertwiner] \label{UV}
Let $u(e,p)$ be as above in $d=3$ or $4$. 
Then its norm in $\CC^{2s+1}$, or modulus in $\Bc$, respectively, is given as  
\begin{align} \label{equpeBd}
    \|u(p,e)\|^2  = c \, |F(e\cdot p)|^2\,\big(m^2 + (e\cdot p)^2\big)^s, 
\end{align}
where $c>0$ depends on $m$ and $s$.
\end{Prop}
Note that string-localization requires $F$ analytic (only) on the upper half plane, cf.\  Proposition~\ref{um>0}. This is compatible with $F$ vanishing at real infinity with any given order, and hence with a bounded norm of $u$ (i.e., good UV behavior) --- in contrast to the point-localized case where $F$ must be a polynomial (cf.\ Remark  1 after Proposition~\ref{um>0}).


\section{String--Localized Fields for Photons} \label{Photons}

It is well-known that the free electromagnetic field $F_{\mu\nu}$ has a quantized version which complies with the requirements of (point-like) localization, covariance and Hilbert space positivity. Namely, it
transforms covariantly according to
\begin{align} \label{eqFTens}
U(a,\lor)\,F_{\mu\nu}(x)\,  U(a,\lor)^{-1} =  
F_{\rho\sigma}(a+\lor x)\,\lor^\rho{}_\mu \,\lor^\sigma{}_\nu,  
\end{align}
and acts on the Fock space over the single particle space of the
photon, which is the direct sum of helicity $\lambda=+1$ and
$\lambda=-1$ spaces.  In order to introduce interactions with matter
fields one needs a description in terms of vector potentials.  Whereas
in the classical setting this is straightforward, it is well-known
that the Wigner photon description does not allow a representation in
terms of a covariant vector potential.  This is the point of departure
of the gauge theory formalism: by allowing indefinite metric (and
corresponding ``ghosts'') one embeds the Wigner photon representation
into an unphysical formalism which formally maintains the point-like
local nature of a vector-potential and its milder short distance
property.  Of course such a construction would lead into the
unphysical blue yonder if at the end of calculations in the presence
of interactions one would not return to the physical setting by
removing the ghosts, which is accomplished by the BRST formalism. 
Though this ``quantum gauge formalism'' has been quite successful,
there are several reasons why the gauge formalism in the quantum
setting should be considered as a transitory prescription of an
incompletely understood physical situation.  Firstly, a formulation
completely in physical terms seems more desirable.  The second
observation is that if one invokes the renormalizability requirement
as a (formally not yet completely understood) quantum principle, spin=1
interacting theories are nailed down uniquely in terms of one coupling
parameter.  In particular for interacting massive vector-mesons the
necessary existence of additional physical degrees of freedom (usually
realized as Higgs mesons \footnote{But if one starts the perturbation
with massive vector-mesons these mesons do not possess non-vanishing
vacuum expectation values.}) within the perturbative setting is not an
input, but follows from consistency \cite{DS}.  Since this theory
selected by the renormalization principle is unique, no further
selection by a gauge principle is necessary; in fact the
quasi-classical approximation reveals that the classical gauge
selection principle follows from the geometrically less beautiful and
less understood, but in the long run probably more fundamental quantum
renormalization principle.

The present setting of string localization offers a much more mundane
ghostfree and covariant description: photons can be described by
string-localized vector potentials $A_{\mu }(x,e)$. 
These fields, whose construction we describe in the following, are distributions in $x$ and in $e\in \Spd$ transforming as  
\begin{equation} \label{eqPhotCov}
U(a,\Lambda )\,A_{\mu }(x,e)\,U(a,\Lambda )^{\ast }=A_{\nu }(a+\Lambda x,\Lambda e)\Lambda^{\nu }{}_{\mu }. 
\end{equation}
Actually a particular example of these fields 
has appeared in the literature under the heading of ``axial gauge''. 
But the direction $e$ has been considered fixed so that their Lorentz
transformation property had to be ``regauged'' according to 
\begin{equation}   
U(\Lambda )A_{\mu }(x,e)U(\Lambda )^{\ast }=A_{\nu }(\Lambda x,e)\Lambda^{\nu }{}_{\mu }+\text{gauge term}. 
\end{equation} 
As a result of cumbersome divergences at momenta orthogonal to $e$,
the axial gauge became unpopular in perturbative calculations.  In the
present setting these difficulties are overcome by considering
$A_\mu(x,e)$ as a distribution in $e$, with the nice transformation
behavior~\eqref{eqPhotCov} which had apparently been overlooked.  This
opens up the possibility of a perturbative, covariant, implementation
of interaction, where the weaker localization (in space-like cones)
requires new techniques but promises better UV behavior.  Here we only
describe their construction as free fields; the issue of interactions
of string-localized fields will be taken up in a separate paper.

We now define the string-localized vector potential in such a way that its physical nature within the Wigner setting is manifest, as well as the transformation property~\eqref{eqPhotCov}:  
\begin{align} \label{eqA}
A_\mu(x,e):= \int_0^\infty dt\, f(t)\,F_{\mu\nu}(x+te) \, e^\nu,     
\end{align} 
where $f$ is supported in $[0,\infty)$. 
By Maxwell's equations and the antisymmetry of $F_{\mu\nu}$ the vector 
field $A_\mu(x,e)$ satisfies the Lorentz and axial ``gauge'' conditions 
\begin{align} \label{eqGauge}
\partial^\mu A_\mu(x,e)=0,\quad e^\mu A_\mu(x,e)=0. 
\end{align}
It is noteworthy that these conditions are satisfied by {\em every}
free vector field $A_\mu(x,e)$ for photons acting in the physical Hilbert space
and transforming as in Eq.~\eqref{eqPhotCov}; hence they cannot be
regarded as additional gauge conditions in our context.  This fact will be shown
in the following proposition.

A distinguished choice for the function $f$, which yields our ``covariant'' version of an axial gauge potential, is the Heaviside function. 
With this choice our $A_\mu$ is indeed a potential for $F_{\mu\nu}$: Namely,  
\begin{equation} \label{eqFdA} 
\partial_\mu A_\nu(x,e)-\partial_\nu A_\mu(x,e) = F_{\mu\nu}(x) 
\end{equation} 
holds in the sense of matrix elements between states which are locally
generated from the vacuum. 
This choice also yields a dilatation
covariant vector potential.  Namely, as is well-known the field
strength is covariant under an extension of the representation $U$ of
$\Po$ to the dilatations $d_\lambda$, $\lambda>0$:
$U(d_\lambda)F_{\mu\nu}(x) \, U(d_\lambda)^{-1}=\lambda^2
F_{\mu\nu}(\lambda x)$.  With $f(t)$ the Heaviside distribution, our
vector potential satisfies
\begin{equation} \label{eqADil}
 U(d_\lambda) \, A_{\mu}(x,e) \, U(d_\lambda)^{-1}=\lambda \, A_{\mu}(\lambda x,e). 
\end{equation} 
However, our potential is not covariant under the entire conformal
group, since special conformal transformations take space-like
infinity to finite points.  Thus, under such transformations the
formula \eqref{eqA} changes its form and goes over into an integral
over a finite line segment.

It is noteworthy that the scaling behavior~\eqref{eqADil} implies that the scale 
dimension of our $A_\mu(x,e)$ is one, whereas that of the field itself is two. 
Thus, our potential shares with the usual (indefinite metric) potential a better UV-behavior than the field strength. 

To get more explicit expressions, let us recall the representation of the field strength in the Fock space over the single particle space of the photon. The latter is the direct sum of helicity $\lambda=+1$ and $\lambda=-1$ spaces, corresponding to the representations $D_\lambda(c,R_\phi):=e^{i\lambda\phi}$ of the little group $G\cong E(2)$.  
Denoting by $a^*(p,\lambda)$ the creation operator in Fock space, the field strength is given by~\cite{Wei-book} 
\begin{align} \label{eqFa*a} 
F_{\mu\nu}(x) = \int_{\HypNull}\d\mu(p)\sum_{\lambda=\pm1} 
\left\{  \; e^{ip\cdot x}\, u(p)_{\lambda,\mu\nu}\, a^*(p,\lambda) 
+ e^{-ip\cdot x} \; \overline{u(p)_{\lambda,\mu\nu}} \, a(p,\lambda)\; \right\},    
\end{align}
where $u(p)_{\lambda,\mu\nu}$ are the intertwiner functions 
$u(p)_\pm = i \,p\wedge \hat e_\mp(p)$, 
with $\hat{e}_\pm(p):= B_p\,\hat{e}_\pm$ and $\hat{e}_\pm:=(0,1,\pm
i,0)$.  
Then $A_\mu(x,e)$  may be written as 
\begin{align} \label{eqAa*a}
A_\mu(x,e) = \int_{\HypNull}\d\mu(p)\sum_{\lambda=\pm1} 
\left\{  \; e^{ip\cdot x}\, u(e,p)_{\lambda,\mu} \, a^*(p,\lambda) 
+ e^{-ip\cdot x} \; \overline{u(e,p)_{\lambda,\mu}} \, a(p,\lambda)\; \right\},    
\end{align}
with ``intertwiner function''
\newcommand{\upoint}{u_{\text{point}}} 
\newcommand{\upointNull}{\upoint{}_{,0}}
\begin{align} 
u(e,p)_{\pm} & = F(e\cdot p) \, \upoint(e,p)_\pm \label{equep}, \\ 
\upoint(e,p)_\pm  &= i\, \big\{(\hat{e}_\mp(p)\cdot e) \,p - (p\cdot e)\, \hat{e}_\mp(p) \big\} 
\label{equepPt} 
\quad \in\CC^4, 
\end{align}
where $F$ is the Fourier transform of $f$.    
Note that $u(e,p)$ satisfies the intertwiner relation 
\begin{equation} \label{equInt} 
D_\lambda(\WigRot(\lor,p)) u(e,\lor^{-1}p)_\lambda = 
\lor^{-1}\,u(\lor e, p)_\lambda,   
\end{equation}
which in turn implies the transformation property~\eqref{eqPhotCov} of $A_\mu(x,e)$ directly. 
Also the Lorentz and axial gauge conditions~\eqref{eqGauge} follow directly from 
$p\cdot u(e,p)_{\lambda}=0$ and $e\cdot u(e,p)_{\lambda}=0$, respectively. 
Returning to the viewpoint of Section~\ref{intertwiner}, the quantum field $A_\mu(x,e)$ is a distribution in $x$ and $e\in\Spd$ with certain properties, and we now make a statement on its uniqueness, analogous to the one in  Theorem~\ref{StringField}.  
\begin{Prop}[Uniqueness of string-localized vector potential] \label{StringPhoton}
Let $A_\mu(x,e)$ be a hermitian string-localized vector field for the free photon 
(i.e., it creates single photon states from the vacuum) transforming as in Eq.~\eqref{eqPhotCov} and satisfying the Bisognano-Wichmann property. Then the field $A_\mu(x,e)$ satisfies the following. 

i) It is the form~\eqref{eqAa*a}, with $u(e,p)$ as in~\eqref{equep} 
and where $F$ enjoys the following properties. $F$ is holomorphic in the upper half plane, of moderate 
growth near the reals, polynomially bounded at infinity, 
and satisfies $\overline{F(-\omega-i0)}=F(\omega+i0)$,  $\omega\in\RR$.   
$A_\mu$ is point-like localized if and only if $F$ is a polynomial, i.e.\ its inverse Fourier transform is the delta distribution or a derivative thereof. 

ii) It is a potential for the free field strength $F_{\mu\nu}(x)$ in the sense of Eq.~\eqref{eqFdA} if, and only if, the Function $F$ is the Fourier transform of the Heaviside distribution, i.e.\ $F(\omega)= i/\omega$ for $\omega \in \RR+i\RR^+$. 

iii) It satisfies the Lorentz and axial ``gauge'' conditions~\eqref{eqGauge}. 
\end{Prop}    
\begin{Proof}
Ad $i)$. As in the proof of $iii)$ of Theorem~\ref{StringField}, one concludes that $A_\mu(x,e)$ is of the form~\eqref{eqAa*a} for some $u(e,p)_{\lambda,\mu}$ which satisfies the intertwiner property~\eqref{equInt}. 
The Bisognano-Wichmann property implies that $u(e,p)_{\lambda}$ and is analytic in $\Tub$ with moderate growth near $\Spd$, and satisfies the self-conjugacy condition 
\begin{equation} \label{equConjPhot} 
j_0\,\overline{u(j_0e,-j_0p)_{\pm}}  = u(e,p)_{\mp}. 
\end{equation}
(Here we have used that the anti-unitary representer of $j_0$ on the single particle space is given by 
$(\Uirr(j_0)\phi)_\pm (p)=\overline{\phi(-j_0p)_\mp}$.)
The proof that $u(e,p)$ is as in Eq.~\eqref{equep} goes analogous to the proof of the uniqueness statement in $ii)$ of Theorem~\ref{StringField}: One first concludes that $u(e,p)$ is fixed by $u(e,\bar p)$ via the relations 
\begin{equation} \label{equu0Phot}
u(e,p)_\lambda = B_p \, u_0(B_p^{-1}e)_\lambda, \quad 
u_0(e)_\lambda:=u(e,\bar p)_\lambda,   
\end{equation} 
and $u_0$ satisfies the intertwining property 
\begin{equation} \label{equ0CovPhot}
 \Lambda  u_0(e)_\pm = 
 e^{\mp i\phi}\,u_0(\Lambda e)_\pm,\quad \text{if } \Lambda=\Lambda(c,R_\phi)\in G.
\end{equation}
But $\upointNull(e)$ (corresponding to $\upoint(e,p)$ from
Eq.~\eqref{equepPt}) also satisfies this equation. 
Lemma~\ref{StabPhot} implies that $u_0(e)_\pm$ and
$\upointNull(e)_\pm$ are linearly dependent for all $e\in\Spd$ with
$e_0\neq e_3$, and hence, by analyticity, for all $e\in \Spdc$.  One
then concludes precisely as in the proof of Theorem~\ref{StringField}
after Eq.~\eqref{equ0Cov} that $u_0(e)_\lambda=F(\bar p\cdot
e)\upointNull(e)_\lambda$, where $F$ is analytic on the upper half
plane except, possibly, at those $ \bar p\cdot e$ with
$\upointNull(e)_\lambda=0$.  But the latter equation is satisfied if
and only if $e$ is of the form $(e_0,e_1,\pm ie_1,e_0)$.  Now such $e$
satisfies $e\cdot e=0$ and is not in $\Spdc$.  Hence $\upointNull$ has
no zeroes in $\Spdc$, and $F$ must be analytic.  Finally, one checks
that $\upoint(e,p)$ satisfies the self-conjugacy
condition~\eqref{equConjPhot}.  Then $u(e,p)$ satisfies this condition
if and only if $\overline{F(-\omega-i0)}=F(\omega+i0)$,
$\omega\in\RR$.  The statements about moderate growth, polynomial
boundedness and on point-like localization follow as in the proofs of
Theorem~\ref{StringField} and Proposition~\ref{um>0}.

Ad $ii)$. Given any string-localized $A_\mu(x,e)$ as in $i)$, {define} $ F_{\mu\nu}(x,e)$ by 
\begin{equation} \label{eqdAdA} 
\partial_\mu A_\nu(x,e)-\partial_\nu A_\mu(x,e). 
\end{equation} 
Suppose this field  is independent of $e$. Then it transforms as in equation~\eqref{eqFTens}, and the Jost-Schroer-Pohlmeyer theorem implies that it coincides, up to unitary equivalence, with the free field strength $F_{\mu\nu}(x)$ from Eq.~\eqref{eqFa*a}. It follows that $A_\mu(x,e)$ is a potential for the free field strength $F_{\mu\nu}(x)$ in the sense of Eq.~\eqref{eqFdA} if, and only if, the above expression~\eqref{eqdAdA} is independent of $e$. The latter condition translates to $e$-independence of the expression 
$$
p_\mu u(e,p)_{\pm,\nu} -  p_\nu u(e,p)_{\pm,\mu}\, \equiv \, -i F(e\cdot p) \,(e\cdot p) 
\big\{\hat e_{\mp}(p)_\mu p_\nu - \hat e_{\mp}(p)_\nu p_\mu\big\}.   
$$
Clearly, this is independent of $e$ if and only if $F(\omega)=$const.$/\omega$. By hermiticity of $A_\mu$ and the correct normalization, the constant must equal the imaginary unit $i$. This proves the claim.  

Ad $iii)$. As mentioned, one checks that $\upoint(e,p)_\pm$ from Eq.~\eqref{equepPt} is orthogonal to $e$ and to $p$. Hence, by the uniqueness statement $i)$, the same holds for the intertwiner corresponding to the field $A_\mu(x,e)$ at hand. This implies the ``gauge'' conditions. (A direct argument, without the special intertwiner $\upoint$, goes as follows. Equation~\eqref{equ0CovPhot} implies that $u_0(e)$ is an eigenvector for all $\Lambda(c,R_\phi)\in G$ which leave $e$ invariant, with eigenvalue $e^{\mp i\phi}$. Multiplying with $e$ yields that either $\phi=0$ or $e\cdot u_0(e)=0$. But the proof of Lemma~\ref{StabPhot} shows that for all $e\in\Spd$ with $e_0\neq e_3$ there is a $\Lambda(c,R_\phi)$ leaving $e$ invariant and which has $\phi\neq0$. Hence $e\cdot u_0(e)=0$ for such $e$, and 
by analyticity for all $e\in\Spdc$. The same goes through for $p$.) 
\end{Proof}

Similarly constructed string-localized analogs of potentials for
point-like ``field strengths'' can be incorporated into the higher
helicity Wigner representations.  A particularly interesting case is
the string localized metric tensor as the potential for the 
field strength in the case of helicity 2, the latter being a tensor of rank 4.  The answer to the
question of whether these objects offer a useful alternative to the
gauge formalism (which saves the point-like nature of potentials at
the expense of introducing unphysical ``ghosts'' in intermediate
steps) depends on whether it will be possible to extend perturbation
theory to include string-like localized fields.

Our string-localized vector-potential construction has an interesting
connection with the breakdown of Haag duality for non-simply connected
localization regions as pointed out by Leyland, Roberts and Testard
in~\cite{LRT}.  These authors show that the flux of the
electromagnetic field through a torus commutes with every observable
localized in the causal complement of the torus, but is not localized
in the (causal completion of the) torus.  The present viewpoint helps
to understand this mismatch.  Namely, since $F=dA$, the flux through a
torus $T$ can be expressed by an integral of $A_\mu(x,e)$ over the
torus and hence is localized in $T''+\RR^+_0 e$, where $T''$ is the
causal completion of $T$ and $e$ can be chosen at will.  Given an
observable $B$ localized in a double cone $\calO$ causally disjoint
from the torus, one can choose the direction $e$ such that
$T''+\RR^+_0 e$ is causally disjoint from $\calO$, and hence the flux
commutes with $B$.  (The same can be achieved, of course, if one
defines a vector potential as in the classical proof of the Poincar\'e
Lemma via line integrals starting from a common finite base point
instead of space-like infinity as in \eqref{eqA}.)

\section{Massless Infinite Spin Particles} \label{massless}
Here we construct a family of intertwiners $\intfctdeg{e}{p}$ along the lines of Section~\ref{Construction} for the massless infinite spin particles, labeled by $\degree\in\CC$ with $\Re\degree<0$.  
In $d=4$ (Subsection~\ref{massless4d}), it turns out that for $\Re\degree\in[-2,-\half)$, they have mild $UV$ behavior, namely after smearing with a test function $h\in\calD(\Spd)$ they are bounded in $p$.   
We also find intertwiners which are {\em functions} on $\Spd$, leading 
to localization in space-like half-cylinders. This improves the result of the abstract analysis~\cite{BGL} which guarantees only localization in space-like cones. In $d=3$ (Subsection~\ref{massless3d}), and for $\Re \degree\in(-1,0)$, our intertwiners are uniformly bounded in $e$ and $p$. This leads to fields which are well-behaved with respect to UV-behavior 
{\em and} to localization (in that they are localizable in space-like half-cylinders instead of cones). 

We first recall the irreducible representations ${D}$ of the
little group $G$ corresponding to these particle types, 
and of ${j_0}$. \label{Dm=0}
For $m=0$, $G$ is isomorphic to the euclidean group $E(d-2)$. 
Recall that the irreducible representations 
of $E(1)=\RR$ are labeled by $\PauliLub\in\RR$ and act in $\CC$ as
$\real\mapsto e^{i\PauliLub\real}$, 
and that the {\em faithful\/} irreducible unitary representations  of $E(2)$ are labeled by
$\PauliLub\in\RR^+$, with $D=D_\PauliLub$ acting on 
$\calh:=L^2(\Bb^2,d\nu_\kappa(k))$, where $d\nu_\kappa(k):=\delta(|k|^2-\PauliLub^2)d^2k$, according to
\begin{align} \label{eqIrrepE(2)}
(D(c,R) u) (k) &:= e^{ic\cdot k} \,u(R^{-1}k),\quad (c,R)\in E(2). 
\end{align}
These representations extend to a representation of the semi-direct product of $G$ and $j_0$ by the anti--unitary involution $D(j_0)$. Namely, $D(j_0)$ is complex conjugation (point-wise in the $4$ dimensional case). 
(We show in Lemma~\ref{Dj} that $D(j_0)$ indeed satisfies the representation properties
$D(j_0)D(\Lambda)D(j_0)=D(j_0\Lambda j_0)$, $\Lambda\in G$.)  

We now specify the general construction of Section~\ref{Construction}. 
For $m=0$, and base-point $\bar p:=(1,0,0,1)$ or $(1,0,1)$ in $\HypNull$, the set $\orbit$ of all $q\in\HypNull$ with $q\cdot \bar p=1$, cf.~\eqref{eqOrbit}, is 
isometric to the euclidean space $\RR^{d-2}$ via the parametrization of $\orbit$ given by 
\begin{align} \label{eqxi1} 
\xi(r)&:= 
\big(\frac{1}{2}\left(r^{2}+1\right),r,\frac{1}{2}\left(r^{2}-1\right)\big),
\quad r \in\Bb, & d&=3,\\
\xi(z)&:=\big(\frac{1}{2}\left(z^{2}+1\right),z_1,z_2,\frac{1}{2}\left(z^{2}-1\right)\big),  
 \quad z\in \RR^2, & d&=4, \label{eqxi} 
\end{align} 
where $z^2:=z_1^2+z_2^2$ (cf.~Lemma~\ref{Orbit}). The isomorphism $\xi$ from $\RR^{d-2}$ onto $\orbit$ identifies the action of $G$ in $\orbit$ with the action of $E(d-2)$ in $\RR^{d-2}$, and $-j_0$ acts as $z\mapsto -z$. 
One gets generalized intertwiners
$V=V_{\PauliLub}$ from the representation $\RepPulBac$, cf.~\eqref{eqRepPulBac}, 
to the irreducible representation $D=D_{\PauliLub}$ as 
\begin{align} 
 V_{\PauliLub}v 
  &:= \tilde{v}(\kappa)= \int_{\Bb}dr e^{i\kappa r } \, v(\xi(r)),& d=3, \label{eqIntm=0} \\
(V_{\PauliLub}v)(k) 
  &:= \tilde{v}(k)= \int_{\Bb^2}d^2z e^{ik\cdot z}\,v(\xi(z)), 
  \quad |k|^2=\kappa^2,& d=4.   \label{eqIntm=0'} 
\end{align}
(This is of course only defined on the dense sets where the 
restrictions of the Fourier transforms to a fixed value $\kappa$ or to
$|k|^2=\kappa^2$, respectively, make sense.) 
Thus, our construction \eqref{equF} and \eqref{equExp} leads {\em formally} to the following
intertwiners, defined for $e\in\Tub$ and $p\in\HypNullMinus$ :  
\begin{align}
\intfctdeg{e}{p} &= e^{-i\pi\degree/2} \,\int_{\Bb} dr\, e^{i\kappa r} \, 
  (B_p\xi(r)\cdot e)^\degree, & d=3, \label{equm=0d=3} \\
 \intfctdeg{e}{p}(k) &= e^{-i\pi\degree/2}\, \int_{\Bb^2} d^2z\, e^{ik\cdot z} \,(B_p\xi(z)\cdot e)^\degree, \quad
|k|=\kappa,\; &d=4.     \label{equm=0d=4}
\end{align}
(We have again introduced a factor $e^{-i\pi\degree/2}$ for later convenience.) 
Here, $\degree$ is a complex number with 
$$
\ReDeg:= \Re \degree < 0.  
$$
In the following subsections, we will make these expressions precise and prove that they
actually  have the properties they should formally have. 
\subsection{Intertwiners for d=4.}  \label{massless4d}
The function  $z\mapsto B_{p}\xi(z)\cdot e$ is a  polynomial in
$z$ without any real zeroes if $\spd\in\Tub$, cf.~Eq.~\eqref{eqP(z)}. 
It follows that the integral in Eq.~\eqref{equm=0d=4} exists and defines a continuous function 
$\intfctdeg{\spd}{p}$ of $k$. We show in Proposition~\ref{IntMassless4d} that it has indeed the required properties and that,  after smearing with a test function $h\in\calD(\Spd)$, it is bounded in $p$ for $\ReDeg\in[-2,-1/2)$.

Considering the limit of \eqref{equm=0d=4} for $\spd$ approaching the real ``boundary'' $\Spd$, one has to note that 
the polynomial $B_{p}\xi(z)\cdot e$ is linear if $\spd\cdot p=0$ and quadratic if $\spd\cdot p\neq 0$, cf.\ Eq.~\eqref{eqpxie}. Hence for real 
$\spd\in\Spd$ with $\spd\cdot p=0$ the integral diverges. On the other hand, if  $\spd\cdot p\neq 0$ then the  polynomial is of the form $(z-z_0)^2-(\spd\cdot p)^{-1}$, cf.\ Eq.~\eqref{eqP(z)}. The corresponding pole, for  real $\spd\in\Spd$, of the integrand is therefore integrable iff $\ReDeg>-1$. 
It follows that for $\ReDeg>-1$ the singular set on $\Spd$ of the distribution $u^\degree(\spd,p)$ consists precisely of those $\spd$ with $\spd\cdot p=0$. This can be cured by multiplying this distribution with a suitable power of $\spd\cdot p$. Then one ends up with an intertwiner which is a {\em function} on $\Spd$, thus leading to localization in space-like half cylinders, and to fields which do not have to be smeared in $\spd$.
In fact, it turns out that 
\begin{equation}\label{eqIntFctString}
\hat{u}^\alpha(e,p):= (e\cdot p)^2 \,\intfctdeg{e}{p}
\end{equation}
enjoys the mentioned properties if  $-1<\ReDeg<-\half$. 
\begin{Prop}[Intertwiners for infinite spin, $\boldsymbol{d=4}$] \label{IntMassless4d}
$\intfctdeg{e}{p}$ is an intertwiner function in the sense of
Definition~\ref{DefIntFct}. The ``conjugate'' intertwiner function $(u^\degree)_c$
defined in Eq.~\eqref{eqv} coincides with $u^{\bar\degree}$. 
If $\ReDeg\in[-2,-\half)$, then for given $h\in\calD(\Spd)$ the norm of $\intdistdeg{h}{p}$ is bounded in $p$. 

Further, if $\ReDeg\in(-1,-\half)$, then $\hat{u}^\alpha(e,p)$ as defined in Eq.~\eqref{eqIntFctString} is an  intertwiner function with growth order $N=0$ in~\eqref{eqModGrowth}, and whose norm is bounded by const.$\times |\spd\cdot p|^2$. 
\end{Prop}
The rest of this subsection is concerned with the proof of the proposition. 
\begin{Proof} 
Let $\Lambda\in G$ correspond to $(c,R_\vartheta)\in E(2)$ under the identification 
$G\cong E(2)$. 
Then, for $e\in\Tub$, 
\begin{align} \label{eqIntLor'}
& D_{\kappa}(\Lambda)u^{\alpha}(e,\bar{p})(k)
=e^{ic\cdot k}u^{\alpha}(e,\bar{p})(R_{\vartheta}^{-1}k)\,\nonumber\\
& =e^{-i\pi\alpha/2}
\int d^{2}ze^{ik\cdot z}\left(\xi\left(R_\vartheta^{-1}(z-c)\right) \cdot e\right)  ^{\alpha}\nonumber\\
& =e^{-i\pi\alpha/2}\int d^{2}ze^{ik\cdot 
z}\left(  \Lambda^{-1}\xi (z) \cdot e\right)  ^{\alpha}=u^{\alpha}(\Lambda e,p)(k). 
\end{align}
This implies Eq.~\eqref{equ0Cov} and thus establishes
the intertwiner property~\eqref{equCov}. 
To prove that $(u^\degree)_c$ coincides with $u^{\bar\degree}$, we consider 
\begin{align} \label{eqIntJ'}
\intfctdegconj{e}{-j_0p}(k)
&=e^{-i\pi\bar\alpha/2} \int d^{2}ze^{ik\cdot z}
\left(  B_{-j_0p}\xi(z) \cdot e\right)^{\bar\alpha}\nonumber\\
& =e^{-i\pi\bar\alpha/2}\int d^{2}ze^{ik\cdot z}
\left(-B_{p}\xi(-z) \cdot j_0 e \right)^{\bar\alpha} \nonumber\\
&=e^{i\pi\bar\alpha/2}\int d^{2}ze^{-ik\cdot z}
\left(B_{p}\xi(z) \cdot {j_0 e}\right)^{\bar\alpha}=
\overline{\intfctdeg{j_0e}{p}(k)}  \nonumber\\
&= \big(D(j_0)\intfctdeg{j_0e}{p}\big)(k) . 
\end{align}
In the second line we have used the fact that $j_0\xi(z)=-\xi(-z)$ and
equation~\eqref{eqjBpj} to conclude that
$B_{-j_0p}\xi(z)=-j_0B_p\xi(-z)$. In the third line we have used the facts that
$(-w)^\degree=e^{i\pi\degree}w^\degree$ for $w\in\Bb+i\Bb^-$, and that 
$\bar w^{\bar\degree}=\overline{w^{\degree}}$ for $w\in\Bc\setminus \Bb_0^-$. 
This implies that 
$(u^\degree)_c=u^{\bar\degree}$, as claimed. 

As to analyticity, we already know that the integrand in the definition~\eqref{equm=0d=4} of the intertwiner, is analytic on the tuboid $\Tub$. It turns out that this property survives after the integration, hence $e\mapsto\intfctdeg{e}{p}(k)$ is analytic, point wise in $p$ and $k$. This is made rigorous in Lemma~\ref{IntFctAna}. 
Now Lemma~\ref{IntFctEst} implies that (for fixed $p$) the continuous functions 
$k\to \intfctdeg{e}{p}(k)$ are dominated by a suitable constant, uniformly for 
$e$ in a compact set in $\Tub$. It follows that $\intfctdeg{e}{p}$ is analytic as an $L^2(\RR^2,d\nu_\kappa)$-valued function. 
  
The main work in establishing the bound~\eqref{eqModGrowth} is done in Lemma~\ref{IntFctEst}, where we show that 
for all $\spdc=\spdr+i\spdi\in\Tub$, $p\in\HypNullMinus$ and $k$ with $|k|=\kappa$ holds  
\begin{equation} 
|\intfctdeg{\spdc}{p}(k)| \leq   c \, |p\cdot\spdc|^{-\ReDeg+n-2} + \sum_{\nu=0}^{[n/2]}  
c_\nu \, (\spdi^2)^{\ReDeg-n+\nu +1} \, (p\cdot\spdi)^{-\ReDeg+n-\nu-1} \, |p\cdot\spdc|^{\nu-1},\label{equEst'}
\end{equation}
where $n$ is any natural number strictly larger than $2\ReDeg+2$. 
This estimate implies the bound~\eqref{eqModGrowth} as follows.  
Consider the canonical norm in $\RR^{d}$ given by $|e|^2:=e_0^2+\sum_{k=1}^{d-1}e_k^2$.  
Let $\Theta$ be a subset of $\Tub$ as in \eqref{eqTheta}. 
We claim that there are positive constants  $c_1$ and $c_2$ (depending on $\Theta$) 
such that for all $e=e'+ie''\in\Theta$ the following inequalities hold: 
\begin{eqnarray}
c_1  |e''|^2  \,\leq&  (e'')^2              &                   \label{eqy2}\\
c_1 \, p_0 \,|e''| \, \leq & p\cdot e'' & \leq p_0 \,|e''| \label{eqpy}\\
                    &     |p\cdot e'|            &\leq \,c_2\, p_0 .      \label{eqpx}
\end{eqnarray}
As to the first inequality, note that $e''$ is contained in the cone
$\RR_0^+\Omega_2$, cf.~\eqref{eqTheta}, which implies that 
$$ 
(\spdi_0)^2\geq
(1+\eps)\,|\underline{\spdi}|^2, 
$$ 
for some $\eps>0$ depending on $\Omega_2$. Here we have written 
$|\underline{e''}|^2:=\sum_{i=1}^3(e'')_i^2$.   
This implies that $\spdi^2\geq \eps |\underline{\spdi}|^2 $ and  hence
Eq.~\eqref{eqy2}, with $c_1:= (1+2/\eps)^{-1}$. 
Next, note that the Cauchy Schwartz inequality implies that 
\begin{align}  \label{eqCSU}
  p_0 ( e''_0-|\underline{e''}|) 
 \leq p\cdot e'' \leq p_0 (e''_0+|\underline{e''}| )\leq p_0 |e''| 
\end{align}
holds for $p\in\HypNullMinus$. 
Now $\spdi^2=(\spdi_0-|\vec{\spdi}|)(\spdi_0+|\vec{\spdi}|)\leq 
(\spdi_0-|\vec{\spdi}|)|\spdi|$, hence Eq.~\eqref{eqy2} and the
l.h.s.\ of \eqref{eqCSU} imply the l.h.s.\ of \eqref{eqpy}. 
Similarly, the Cauchy Schwartz inequality implies that $|p\cdot e'|$ 
$\leq p_0(e'_0+|\underline{e'}|)$,  which proves ineq.~\eqref{eqpx}
since $e'$ has been taken from a compact set. 
The inequalities~\eqref{eqy2} to \eqref{eqpx} imply that 
\begin{align}
 (e''^2)^{s} &\leq c \, |e''|^{2s},&  s<0, \label{eqy2'} \\
 (p\cdot e'')^{s}&\leq c\, (p_0 |e''|)^{s},& s\in\RR, \label{eqpy'} \\
|p\cdot e|^{s}&\leq c \, p_0^{s} ,& s \geq 0.\label{eqpx'}
\end{align}
Using these inequalities, and $|\spdi|\leq c$ (which follows from~\eqref{eqy2} since $(\spdi)^2\leq1$ for $\spdc\in\Tub$), one gets from ineq.~\eqref{equEst'} the bound 
\begin{equation} \label{eqx_vs_e}  
|\intfctdeg{e}{p}(k)|\leq c_n p_0^{-\ReDeg+n-2} |\spdi|^{\ReDeg-n} 
\quad \text{ for }  n>2\ReDeg+2, 
\end{equation} 
and hence a similar bound for $\|\intfctdeg{e}{p}\|$. Choosing $n$
large enough, one concludes that the claimed
bound~\eqref{eqModGrowth} is satisfied, with $M(p)=p_0^{-\ReDeg+n-2}$ and growth order $N$ smaller or equal to  $n-\ReDeg$. 

In order to prove boundedness of $\intdistdeg{h}{p}$ for $\ReDeg\in[-2,-\half)$, we consider the best bounds contained in \eqref{eqx_vs_e}, corresponding to the smallest $n>2\ReDeg+2$. 
For $\ReDeg<-1$ we may take $n=0$, hence $M(p)=p_0^{-\ReDeg-2}$, $-\ReDeg-2>-1$. 
For $\ReDeg\in[-1,-\half)$, we may take $n=1$, hence $M(p)=p_0^{-\ReDeg-1}$, $-\ReDeg-1\in(-\half,0]$. 
For $\ReDeg\in[-\half,0)$, we may take $n=2$, hence $M(p)=p_0^{-\ReDeg}$, $-\ReDeg\in(0,\half]$. 
Hence, for $\ReDeg\in[-2,-\half)$ one has $M(p)=p_0^r$ for some $r\in(-1,0]$. Then Eq.~\eqref{eqIntDistBound} implies that the norm of $\intdistdeg{h}{p}$ is bounded by $p_0^r$ (times a constant depending on $h$), hence it is bounded for large $p$. But increasing $n$ by $1$ in the above considerations, one also gets the bound $p_0^{r+1}$, where $r+1\in(0,1]$, hence the norm is also bounded for small $p$. This implies that the norm of $\intdistdeg{h}{p}$ is bounded for given $h$ if $\ReDeg\in[-2,-\half)$. 

The function $\hat{u}^\degree$ inherits the intertwiner property~\eqref{equCov} and analyticity from $u^\degree$. 
As to the claim on the vanishing growth order for $\ReDeg\in(-1,-\half)$, we show in Lemma~\ref{IntFctEst'} that for all $\spdc\in\Tub$, 
$p\in\HypNullMinus$, $k\in\RR^2$ with $|k|=\PauliLub$ the following estimate holds: 
\begin{align} \label{eqEstIntFct''}
|\intfctdeg{\spdc}{p}(k)| &\leq  c_1 \,|p\cdot \spd|^{-\ReDeg-1}+c_2\,|p\cdot \spd|^{-\ReDeg-2}. 
\end{align}
Inequality~\eqref{eqpx} then implies that $\|\hat{u}^\alpha(e,p)\|\leq c_1 p_0^{-\ReDeg} + c_2
p_0^{-\ReDeg+1}$ for all $e\in\Theta$, where $\Theta$ is a subset of $\Tub$ as in \eqref{eqTheta}. 
This proves the bound~\eqref{eqModGrowth}, with growth order $N=0$. Ineq.~\eqref{eqEstIntFct''} also implies 
that the norm of $\hat{u}^\degree(e,p)$ is bounded by $|\spd\cdot p|^2$. This completes the proof.  
\end{Proof}

\subsection{Intertwiners for d=3} \label{massless3d}
Recall that we have defined intertwiners, for $\spd\in\Tub$ and $p\in\HypNullMinus$, by  
\begin{align} \label{eqIntFctDef3d}
\intfctdeg{\spd}{p}:= e^{-i\pi\degree/2}
\int_{\RR}\d \real  e^{i\PauliLub \real }\,(B_p\lc(\real )\cdot e)^\degree.   
\end{align}
Here, $\kappa$ is the real number characterizing the representation of the 
reals (and hence of the Poincar\'e group) at hand.  
We will consider the case 
\begin{equation}  \label{eqRealDegree} 
\ReDeg := \Re  \alpha \in (-1,0). 
\end{equation}
We show in Lemma~\ref{IntFct3d} that for these values of $\degree$, 
$\intfctdeg{\spd}{p}$ is a {\em bounded}  function on $\Tub\times\HypNullMinus$. 
With the same methods as used in the proof of Proposition~\ref{IntMassless4d}, 
this implies the following facts: 
\begin{Prop}[Intertwiners for `infinite spin', $\boldsymbol{d=3}$] \label{IntMassless3d} 
$\intfctdeg{e}{p}$ is an intertwiner function in the sense of
Definition~\ref{DefIntFct}. It is uniformly bounded in $e$ and $p$, in particular 
has growth order $N=0$ in \eqref{eqModGrowth}.  
Further, the ``conjugate'' intertwiner function $u_c$
defined in Eq.~\eqref{eqv} coincides with $u^{\bar\degree}$. 
\end{Prop}
\subsection{Compactly Localized Two-Particle States}  
We now address the important question of the existence of compactly localized 
observables. 
If $A=A^*$ is localized in a region $\calO$, then $A\Omega$  is modular-localized in 
$\calO$, i.e. 
$$
A\Omega\in {\mathcal K}(\calO),
$$
where ${\mathcal K}(\calO)$  
is defined as in eqs.~\eqref{eqModOp} -- 
\eqref{eqKW} and  \eqref{int}, with $\Uirr$ replaced by its second quantization. 
This is a consequence of the Bisognano-Wichmann property which holds in our model.  
Thus the existence of compactly modular-localized vectors is a necessary condition for
the existence of compactly localized operators. Now the spaces ${\mathcal K}(\calO)$ are, in principle, known to us and we can therefore decide whether or not this necessary condition is satisfied. The answer is positive, and  
we shall, as an example, exhibit two-particle state vectors 
which are compactly localized in the sense of modular localization.  We restrict to the
$4$-dimensional case. 
 
Let $F\in\calS(\RR)$. We define, for $p,q$ in $\HypNullMinus$
and $k,l$ in $\Bb^2$, a ``two--particle intertwiner function''  
\begin{equation}\label{eqIntFctDef2}
u_2(p,q)(k,l) := \int d^{2}z d^2w \, e^{i(k\cdot z+l\cdot w)} \;
F\left(  B_{p} \xi(z)\cdot B_{q} \xi(w)\right).
\end{equation}
A straightforward calculation yields: 
\begin{Lem}  \label{IntFct2Cov}
For fixed $p$ and $q$, the function $(k,l)\mapsto u_2(p,q)(k,l)$ is in 
$L^2(\Bb^2,d\nu_\kappa)^{\otimes 2}$ and has the following
intertwining property: 
\begin{align}
\big(D_\kappa(R(\lor,p))\otimes D_\kappa(R(\lor,q))\big)\,
u_2(\lor^{-1}p,\lor^{-1}q) = u_2(p,q),\quad\lor\in\Lor. 
\end{align}
\end{Lem}
We then define, for $f_1,f_2$ in $\calS(\Min)$, a covariant
two--particle wave function $\covfcttwo{f_1}{f_2}$ by 
\begin{align}   \label{eqCovfcttwo2}
\covfcttwo{f_1}{f_2}(p,q;k,l):= (\FT f_1\otimes_{\rm s}\FT f_2)(p,q)\;u_2(p,q)(k,l),  
\end{align}
where again $\FT f$ is the restriction of the Fourier transform of $f$ to the mass shell and $\otimes_{\rm s}$ denotes that symmetrized tensor product. 
These wave functions have the following covariance and localization
properties. (We denote by $U_2$ the two-fold symmetric tensor power of $\Uirr$, 
and define $K_2(\calO)$ as the intersection of ${\mathcal K}(\calO)$ with the two-particle space.) 
%
\begin{Prop}[Localization of two-particle state vectors]\label{CovLoc2Part} 
Let $f_1$ and $f_2$ be in $\calS(\Min)$. 

i) $\covfcttwo{f_1}{f_2}$ is in the two-particle space $L^2(\HypNull\times\Bb^2,d\mu
d\nu_\kappa)^{\otimes_{\rm s}2}$. 

ii) Let $g\in\Po$ and let $j\in\Poj$ be the reflection at the edge of
some wedge. Then\begin{align} 
U_2(g) \,\covfcttwo{f_1}{f_2} &= \covfcttwo{g_*f_1}{g_*f_2}, \label{eqCovLoc}\\
U_2(j) \,\covfcttwo{f_1}{f_2} &= \covfcttwo{j_*\bar{f_1}}{j_*\bar{f_2}}. \label{eqCovLocj}
\end{align} 

iii )
Let $f_1$ and $f_2$ be real valued test functions with support in a compact set
$\calO\subset\Min$. Then the vector $\covfcttwo{f_1}{f_2}$ is in $K_2(\calO)$. 
\end{Prop}
\begin{Proof}
Equation~\eqref{eqCovLoc} of $ii)$ follows from
Lemma~\ref{IntFct2Cov}. This equation in turn implies that the
analyticity properties of $t\mapsto U_{(2)}(\Boo{W}{t})\,\covfcttwo{f_1}{f_2}$ depend entirely on
those of $t\mapsto \FT\Boo{W}{t}_*f_i$, $i=1,2$, and hence of the scalar
representation. This implies that $\covfcttwo{f_1}{f_2}$ is in the  domain of $\Delta_W^{1/2}$ whenever $W$ contains $\calO$, and that 
\begin{align} 
\Delta_{W}^\half \, \covfcttwo{f_1}{f_2} &=\covfcttwo{(j_W)_* f_1}{(j_W)_*f_2}\,.  \label{eqDeltaPsi2}
\end{align}
Together with Eq.~\eqref{eqCovLocj}, this implies that 
$ 
S_W\covfcttwo{f_1}{f_2} =\covfcttwo{\bar f_1}{\bar f_2} 
$
whenever $W$ contains $\calO$. This shows $iii)$ and completes the proof. 
\end{Proof}
Since our results on compact localization are not yet conclusive, it may be
interesting to comment on how one would proceed to settle this problem. The
compact localized wave function in the two-fold tensor product of the
indecomposable infinite spin representation suggests to look for an operator
of the form%
\begin{align}
&B(x,y)=\int d\nu(k)d\nu(l)d\mu(p)d\mu(q)\,e^{ipx+iqy}u_{2}(p,q)(k,l)a^{\ast}(p,k)a^{\ast}(q,l)+...\\
&\omega_0\left(\left[  B(x,y),B(x^{\prime},y^{\prime})\right]\right)
=0,\;\text{ if }\,x,y \,\text{space like to }\; x^{\prime},y^{\prime},%
\end{align}
where the pure creation component has to be complemented by its hermitian
adjoint and a mixed (normal ordered) $a^{\ast}a$ component. The previous
construction then guarantees that the bilinear $B(x,y)$ is local within the
vacuum state if, as written in the second line, the pair $x,y$ is space like
with respect to $x',y'$. This is a slight generalization of the well-known
statement that locality within the vacuum state is an automatic
consequence of covariance within the vacuum state. The locality of the
mixed contributions on the other hand is equivalent to the vanishing of the matrix elements of
the commutator between
particle states. The latter property is not guaranteed by covariance and the
appropriate tool to decide whether it can be achieved is the validity of the
Jost-Lehmann-Dyson representation. As a result of the Fock space structure the problem of locality can be systematically investigated by starting with degree two monomials in Wigner creation/annihilation operators and, in case there is no solution, to extend the calculation to include higher degree polynomials. Although a clarification is important for a physical use of these Wigner representations we will defer a systematic search for local observable sub-algebras to future work.  

A negative result would mean that these string-localized fields do not admit
local observable sub-algebras, which would lend theoretical support to the idea that 
``nature cannot make
direct use'' of such representation. However even if they do not appear as
indecomposable asymptotic states in collision theory they might play a more
hidden role; unlike tachyons they are positive energy objects and share the
stability and localization properties which are common to all positive energy representations.


%
\section{String-Localized Fields and String (Field) Theory
}

In this section an attempt will be made to compare the concept of
string-localized fields with String Theory 
\cite{Polchinski}. Despite
the shared use of the word ``string'' this is not an easy task since the
conceptual position of string theory in particle physics, its historical roots
in the dual model and its ability in catalyzing new mathematical ideas
notwithstanding, is not anywhere close to the firm embedding of
string-localization (in the sense of the present paper) 
in the general principles underlying QFT. Our conclusion will be
that contrary to the intuitive appeal of the common word ``string'', the two
concepts have little in common. This said, the reasons behind this negative
conclusion are actually quite interesting and worthwhile to be presented. In
the present case it turns out that they reveal a lot about the role of
classical versus quantum localization and the limits of Lagrangian
quantization in particle physics.

An appropriate conceptual understanding of the aims of String (Field)
Theory is difficult to gain by looking only at the highly technical
actual computations, a glance at its history on the other hand is less
confusing.  String Theory started with the observation that
Veneziano's proposal for a crossing symmetric (but not yet unitary)
$S$-matrix permits an auxiliary description in terms of a classical
string Lagrangian.  After a classically permitted reformulation and a
subsequent canonical quantization this Nambu-Goto Lagrangian
reproduces the Veneziano amplitude, including the underlying
mass-tower spectrum.  However the operator formulation of this
auxiliary string theory led to a Poincar\'e Lie algebra only for very
special values of the string's ambient space-time
dimension\footnote{There is however apparently no proof in the
existing literature that the joint domain properties of the unbounded
Lie algebra generators in the light-front quantization allow an
exponentiation to a unitary representation of the Poincar\'e group
.}.  This result is surprising and goes somewhat against common sense
since one would not expect the answer to such a deep problem (why
nature selects a particular dimension for the space-time vessel for
quantum matter) to be obtained simply from the quantization of a
classical relativistic string.  It was certainly not a restriction in
the logic of Veneziano's dual $S$-matrix theory (crossing and
unitarity are not capable to select a preferred space-time dimension). 
There is of course the alternative to treat the quantization problem
of the Nambu-Goto string in compliance with its nature as an
integrable system, which would not lead to a limitation in space-time
dimension.  But it has been shown by Bahns \cite{Bahns} that the
result would be inequivalent to that of the canonical quantization
(which is the one needed to reproduce the dual model).  As the
quantization compatible with the integrability, the N-G string and the
string-localized fields of the present work share the property to be
consistent with any space-time dimension $d\geq 3.$ However, as
mentioned before there are good reasons to believe that quantized
classical relativistic strings do not lead to string-localization in
the sense of the present article \cite{Dimock, Erler-Gross}.  The main
point of our brief excursion into the history of string theory was to
make clear that there is also no reason why a string-$S$-matrix theory
(where the terminology ''string'' has an auxiliary meaning) should
comply with the weakening of compact localization for certain positive
energy representations of the Poincar\'e group.  Whereas the auxiliary
role of the canonically quantized N-G string in the the dual
$S$-matrix prescription does not impose any physical localization
properties (such properties only apply to interpolating fields of an
$S$-matrix), the conjectured absence of string localization in the
more intrinsic quantization of the N-G Lagrangian based on conserved
charges would be more surprising.

It is a remarkable fact that the solutions to the bootstrap
program for two-dimensional purely elastic $S$-matrices (factorizing
models) permit a classification.  The existence of an abundance of
solutions shows that, contrary to the original expectations, the
$S$-matrix bootstrap principle is in no way more restrictive than the
axioms of QFT. In fact the bootstrap-formfactor program associates a
unique QFT (in terms of its generalized formfactors) with a given
factorizing crossing symmetric unitary $S$-matrix, and modular
localization permits to understand the associated computational
recipes on a fundamental level of local quantum physics 
\cite[and earlier papers cited therein]{Lechner, crossing}.  
The bootstrap formfactor
approach is an example of a field theoretic construction outside the
Lagrangian quantization scheme \cite{Babujian}.  In contrast to the dual model and contemporary string
theory the bootstrap $S$-matrices do not contain infinite particle
towers, rather the implementation of crossing is achieved through a
delicate interplay of one-particle poles with the scattering
continuum.  The formulation of the bootstrap-formfactor constructions
in the setting of modular wedge localization and their relation to the
Zamolodchikov-Faddeev algebra structure suggests to view these
constructions as a special case of an extension of the ideas
underlying the Wigner particle-based representation theory to
interacting particles and their local fields.

There are some similarities between string-localized fields and string theory
which have attracted our attention at the beginning of our research. The
helicity tower of the infinite spin representation but also the Lorentz-spin
tower (for fixed physical spin) of massive strings suggests an analogy with
the infinite mass tower. Also the improvement of short distance behavior and
in particular the somewhat surprising fact that the short distance behavior of
e.g. massive string-localized fields is independent of the physical spin (for
point-like free fields it gets worse with higher spin) resembles the improvement of the
ultraviolet behavior in string theory. But in view of the before-mentioned
significant conceptual differences concerning localization we think that these
similarities are superficial.

\section{Concluding Remarks}

In this paper we have analyzed in detail the concept of string
localized quantum fields within the setting of free fields. There are
two main motives for studying such objects.  One reason is their
natural occurrance in concrete realizations of the B-G-L theorem
\cite{BGL} which assures the existence of quantum fields localized in
space-like cones for all positive energy representations of the
Poincar\'{e} group. This incorporates Wigner's infinite spin
representations as well as the description of photons in terms of a
covariant, string-localized vector potential that operates on the
physical particle space.

Another motive is the theorem of Buchholz and Fredenhagen\cite{BF}
which states that in a theory of local observables and massive
particles separated by a mass gap, the charge-carrying fields are not
worse than string-localized. Understanding the interaction free
situation is a necessary preparatory step towards (possibly
perturbative) constructions of interacting string-localized objects.
Since our free string-fields have milder short-distance behavior of
their two-point functions (independent of spin!) than point-fields
they potentially widen the framework of perturbatively admissible
interactions.

The modular setting and in particular the distinguished role of wedge-localized
algebras as the starting objects of an algebraic approach suggest to aim for
generators of wedge algebras (even if a simple algebraic characterization in
terms of PFGs as in the case of factorizable models is not possible). It is
not unreasonable to expect that by specifying an interaction through its
lowest order (tree graph) $S$-matrix one obtains a first order deviation of
the modular conjugation $J$ from its free value $J_{0}$. The imbalance
between the new commutant formed from $J$-transformed free field generators
and the original free field wedge generators would then require a first order
correction such that the relation between the modified generators and the
commutant is correct up to first oder but violated in the next order. In this
way one may arrive at an iterative scheme (for the wedge generators as well as
for the $S$-matrix) not unlike those existing perturbative schemes for the
iterative determination of local fields. The fact that the $S$-matrix enters in
the definition of the commutant of the wedge algebra is certainly a restriction
on the would-be generators of wedge algebras and whether this can be explored
in an iterative scheme and how many iterative solutions one obtains from a
lowest order $S$-matrix input are important unexplored problems. Unlike the
standard approach based on the computational use of point-like fields, a
construction of wedge generators remains ``on-shell'' (no short distance
correlations) and hence free of ultraviolet problems. It is therefore expected
to reveal the true frontiers created by the physical principles of QFT beyond
those which are generated by the computational use of unavoidably singular
point-like fields (non-/renormalizable). 

The proposal to permit string-like interactions  is
conceptually somewhere between the standard approach and the radical idea of
aiming at wedge generators and obtaining improved localizations and their
possible string- or point- like field generators via intersections of
algebras.


\appendix
\setcounter{equation}{0}
\renewcommand{\theequation}{\thesection.\arabic{equation}}
\setcounter{Thm}{0}
\renewcommand{\theThm}{\thesection\,\arabic{Thm}}
\section{Proofs} \label{Proofs}
\subsection{Proof of Proposition~\ref{Cov}.}
\newcommand{\sn}{\text{\rm p}}   
We prove Proposition~\ref{Cov}, after establishing three lemmas. The first one states a necessary and sufficient condition for a string   
\newcommand{\String}[2]{S_{#1,#2}} 
\begin{equation} \label{eqStringDef} 
\String{x}{e} := x+\RR_0^+\,e
\end{equation}
to be contained in a wedge. 
\begin{Lem} \label{StringWedge}
i) Let $x\in\RR^{d}$, $e\in\Spd$ and $W$ be a wedge region. The string $\String{x}{e}$ is contained in $W$ if and only if $x\in W$ and $e$ is in the closure of $W_H$. 

ii) If $\String{x}{e}$ is causally disjoint from $\String{x'}{e'}$ then $(x-x')^2 < 0$ and $(e-e')^2\leq0$.   
\end{Lem}
($W_H$ has been defined in Eq.~\eqref{eqWH}.)  
\begin{Proof}
Ad $i)$. It suffices to consider $W=W_0$ as defined in Eq.~\eqref{eqW0}, and we suppose $d=4$.  Then $\String{x}{e}\in W_0$ if and only if $|x_0+te_0|< x_3+te_3$ for all $t\geq0$. This condition implies that $x\in W_0$ and $e_3\geq0$. 
We may hence assume in the following that $x\in W_0$ and $e_3\geq0$, since these are consequences of both conditions whose equivalence we want to establish. We have to show that then $\String{x}{e}\subset W_0$ iff $e$ is in the closure of $W_0$. (Note that $(W_0)_H=W_0$.)  
Under our assumptions, 
$\String{x}{e}\subset W_0$ if and only if for all $t\geq0$ holds $f(t)>0$ with  
\begin{equation} \label{eqspacelike}
f(t):= (x_3+te_3)^2-(x_0+te_0)^2= (e_3^2-e_0^2)t^2+2(x_3e_3-x_0e_0)t+x_3^2-x_0^2.
\end{equation} 
Suppose first that $e_3^2-e_0^2=:a\neq0$. Then $f(t)$ is a quadratic polynomial with zeroes $t_\pm=-a^{-1}(x_3\pm x_0)(e_3\mp e_0)$. Thus $f(t)>0$ for all $t\geq0$ iff $a>0$ and both zeroes are strictly negative. Since $x\in W_0$ (hence $x_3\pm x_0>0$) by assumption, this is equivalent to $e_3+e_0>0$ and $e_3-e_0>0$, hence to $e\in W_0$. 
Suppose now that $e_3^2-e_0^2=0$. We show that this implies both $f(t)>0$ and $e\in W_0^\clo$. Namely, $e_3^2-e_0^2=0$ implies that $f(t)=2e_3(x_3\pm x_0)t+x_3^2-x_0^2$ is strictly positive since $x\in W_0$ and $e_3\geq0$. But $e_3^2-e_0^2=0$ (together with the hypothesis $e_3\geq0$) also implies $e_3=|e_0|$, i.e.\ $e\in \partial W_0$. This completes the proof of $i)$.  
Ad $ii)$. The hypothesis implies~\cite[Prop.\ x]{BGL} that there is a wedge $W$ such that $\String{x}{e}\subset W$ and $\String{x'}{e'}\subset W'$, where $W'$ denotes the causal complement of $W$. By $i)$, it follows that 
$x\in W$, $x'\in W'$, $e\in W_H^\clo$ and $e'\in (W'_H)^\clo$. This implies that claim. 
\end{Proof} 
The next two lemmas concern the properties of $\intdist{h}{p}$, defined in \eqref{eqIntDist}.  
\begin{Lem} \label{IntDist}
i) Let  $h$ be a smooth function with compact support in some ``wedge region'' $W_H$.\footnote{$W_H$ has  been defined in \eqref{eqWH}.} Then for
almost all fixed $p$ the $\calh$-valued function 
\begin{align} \label{eqIntDistAna}
t\mapsto \intdist{\Boo{W}{t}_*h}{p}
\end{align}
is the boundary value of an analytic function on the strip $\strip$, which is 
weakly continuous on the closure $\strip^\clo$ and satisfies the boundary condition 
\begin{align} \label{eqIntDistBV}
\intdist{\Boo{W}{i\pi}_*h}{p}= \intdist{({j_W})_*h}{p}. 
\end{align} 
Further, for given compact subsets $\Omega$ of $\Spd$ and $\strip_0$ of
$\strip^\clo$, there is some $c>0$ such that for
all $h$ with $\supp h\subset \Omega$ and $z\in \strip_0$ holds 
\begin{equation} \label{eqIntDistBound}
\|\intdist{\Boo{W}{z}_*h}{p}\| \leq c \, M(p) \,\sn_\Omega(h), 
\end{equation} 
where $M$ is the dominating function from \eqref{eqModGrowth}, and $\sn_\Omega$ is the semi-norm on  $\calD(\Omega)$ defined by $\sn_\Omega(h)=\sum_{|\alpha|\leq N+1}\|\partial^\alpha h\|_\infty$, $N$ as in \eqref{eqModGrowth}. 

ii) If the growth order $N$ of $e\rightarrow u(e,p)$ in \eqref{eqModGrowth} is
zero, then the analogous statements hold 
with the appropriate replacements $h\rightarrow\spd$, $g_*h\rightarrow
ge$. 
\end{Lem}
(Note that the estimate~\eqref{eqIntDistBound} is claimed to hold also
for $z=0$, i.e.\ for $ \intdist{h}{p}$.)
\begin{Proof}
We shall make use of some details on the entire matrix-valued function $z\mapsto\Boo{W}{z}$, $z\in\CC$. Namely, it satisfies 
\begin{equation} \label{eqBooW} 
 \Boo{W}{t+it'} = \Boo{W}{t}\big(j_{W}(t')+i \sin(t')\,\sigma_{W} \big) \,,
\end{equation} 
where $j_W(t')=\half\cos t'(1-j_W)+\half(1+j_W)$ continuously deforms the unit to $j_W$ when $t'$ runs
through  $[0,\pi]$,\ 
and $\sigma_W$  maps the wedge  $W$ continuously into the
interior of the forward light cone, cf.~\cite{H96}. 
This implies that for $e\in W_H$, the function $z\to\Boo{W}{z}e$ is analytic on 
$\strip$.\footnote{$\Boo{W}{z}e$ refers 
to the action of $\Po$ on $\Spd$ defined in \eqref{eqPoSpd}.}  
Moreover, if $e$ and $z$ are in some compact subsets $\Omega\subset W_H$ and
$\strip_0\subset\strip^\clo$, respectively, then $\Boo{W}{z}e$ is in some $\Theta$ of
the form ~\eqref{eqTheta}. Hence, the bound \eqref{eqModGrowth} implies that
there is a constant $N\geq0$, a function $M(p)$ (locally $L^2$ and
polynomially bounded)  and $c=c_{\Omega,\strip_0}$ such that for
all $e\in\Omega$ and $z\in \strip_0$ holds 
\begin{equation} \label{eqModGrowth'}
\|\intfct{\Boo{W}{z}e}{p}\| \leq c \, M(p) \, \sin(t')^{-N}, 
\end{equation} 
where $t':=\Im z$.

Let now $h\in\calD(\Spd)$ be as in the Proposition, and fix $p$. 
For $z\in \strip$,  denote 
\begin{align} 
F(z)&:= \intdist{\Boo{W}{z}_*h}{p}\\  
 &= \int_W d\sigma(e)h(e) f(z,e) \quad\text{ with } f(z,e) =\intfct{\Boo{W}{z}e}{p}.
\end{align} 
Note that the integration variable $\spd$ is in $W_H$, hence $f(\cdot,e)$
is analytic on the strip $\strip$, as noted after \eqref{eqStrip}. 
Eq.~\eqref{eqModGrowth'} guarantees the existence of a majorizing function for all 
$z$ in a given compact subset of $\strip$. 
This shows that $F(z)$ is analytic on $\strip$. 
Note that Eq.~\eqref{eqBooW} implies that  
$\Boo{W}{t+it'}e$ approaches $\Boo{W}{t}e$ from inside the tuboid
$\Tub$ if $t'\rightarrow0^+$, hence $F(t+it')$ approaches $F(t)$ 
{\em by definition}, cf.\ the remark after equation~\eqref{eqIntDist}. 
This implies that $F$ is continuous on
the lower boundary $\RR$ of the strip. 
We consider now the limit of $F(t+it')$ for $t'\rightarrow\pi^-$. 
Note that equation~\eqref{eqBooW} implies that 
$\Boo{W}{t+ i\pi}=\Boo{W}{t}j_W$. 
Hence, for $e\in W$, $\Boo{W}{t+ it'}e$ approaches
$\Boo{W}{t}j_We$ from $\Tub$. Again, it follows 
that by definition $F(t+it')$ approaches $\intdist{(\Boo{W}{t}j_W)_*h}{p}$ as
$t'\rightarrow\pi^-$. 
This implies  equation~\eqref{eqIntDistBV} and continuity of $F$ on the upper
boundary $\RR+i\pi$ of the strip. 
It remains to prove the bound~\eqref{eqIntDistBound}. If $\strip_0$ is in the
interior of the strip $\strip$, then the estimate \eqref{eqModGrowth'}
immediately implies that 
$|F(t+it')|\leq c\, M(p) \,(\sin t')^{-N}\int_\Omega |h|$. This
implies~\eqref{eqIntDistBound}, since $\sin t'$ is bounded away from zero. 
We now discuss the boundaries of $\strip$, considering first the lower
boundary $\RR$, namely $t'\in[0,1]$.  
To this end, we control $|F(t+it')|$ in the limit $t'\rightarrow0^+$, 
following standard arguments, cf.~\cite[Thm.~2-10]{SW} 
and \cite[Thm.~IX.16]{ReSi}. 

We first introduce Lagrangian coordinates on $W_H$
as follows. The flow of $\Boo{W}{t}$ on $W_H$ 
is time-like and complete. Hence, $\Sigma:=\{0\}\times S^{d-2}
\cap W_H$ is a Cauchy surface for $W_H$ and every point in $W_H$ is of the
form $e=\Boo{W}{\tau}\hat e$ for some unique $\tau\in\RR$, $\hat
e\in\Sigma$. Putting $\phi:e\mapsto (\tau,\hat e)$  
establishes a diffeomorphism $\phi:W\rightarrow \RR\times \Sigma$. 
We now observe that $f(z,\Boo{W}{\tau}\hat e)=f(z+\tau,\hat e)$ for
$z\in\strip, \tau\in\RR$, and get 
$$ 
F(z)=\int_{\RR\times \Sigma} d\tau d\Sigma(\hat e)\, 
f(z+\tau,\hat e) \,\hat{h}(\tau,\hat e).
$$
Here $d\Sigma$ denotes the canonical volume form on $\Sigma\cong S^{d-2}$, and
$\hat{h}(\phi(e))=h(e) \frac{\phi_*d\sigma}{d\tau d\Sigma}$ (where  
$\frac{\phi_*d\sigma}{d\tau d\Sigma}$ denotes the Radon Nikodym 
derivative). 
With the same method as in \cite[Thm.~IX.16]{ReSi}, one now shows that 
for $t\in\RR$, $t'\in (0,1]$, $\nu\geq 2$ the bound \eqref{eqModGrowth'} 
implies the following estimate: 
\begin{align*} 
|F(t+it_1) | \leq  c'\,M(p)\,\Big\{ 
\|(\partial_\tau )^{\nu-1}\hat{h}\|_\infty  
\int_{t_1}^{1}dt_2\ldots\int_{t_{\nu-1}}^{1}dt_\nu\,t_\nu^{-N} + 
\sum_{j=0}^{\nu-2}\|(\partial_\tau)^j\, \hat h\|_\infty\,|P_j(t_1)|  \Big\}.
\end{align*}
Here  $\|\cdot\|_\infty$ denotes the supremum norm, and $P_j$ is a polynomial (of degree $j$). 
Let now $\nu$ be strictly larger than $N+1$. Then the multiple integral 
over $t_\nu^{-N}$ has a finite limit for $t_1\rightarrow0$, and hence
extends to a continuous function  of $t_1$ on $[0,1]$, 
which we denote by $P_{\nu-1}$. 
Then the above inequality reads 
$$
|F(t+it_1)| \leq  c'\,\hat c(t_1)\,M(p)\,
\sum_{j=0}^{\nu-1} \;\|(\partial_\tau)^j\, \hat h\|_\infty,\quad t_1\in[0,1],
$$
where $\hat c(t_1):=\max_{j=0\ldots \nu-1}|P_j(t_1)|$. 

This proves the claimed bound~\eqref{eqIntDistBound} near the lower boundary $z=t\in\RR$. 

For $z$ near the upper boundary, $\RR+i\pi$, one may write
$\lim_{t'\rightarrow\pi^-}\Boo{W}{t+it'}=\lim_{\eps\rightarrow0^+}
\Boo{W}{t-i\eps}j_W$ and apply an analogous argument. 
This completes the proof of $i)$. $ii)$ is shown analogously.  
\end{Proof}
We now show that the distribution $\intdist{h}{p}$ inherits the
covariance properties \eqref{equCov} from its defining analytic function $\intfct{e}{p}$. 
\begin{Lem}   \label{Int}
The family $\intdist{h}{p}$ has the following intertwining properties: 
\begin{align} 
D(R(\lor,p))\,\intdist{h}{\lor^{-1}p} &=\intdist{\lor_*h}{p},\quad\lor\in\Lor,
\label{eqIntLor} \\
{\intdistconj{h}{p}} &= D(j_0)\,\intdist{(j_0)_*\bar{h}}{-j_0p}, \label{eqIntJ} 
\end{align}
where $(\lor_*h)(e):=h(\lor^{-1}e)$. 
\end{Lem}
\begin{Proof}
We choose a continuous map $\phi: \Spd\times [0,\eps)\rightarrow \Spdc$ such that $\phi(e,t)\in\pm\Tub$ for
$t\gtrless0$, respectively, and $\phi(e,0)=e$ for each $e\in\Spd$. 
Then, as remarked after  Eq.~\eqref{eqIntDist}, 
$$ 
\intdist{h}{p}= \lim_{t\to 0^+} \int d\sigma(e)\,h(e)\,\intfct{\phi(e,t)}{p}.
$$
Equation~\eqref{eqIntLor} is a straightforward consequence of the covariance~\eqref{equCov} and the fact that 
$(e,t)\mapsto \lor\phi(\lor^{-1}e,t)$ satisfies the same conditions
as $\phi$. 
Further, by definition~\eqref{eqv} of $u_c$ we have  
\begin{align*} 
{\intdistconj{h}{p}}  
&= \lim_{t\to 0^-}  \int d\sigma(e) \, h(e)  \,
   {\intfctconj{\phi(e,t)}{p}}\\
&=  \lim_{t\to 0^-} \int d\sigma(e) \, h(e) \,D(j_0)\, \intfct{j_0\phi(e,t)}{-j_0p}\\
&=  \lim_{t\to 0^+} \int d\sigma(e) \, h(j_0e)\,D(j_0)\,  \intfct{j_0\phi(j_0e,-t)}{-j_0p} 
=D(j_0)\, \intdist{(j_0)_*\bar h}{-j_0p}.   
\end{align*}    
In the last equation we have used that 
$(e,t)\mapsto j_0\phi(j_0e,-t)$ has the same properties as $\phi$. 
This completes the proof of the lemma. 
\end{Proof}
We are now prepared to prove Proposition~\ref{Cov}. 
\begin{Proofof} {\em of Proposition~\ref{Cov}.}  
Ad $0)$. $\psi(f,h)\in\Hirr$  follows from the bound~\eqref{eqIntDistBound} for $t=0$. As to the ``single particle 
Reeh-Schlieder'' property, note that the span of $\{u_0(e),e\in\Spd\}$ carries a representation of the little group due to Eq.~\eqref{equ0Cov}. Since $D$ is irreducible, this set spans the little Hilbert space $\calh$. By going over to the Lie group (or using analyticity of $u_0$), the same holds if one restricts $e$ to some open neighborhood $U$. This implies the Reeh-Schlieder property straightforwardly. 

$i)$ is a straightforward consequence of Lemma~\ref{Int}, the equations 
$e^{ia\cdot p} (\FT f)(\lor^{-1}p)=(\FT (a,\lor)_*f)(p)$ and 
$\overline{(\FT f)(-j_0p)}=(\FT (j_0)_*\bar f)(p)$, and the fact that
every $j\in\Poj^{\downarrow}$ is of the form $gj_0g^{-1}$. 

Ad $ii)$. 
By Lemma~\ref{StringWedge}, the condition $\calO+\RR_0^+\Omega\subset W$ holds if and only if 
$\calO\subset W$ and $\Omega$ is contained in the {\em closure} of $W_H$. 
Suppose first that ($\calO\subset W$ and) $\Omega$ is contained in $W_H$. 
We shall consider the $\Hirr$-valued function 
\begin{align} \label{eqPsit}
t\mapsto \psi_t := 
\Uirr(\Boo{W}{t})\,\covfct{f}{h}\,,\quad t\in\RR.
\end{align}
It follows from the covariance equation~\eqref{eqPsiCov} that
\begin{align}
\psi_t(p)= \psi^0_t(p)\,\intfct{\Boo{W}{t}_*h}{p}\,,  \label{eqPsit'} 
\quad 
\psi^0_t := E\Boo{W}{t}_*f. 
\end{align}
It is well-known that for almost all $p$, $t\mapsto\psi^0_t(p)$ extends to 
an analytic function on the strip $\strip$. The analyticity statement of 
Lemma~\ref{IntDist} then implies that for almost all $p$ 
the $\calh$-valued function $ z\mapsto\psi_z(p)$ is analytic on 
$\strip$ and weakly continuous on $\strip^\clo$. Further, it is well-known~\cite[Eq.~(4.58)]{M02a} 
that for any  given compact subset $\strip_0$ of $\strip^\clo$, one can find a 
dominating function $\Psi^0$ of fast decrease in $p$ such that 
\begin{equation} \label{eqDomin0} 
|\psi^0_z(p)|<\Psi^0(p),\quad z\in\strip_0, \,p\in\HypNullMinus. 
\end{equation}
The bound \eqref{eqIntDistBound} then implies that $\Psi^0(p) M(p)$ is a dominating function for $\psi_z$. 
These facts imply that $z\mapsto \psi_z$ is analytic on $\strip$ as a $\Hirr$-valued function, and  
weakly continuous on $\strip^\clo$.  

It follows from these facts that $\psi_0$ is in the domain of $\Delta_{W}^{\half}$,
and that $\Delta_{W}^{\half}\psi_0=\psi_{i\pi}$, cf.~Lemma~\ref{ModOp}.  
But equation~\eqref{eqIntDistBV} and the equation
$  \psi^0_{i\pi}= \FT(j)_*f$,  
which holds as a consequence of $\Boo{W}{i\pi} =j_W$, imply that  
\begin{align} 
\psi_{i\pi}&= \covfct{(j_W)_*f}{(j_W)_*h}\,. \label{eqPsiipi}
\end{align}
Hence we have shown that $\Delta_{W}^\half \psi({f},{h})$ coincides with the r.h.s.\ 
of the above equation. This implies, by Eq.~\eqref{eqPsiCovj}, that $S_W$ acts as in 
equation~\eqref{eqSPsi} of the Proposition. 
It remains to show that this equation holds also if $\Omega$ is only contained in the closure of $W_H$. 
But then there is a sequence of vectors $\psi_n$ of the above form (i.e.\ for which \eqref{eqSPsi} holds), which converges to $\psi(f,h)$. ($\psi_n$ may be constructed via a suitable curve in the Poincar\'e group, or from functions $h_n$ with support bounded away from the boundary of $W_H$.)   
Since $S_W$ is a closed operator, Eq.~\eqref{eqSPsi} is also valid for $\psi(f,h)$. 
This completes the proof of $ii)$. 

$iii)$ is shown in complete analogy. 
\end{Proofof}

\subsection{Proofs of Propositions~\ref{IntMassless4d} and \ref{IntMassless3d}.} 
The proof of Proposition~\ref{IntMassless4d} makes use of the Lemmas~\ref{IntFctEst} through \ref{IntFctEst'}, which we now state and prove. 
\begin{Lem} \label{IntFctEst}
Let $n\in\NN_0$ be strictly larger than $2\ReDeg+2$. Then there are constants $a_\nu,
b_\nu, c_\nu$, $\nu=0\ldots [n/2]$, such that 
for all $\spdc=\spdr+i\spdi\in\Tub$, $p\in\HypNullMinus$,
and $k$ with $|k|=\PauliLub$ the 
following estimate holds: 
\begin{equation} 
|\intfctdeg{\spdc}{p}(k)| \leq   c \, |p\cdot\spdc|^{-\ReDeg+n-2} + \sum_{\nu=0}^{[n/2]}  
c_\nu \, (\spdi^2)^{\ReDeg-n+\nu +1} \, (p\cdot\spdi)^{-\ReDeg+n-\nu-1} \, |p\cdot\spdc|^{\nu-1}.\label{equEst}
\end{equation}
\end{Lem} 
(We have written $\spdi^2:=\spdi\cdot\spdi$.)
\begin{Proof} 
We write the scalar product in Minkowski space as $x\cdot p=\half
(x_+p_-+x_-p_+)-x_1p_1-x_2p_2$, where $x_\pm\doteq x_0\pm
x_3$. 
Now for $z\in\RR^2$, the components of $\xi(z)$ are $\lc(z)_+=z^2$, 
$\lc(z)_-=1$, $\lc(z)_1=z_1$ and $\lc(z)_2=z_2$. 
Further, $(B_p^{-1}\spd)_-=\bar p\cdot B_p^{-1}\spd = p\cdot \spd$. 
We therefore have 
\begin{align}  
B_p\lc(z)\cdot e &= a z^2+b\cdot z+c,\quad \text{with} \label{eqpxie}\\
a&= \half(p\cdot\spd),\nonumber  \\
b&= -\big((B_p^{-1}\spd)_1,(B_p^{-1}\spd)_2\big)\;\in\CC^2 , \nonumber\\
c&= \half(B_p^{-1}\spd)_+\,.\nonumber
\end{align}
Here, $b\cdot z$ denotes the standard scalar product in $\Bb^2$, and $z^2:= z\cdot z$. 
(Below we shall adapt the same notation for vectors in $\Bc^2$ by {\em bilinear} extension: 
$(z_1,z_2)\cdot (w_1,w_2):= z_1w_1+z_2w_2$.) 
Taking account of $4ac-b^2=\spd^2=-1$, and of $2a= p\cdot\spd >0$, we have 
\begin{equation}  \label{eqP(z)} 
B_p\xi(z)\cdot \spd=a(z+b/(2a))^2-1/(4a).
\end{equation}
We denote the real and imaginary parts of
$b/(2a)$ by $w'$ and $w''$, respectively. Then we have, after
substituting $z+w'\rightarrow z$, 
\begin{align}  \label{eqIntFct'}
\intfctdeg{\spd}{p}(k) = e^{-i(\pi\degree/2+k\cdot w')}\, 
\int_{\RR^2} d^{2}ze^{ik\cdot z}\big(P(z) \big)^{\degree},\quad 
P(z):=a(z+iw'')^2-\frac{1}{4a}.
\end{align}
To evaluate  this integral, we shall assume that the vector $k$ points in  
$1$-direction, so that $k\cdot z=\PauliLub z_1$. (The general case is obtained by
replacing $b$ in Eq.~\eqref{eqpxie} $Rb$, where $R\in SO(2)$
rotates $\PauliLub (1,0)$ into $k$.)
$n$-fold partial integration then yields 
\begin{equation}\label{equP}
\intfctdeg{\spd}{p}(k) =  
c\, \int d^{2}ze^{ik\cdot z} 
\big(\frac{\partial}{\partial z_1}\big)^n\left( P(z) \right)^{\degree},
\end{equation} 
and we shall use that 
\begin{align} \label{eqDeltaP}
\partial_{z_1}^n P(z)^\degree =\sum_{\nu=0}^{[n/2]} c_\nu \, a^{n-\nu}
P(z)^{\degree-n+\nu} (z_1+iw_1'')^{n-2\nu}, 
\end{align}
where $c_\nu$ is independent of $a,b,c$ and $z$. 

We now establish bounds on $P(z)$. First, note that the imaginary parts of 
$B_p\lc(z)\cdot e $ and of $a$ are strictly positive, since $\spdc\in\Tub$, i.e.\
$e''$ is in the interior of the forward light cone. 
In particular, we have from equation~\eqref{eqpxie} 
\begin{align} \label{eqImf}
 \Im (B_p\lc(z)\cdot e) &= a''(z+\frac{b''}{2a''})^2+d \, \geq \, d,\quad
 d:=\frac{\spdi\cdot \spdi}{4a''} >0, 
\end{align}
where $b''\in\RR^2$ and $c''$ denote the real parts of $b$ and
$c$, respectively. (We have used that $4a''c''-(b'')^2=\spdi\cdot \spdi$.) Since $P(z)$ is by definition just $B_p\lc(z-w')\cdot e$, it 
follows that 
$ \Im P(z) \geq  d$. 
Secondly, we observe that 
\begin{align} \label{eqzda}
|P(z)| 
\geq |a|\,\big|\Re\big((z+iw'')^2-\frac{1}{4a^2}\big)\big| 
= |a|\,|z^2-\rho|,\quad \rho:=w''^2 +\Re\big(\frac{1}{4a^2}\big). 
\end{align}
It follows that 
\begin{equation} \label{eqPBound}
|P(z)| \geq \half\big(|a|\,|z^2-\rho|+d\big).  
\end{equation} 
Let now  $s:=\ReDeg-n+\nu$ and $m:=n-2\nu$. Note that $s<-1$ and $m\geq0$.  By Eq.~\eqref{eqPBound}, we have  
\begin{align}  \label{eqIntPL}
\int d^2z |P(z)|^{s} |z_1+iw_1''|^{m}  
\leq c\, \int_0^\infty dr \,r \, \big( |a|\,|r^2-\rho|+d\big)^s \, \big(r^m+\|w''\|^m\big),
\end{align}
where $\|w''\|:=(w''\cdot w'')^{1/2}$ denotes the euclidean norm of $w''\in\Bb^2$. 
We split the integral into $r<|a|^{-1}$ and  $r>|a|^{-1}$. 
We show, at the end of this proof, that 
\begin{equation} \label{eqRho<}
|\rho| \leq \half |a|^{-2} \;\text{ and } \; \|w''\|\leq \half |a|^{-1}.
\end{equation}
Hence for $r>|a|^{-1}$ holds $r^2-\rho \geq \half \,r^2$ and $\|w''\| < r$, and hence 
the integral in \eqref{eqIntPL} over $r>|a|^{-1}$ is bounded by 
\begin{align}
 2^{-s}\,|a|^s \int_{|a|^{-1}}^\infty dr \,r^{2s+m+1} 
 = c'\, |a|^{-s-m-2}. \label{eqPBound1a} 
\end{align}
(Note that $2s+m+1=2\ReDeg-n+1<-1$.) 
The integral in \eqref{eqIntPL} over $r < |a|^{-1}$ is bounded by 
\begin{multline} 
(|a|^{-m}+\|w''\|^m)\,\int_0^{|a|^{-1}}dr \,r\, \big( |a|\,|r^2-\rho|+d \big)^s 
\leq  2 (|a|^{-m} +\|w''\|^m) \, |a|^{-1}\,d^{s+1}.  \label{eqPBound1b} 
\end{multline}
Putting together eqs.~\eqref{eqIntPL}, \eqref{eqPBound1a} and 
\eqref{eqPBound1b}, and using the fact that 
\begin{equation} \label{eqW<}
\|w''\|\leq \frac{a''}{2|a|^{2}}, 
\end{equation}
which we show at the end of this proof, we have 
\begin{align*}  
\int d^2z |P(z)|^{s} |z_1+iw_1''|^{m}  
\leq  c_1\, |a|^{-s-m-2} \,+\, c_2\, |a|^{-m-1} \, d^{s+1} \, + \, c_3\,(a'')^m |a|^{-2m-1}\,d^{s+1} . 
\end{align*}
Putting this inequality into Eq.s~\eqref{equP} and \eqref{eqDeltaP}, one gets the claimed estimate 
\eqref{equEst}, if one recalls that $a= p\cdot \spd/2$ and $d=(\spdi)^2(4a'')^{-1}$, and that  $|p\cdot\spdc|^{-n+2\nu}\leq (p\cdot \spdi)^{-n+2\nu}$. 

It remains to prove equations~\eqref{eqRho<} and \eqref{eqW<}. 
Denoting $B_p^{-1}e=:x+iy\in(\Bb^4+i\Bb^4)\cap \Tub$, we have 
$2a=x_-+iy_-$ and $-b=\vec{x}+i\vec{y}$,  
where we have written e.g.\ $\vec{x}:=(x_1,x_2)\in\RR^2$. 
With this notation, one finds $\Im(\bar{a}b)=x_-\vec{y}-y_-\vec{x}$, and 
$$
\|\Im(\bar{a}b)\|^2 = -y^2x_-^2-x^2y_-^2 \, \leq \, y_-^2\equiv (a'')^2,
$$
where the inequality holds because $y^2=1+x^2>0$ for $e\in\Tub$. 
Since $\|\Im b/a\| = |a|^{-2} \|\Im\bar{a}b\|$, it follows that  
\begin{equation} \label{eqW<'} 
\|w''\| \equiv \|\Im\frac{b}{2a}\| \leq \frac{a''}{2|a|^2}, 
\end{equation}
which is \eqref{eqW<}. 
Now the r.h.s.\ of that inequality is smaller than $(2|a|)^{-1}$ which shows the second 
equality in \eqref{eqRho<}, but also the first one because 
$$
|\rho|\leq \|w''\|^2+|\Re\frac{1}{4a}|\leq \|w''\|^2 + \frac{1}{4|a|^2}\leq \frac{1}{2|a|^2}. 
$$
This completes the proof.   
\end{Proof}

\begin{Lem} \label{IntFctAna}
For fixed $p\in\HypNullMinus$ and $k\in\RR^2$ with $|k|=\kappa$, the function
$\spdc\mapsto\intfctdeg{\spdc}{p}(k)$ is analytic on the tuboid $\Tub$. 
\end{Lem}
\begin{Proof}
We shall show that $\intfctdeg{\spdc}{p}(k)$ is analytic on $\RR^4+iV_+$, where $V_+$ denotes the
forward light-cone. 
As in the proof of Lemma~\ref{IntFctEst}, we pick an integer $n>2\ReDeg+2$ and write 
$\intfctdeg{\spdc}{p}(k)$ as a sum of terms of the form 
\begin{align*} 
c\,\int d^{2}z e^{ik z} f(e,z)\,,\quad  f(e,z):= 
P_e(z)^{\degree-n+\nu}\,Q_e(z)^{n-2\nu}, \quad \nu =0,\ldots,[n/2], 
\end{align*} 
where $P_e$, respectively $Q_e$, is a quadratic, respectively  linear, polynomial with coefficients depending
differentiably on $e$, cf.\ equations~\eqref{equP} and \eqref{eqDeltaP}.  
If $e$ varies in a compact set, each of these polynomials
is bounded, uniformly in $e$, by some continuous 
function with quadratic, respectively linear, behavior for large $|z|$. 
In addition, $P_e(z)$ has no real zeroes and, as a consequence of  Eq.~\eqref{eqImf}, is uniformly bounded below by a strictly positive function with quadratic behavior. Hence  
for each compact subset $\Omega$ of $\RR^4+iV_+$ there is a continuous dominating function
for $f(e,z)$, $e\in\Omega$,  
which goes like $|z|^{2\ReDeg-n}$ for large $z$ and is therefore 
integrable w.r.t.\ $d^2z\sim |z|d|z|$ for $2\ReDeg-n<-2$.  
It follows that the analyticity of the integrand, for fixed $z\in\RR^2$, survives 
after the integration, completing the proof. 
\end{Proof}
\begin{Lem} \label{IntFctEst'}
Let $\ReDeg\in(-1,-\half)$. Then there are constants $c_1, c_2$ such that 
for all $\spdc\in\Tub$,
$p\in\HypNullMinus$, $k\in\RR^2$ with $|k|=\PauliLub$ the
following estimate holds: 
\begin{align} \label{eqEstIntFct'}
|\intfctdeg{\spdc}{p}(k)| &\leq  c_1 \,|p\cdot \spd|^{-\ReDeg-1}+c_2\,|p\cdot \spd|^{-\ReDeg-2}. 
\end{align}
\end{Lem}
\begin{Proof} 
We use the same notations as in the proof of Lemma~\ref{IntFctEst}, and consider 
the intertwiner function as given by the integral~\eqref{eqIntFct'}. 
Here, we partially integrate only over $\|z\|>|a|^{-1}$.  
We obtain 
\begin{align} \label{eqPIBT} 
\int_{\|z\|>|a|^{-1}} d^{2}ze^{ik\cdot z}\big(P(z)
\big)^{\degree} = \frac{i\degree}{\PauliLub}
\int_{\|z\|>|a|^{-1}} d^{2}ze^{i\PauliLub z_1}  
P(z)^{\degree-1}2a(z_1+iw''_1) + (\text{bd. terms}). 
\end{align}
{}From Eq.~\eqref{eqPBound1a}, with $s=\ReDeg-1$ and $m=1$, $\nu=0$, we know that 
the integral on the r.h.s.\ 
is bounded by $c|a|^{-\ReDeg-1}$ if $\ReDeg<-1/2$. 
The boundary terms in Eq.~\eqref{eqPIBT} are given by 
\begin{align*}
\frac{1}{i\PauliLub} 
\int_{-|a|^{-1}}^{|a|^{-1}} dz_2 e^{i\PauliLub \delta(z_2)}P(\delta(z_2),z_2)^{\degree}
\end{align*}
where $\delta(z_2):=\sqrt{|a|^{-2}-z_2^2}$, plus a similar term for
$z_1=-\delta(z_2)$. 
By Eq.s~\eqref{eqzda} and \eqref{eqRho<}, they are bounded by 
\begin{align*}
c |a|^\ReDeg \int_{-|a|^{-1}}^{|a|^{-1}} dz_2
(|a|^{-2}-\rho)^{\ReDeg}
\leq 2^{-\ReDeg} c|a|^\ReDeg \int_{-|a|^{-1}}^{|a|^{-1}} dz_2
|a|^{-2\ReDeg} = c' |a|^{-\ReDeg-1}. 
\end{align*}
Thus, the integral over $\|z\|>|a|^{-1}$ in
Eq.~\eqref{eqIntFct'} is bounded by $c|a|^{-\ReDeg-1}$. 

To evaluate the integral over $\|z\|<|a|^{-1}$, we note that, by \eqref{eqzda}, 
\begin{align} 
\big|\int_{\|z\|<|a|^{-1}} d^{2}ze^{ik\cdot z}\big(P(z) \big)^{\degree}\big|
&\leq |a|^\ReDeg \pi\int_0^{|a|^{-2}} d(r^2)|r^2-\rho|^\ReDeg 
 \leq  c' |a|^{-\ReDeg-2} 
\end{align}
where in the last inequality, valid for $\ReDeg>-1$, we have used \eqref{eqRho<}. 
This completes the proof.  
\end{Proof}

We now prove boundedness of the 3-d intertwiner, used in Proposition~\ref{IntMassless3d}. 
\begin{Lem} \label{IntFct3d}  
Consider the three-dimensional case, with $\ReDeg\in(-1,0)$. 
The function $(\spd,p)\mapsto\intfctdeg{\spd}{p}$ is bounded
on $\Tub\times\HypNullMinus$. 
\end{Lem}
\begin{Proof}
In analogy with the 4-dimensional case, cf.\ the proof of Lemma~\ref{IntFctEst}, we 
have 
\begin{align}
P(\real ):= B_p\lc(\real )\cdot e &= a\real ^2+b\real +c,\quad \text{with} \label{eqpxie3d}\\
a&= \half(p\cdot\spd),\nonumber  \\
b&= -(B_p^{-1}\spd)_1, \nonumber\\
c&= \half(B_p^{-1}\spd)_+\,.\nonumber
\end{align}
Note that the imaginary parts of $B_p\lc(\real )\cdot\spd$  and $a$ 
are strictly positive since $e''\in V_+$. 

The zeroes of the quadratic polynomial~\eqref{eqpxie3d} are
$\real_{\pm}\doteq (-b\pm1)/2a$, and we write the polynomial as 
\begin{equation} \label{eqf-}
P(\real )= a(\real -\real_+)(\real -\real_-) = \big((\real -\real_+)^{-1}-(\real -\real_-)^{-1}\big)^{-1} . 
\end{equation}
(We have used that $\real_+-\real_-=a^{-1}$.) 
For $\real $ close to the zeroes $\real_\pm$, the modulus $|P(\real )|^{\ReDeg}\sim |\real -
\real_\pm|^{\ReDeg}$ is integrable since $\ReDeg>-1$, but for 
large $|\real |$ the modulus $|P(\real )|^\ReDeg \sim  |\real |^{2\ReDeg}$, hence it is only integrable for 
$\ReDeg<-\half$.  
To treat all $\ReDeg\in(-1,0)$ simultaneously, we hence need to keep the oscillating
factor for large $\real $ and partially integrate except for an 
$\eps$-neighborhood around each of the real parts of the poles
$\real_\pm$. To this end, denote the real parts of $\real_+$ and $\real_-$ by
$\kre_\pm$, labeled in such a way that $\kre_-\leq\kre_+$,
and denote by $I_\pm$ the interval
$(\kre_\pm-\eps,\kre_\pm+\eps)$. Here, $\eps>0$ is fixed and 
independent of $\kre_\pm$, so it may happen that the
intervals overlap.  In any case, we get 
\begin{align}  
e^{i\pi\degree/2}\,\intfctdeg{\spd}{p} = &
 \int_{I_+\cup I_-}\d \real  e^{i \PauliLub \real }\,P(\real )^{\degree}\, 
+\, \frac{1}{i\PauliLub} \sum_{\real_l\in\partial(I_+\cup I_-)} (\pm)
e^{i \PauliLub \real_l}\,P(\real_l)^\degree \nonumber \\  
& -\, \frac{1}{i\PauliLub}\int_{\RR\setminus(I_+\cup I_-)}d\real \, e^{i \PauliLub \real }\, 
\partial_\real  P(\real )^\degree.                   \label{equInt'} 
\end{align}
We will use the following estimates. 
\begin{align} \label{eqf<}
|P(\real )^{\degree}|&\leq c 
|(\real -\real_+)^{-1}-(\real -\real_-)^{-1}|^{-\ReDeg} \nonumber \\
&\leq  c \big(|\real -\real_+|^{\ReDeg}+|\real -\real_-|^{\ReDeg}\big)\nonumber \\
&\leq  c \big(|\real -\kre_+|^{\ReDeg}+|\real -\kre_-|^{\ReDeg}\big).
\end{align}
Denoting 
\begin{align} \label{eqk0}
\real_0&\doteq (\kre_++\kre_-)/2,
\end{align}
the last inequality implies 
\begin{align} \label{eqf<'} 
|P(\real )^{\degree}|\leq 2 c |\real-\kre_\pm|^{\ReDeg} ¸\quad\text{ if }\;k\gtrless k_0, 
\end{align} 
since $|k-\kre_+|\lessgtr|k-\kre_-|$ for $k\gtrless k_0$. 
To estimate $(P(k)^\degree)'$, note that equation~\eqref{eqf-} implies 
\begin{align*}
(P(k)^\degree)'(k) &= - \degree
\big((k-k_+)^{-1}-(k-k_-)^{-1}\big)^{-\degree-1}\,\big(-(k-k_+)^{-2}+(k-k_-)^{-2}\big)\nonumber
\\
&= \degree
\big((k-k_+)^{-1}-(k-k_-)^{-1}\big)^{-\degree}\big((k-k_+)^{-1}+(k-k_-)^{-1}\big),
\end{align*}
hence 
\begin{align}
|(P(k)^\degree)'(k)| &\leq c |\degree|\,
\big|(k-k_+)^{-1}-(k-k_-)^{-1}\big|^{-\ReDeg}\,\big|(k-k_+)^{-1}+(k-k_-)^{-1}\big|
\nonumber \\
&\leq c |\degree|\,
\big(|k-k_+|^{\ReDeg}+|k-k_-|^{\ReDeg}\big)\,\big(|k-k_+|^{-1}+|k-k_-|^{-1}\big)\nonumber \\
&\leq c |\degree|\,
\big(|k-\kre_+|^{\ReDeg}+|k-\kre_-|^{\ReDeg}\big)\,
          \big(|k-\kre_+|^{-1}+|k-\kre_-|^{-1}\big)\nonumber \\
&\leq 4c |\degree|\, |k-\kre_\pm|^{\ReDeg-1} \quad\text{if } k\gtrless k_0.
\label{eqf'<}
\end{align} 
We shall first discuss the boundary terms, i.e.\ the second term in
equation~\eqref{equInt'}. From ~\eqref{eqf<'} we get 
\begin{align} 
|P(\kre_++\eps)^\degree|\leq 2 c\eps^{\ReDeg}
\end{align}
since $\kre_++\eps>k_0$, and similarly
$|P(\kre_--\eps)^\degree|\leq2 c\eps^{\ReDeg} $. If the
intervals overlap, then these are the only boundary terms. If they do
not overlap, then $\kre_-+\eps\leq \kre_+-\eps$ which implies 
$
|P(\kre_\pm\mp\eps)^\degree|\leq 2 c\eps^{\ReDeg}
$
by a similar consideration. Hence in either case we have for the boundary
terms the estimate 
\begin{align} 
\sum_{\real_l\in\partial(I_+\cup I_-)}|P(\real_l)^\degree|\leq8c\eps^{\ReDeg}.
\end{align}
Let us discuss the first term in equation~\eqref{equInt'}, first 
assuming that the intervals $I_\pm$ are disjoint. 
Then $\real\in I_-$($I_+$) implies $\real < \real_0$($\real>\real_0$) respectively, and \eqref{eqf<'} 
implies: 
\begin{align}  
\int_{I_+\cup I_-}|P(\real)|^\ReDeg  
&= \int_{I_-}\d \real \,|P(\real)|^{\ReDeg} + \int_{I_+}\d \real \,|P(\real)|^{\ReDeg} \label{eqfInt0}\\
& \leq  c'\,\int_{I_-}\d \real \,|\real-\kre_-|^{\ReDeg}\,+\, c'\,\int_{I_+}\d \real \,|\real-\kre_+|^{\ReDeg}
=c''\, \eps^{\ReDeg+1}. \label{eqfInt}
\end{align}
If the intervals $I_\pm$ do overlap, then a moment's thought shows that we get a ``$\leq$`` 
instead of ``$=$``  in \eqref{eqfInt0}, yielding the same estimate~\eqref{eqfInt}. 
As to the third term in equation~\eqref{equInt'}, let us again first 
assume that the intervals $I_\pm$ are disjoint. Then we write 
\begin{equation}\label{eq4Ints}
\RR\setminus(I_+\cup I_-) = 
(-\infty,\kre_--\eps)\cup(\kre_-+\eps,\real_0) \cup (\real_0,\kre_+-\eps)\cup(\kre_++\eps,\infty), 
\end{equation}
where in the first two intervals $|\partial_\real P(\real)|^\ReDeg \leq c |\real-\kre_-|^{\ReDeg-1}$ 
and in the last two intervals $|\partial_\real P(\real)|^\ReDeg \leq c |\real-\kre_+|^{\ReDeg-1}$  
by  \eqref{eqf'<}. Hence 
\begin{align}  
\int_{\RR\setminus(I_+\cup I_-)}|\partial_\real P(\real)|^\ReDeg  
&\leq c'\,\int_{\RR\setminus I_-}\d k |\real-\kre_-|^{\ReDeg-1} \,+\,
      c'\,\int_{\RR\setminus I_+}\d \real |\real-\kre_+|^{\ReDeg-1} 
\leq c'' \eps^\ReDeg.  \label{eqf'Int}
\end{align}
If the intervals $I_\pm$ overlap, then the second and third intervals in \eqref{eq4Ints} are absent, 
and we have the same estimate \eqref{eqf'Int}. 

Since $\eps$ was arbitrary, we have thus shown that
$|\intfctdeg{\spd}{p}|$ is uniformly bounded, as claimed. 
\end{Proof}
\subsection{A Folklore Lemma.} 
\begin{Lem} \label{ModOp} 
Let $U_t$ be a unitary one-parameter group, with generator $K$. Then
$\psi$ is in the domain of $\exp(-\pi K)$ if, and only if, the
vector-valued map 
$$ 
t\mapsto U_t\psi 
$$
is analytic in the strip $\RR+i(0,\pi)$ and weakly continuous on the
closure of that strip. In this case, $\exp(-\pi K)\psi$ coincides with
the analytic continuation of $U_t\psi$ into $t=i\pi$. 
\end{Lem}
\begin{Proof}
The ``only if'' part is a standard result from functional calculus, and we prove here the ``if'' part for the convenience 
of the reader. Denote $U_t\psi=:\psi_t$. Let $c$ be a smooth function with compact support.
Recall that the bounded operator $c(K)$ may be written as 
$c(K)=\int d t\,\tilde{c}(t)\,U_{t}$, where $({2\pi})^{1/2}\tilde{c}$
is the Fourier transform of $c$, and the integral is understood in the
weak sense.  Then we have for any $\phi\in\calH$, in view of the equation 
$ 
\psi_{t+z}= U_{t}\,\psi_z  
$, that 
\begin{align}  \label{eqDelta*}
\big( \exp(-\pi K)\bar{c}(K)\phi\,,\,\psi_{t=0}\big) &= 
\big( \phi\,,\,c_\pi(K)\psi_0\big)  =\int d t\, \widetilde{c_\pi}(t)
\big( \phi\,,\,\psi_t\big)\,,
\end{align}
where we have written $c_\pi(k):= e^{-\pi k} c(k)$. Now the last 
integral is the limit, for $t'\rightarrow0$, of 
$\int d t\, \widetilde{c_\pi}(t+it') \big(
\phi\,,\,\psi_{t+it'}\big)$.  Since the integrand is analytic and tends to
zero for $|t|\rightarrow\infty$, we may further translate the contour of the 
integral by letting $t'$ approach $\pi$. 
Hence the r.h.s.\ of \eqref{eqDelta*} is equal to 
\begin{align}  \label{eqDelta*'}
\int d t\, \widetilde{c_\pi}(t+i\pi) \big( \phi\,,\,\psi_{t+i\pi}\big)
=\int d t\, \tilde{c}(t) \big( \phi\,,\,\psi_{t+i\pi}\big) 
=\big( \bar{c}(K)\phi\,,\,\psi_{i\pi}\big) ,
\end{align}
where we have used that $\widetilde{c_\pi}(t+i\pi)=\tilde c(t)$. 
We have thus established that 
\begin{align*}  
\big( \exp(-\pi K) c(K)\phi\,,\,\psi_0\big) = 
\big( c(K)\phi\,,\,\psi_{i\pi}\big) \,\quad\phi\in,\,c\in C_0^\infty(\RR).
\end{align*}
But the span of vectors of the form $c(K)\phi$ is  a core for the self-adjoint 
operator $\exp(-\pi K)$, hence the above equation implies that $\psi_0$ is in the domain of 
$\exp(-\pi K)$ and that $\exp(-\pi K)\psi_0=\psi_{i\pi}$, as claimed. 
\end{Proof}
\section{Geometrical Results} 

\subsection{Representation of the Reflections.} 
We prove that the anti--unitary operators $D(j_0)$, defined in sections~\ref{massive_bosons} and \ref{massless}, extend the respective representations of the little groups $G$ to representations of the semi-direct product of $G$ with $j_0$. Namely, in $d=4$, $D(j_0)$ is defined as in Eq.~\eqref{eqDjm} in the massive case or as 
\begin{align} \label{eqIrrepE(2)j}
(D(j_0) u) (k) &:= \overline{u(k)} 
\end{align} 
in the massless case, respectively. In $d=3$, $D(j_0)$ is just complex conjugation for both $m>0$ and $m=0$. 
\begin{Lem} \label{Dj}
In all cases ($m\geq 0$, $d=3,4$), $D(j_0)$ satisfies the representation properties $D(j_0)^2=1$ and 
\begin{align} \label{eqDj} 
D(j_0)D(\lor)D(j_0)=D(j_0\lor j_0),\quad \lor\in G.  
\end{align}
\end{Lem}  
\begin{Proof} 
We first treat the case $m>0$. In $d=4$, one checks that $-j_0q(n)= q(I_3 n)$, where $I_3$ is the inversion 
$(n_1,n_2,n_3)\mapsto (-n_1,-n_2,n_3)$. Hence $j_0$ has the same
commutation relations with the rotations as $I_3$, and a  representer
of $j_0$ is given by $\big( D(j_0)Y_{s,\sz})(n):=
\overline{Y_{s,\sz}(I_3n)}$. But the right hand side coincides with  $(-1)^\sz
Y_{s,-\sz}(n)$, as in~\eqref{eqDjm}. 
In $d=3$, the claim follows from the fact
that $j_0 R(\omega)j_0=R(-\omega)$. 

In the case $m=0$, note that $-j_0\xi(z)=\xi(-z)$, $z\in\RR^{d-2}$, which implies that the adjoint action
of $j_0$ on $G$ corresponds, via the isometry $\xi(\cdot)$, to the 
automorphism $(c,R)\mapsto (-c,R)$ of $E(d-2)$ (put $R=\unity$ in case $d=3$).  This implies the claim. 
\end{Proof}

\subsection{The Orbits of the Little Groups.} \label{orbits} 
One checks that the maps $q$ and $\xi$, defined in equations~\eqref{eqqtheta}, 
\eqref{eqqn} and \eqref{eqxi1}, \eqref{eqxi}, respectively, 
are diffeomorphisms from the sphere $S^{d-2}$ and $\RR^{d-2}$, respectively, 
onto the orbit $\orbit$ defined in equation~\eqref{eqOrbit}. 
Now $\orbit$ is a space-like sub-manifold, hence a Riemannian space with the metric 
$g_\orbit:=-g|_\orbit$. Then $q$ is clearly isometric. To check that the same holds for $\xi$, 
denote by $\partial_i$ the derivatives w.r.t.\ the natural 
coordinates $z_i$ on $\RR^2$, $i=1,2$. Then  
$$
g_\orbit(\xi_*\partial_i,\xi_*\partial_j)=\sum_{k=1}^3 (\partial_i\xi^k)(\partial_j\xi^k)-
(\partial_i\xi^0)(\partial_j\xi^0)=\delta_{ij},  
$$
i.e., $\xi$ is isometric. We therefore have: 
\begin{Lem} \label{Orbit}
The map $\xi$ is an isometry from $\RR^{d-2}$ onto the orbit $\orbit$.  
\end{Lem}

Let again  $G$ denote the stability subgroup of a fixed point $\bar
p\in\Hyp$. In the massive case, $\bar p=(m,0,0,0)$ and $G\cong
SO(d-1)$, while in the massless case $\bar p=(1,0,0,1)$ and $G\cong E(d-2)$. 
\begin{Lem} \label{OrbitsStab}
i) Let $\hat e$, $e\in\Spdc$ or $\Spdc\setminus\{\pm (i,0,0,0)\}$, respectively in the massless or massive case,  satisfy $\bar p\cdot \hat e=\bar p\cdot e$. Then there is a complex Lorentz  transformation\footnote{\label{LorC} That is, a complex linear transformation of $\CC^d$ leaving the bilinear form~\eqref{eqScalProdC} invariant.} in the connected component of the unit, which leaves $\bar p$ invariant and maps $\hat e$ to $e$. 

ii) In $d=4$, consider the stability group, in $G$, of an arbitrary point $e\in \Spd$, and  
a faithful, or scalar, representation $D$ of $G$.  
The restriction of $D$ to this group contains the trivial representation at most once. 

iii) Consider the massless case in $d=4$, and let $D$ be a non--faithful (but
non--trivial) ``helicity'' 
representation of the little group $G\cong E(2)$. That is to say, $D$ acts 
as a direct sum of irreducible representations of the form 
\begin{align} \label{eqDHel}
D(\lor(c,R_\phi))\,v=e^{i n \phi}\,v 
\end{align}
for some integer $n\neq 0$. 
Then the restriction of $D$ to the group mentioned in $ii)$ does not contain the 
trivial representation if $\bar p\cdot e \neq 0$. 
\end{Lem}
\begin{Proof} 
Ad $i)$  We discuss the case $d=4$. To this end, we recall a well-known 2:1 correspondence $(A,B)\mapsto\Lambda(A,B)$ 
between $SL(2,\CC)\times SL(2,\CC)$ and the group $\calL_+^c$ of complex Lorentz transformations (cf.\ footnote~\ref{LorC}) path-connected with the unit. Let $z\mapsto\utilde{z}$ be the isomorphism from $\CC^4$ onto $\text{Mat}(2,\CC)$ given by $\utilde{z}:= z_0+\sum_{i=1}^3 z_i\sigma_i$, $\sigma_i$ the Pauli matrices.  
This map satisfies 
\begin{align} 
\det \utilde{z} & = z\cdot z\quad\text{ and } \label{eqdetz} \\
\trace \utilde{z} &= 2z_0. \label{eqtracez}
\end{align}
Then a pair $(A,B)\in SL(2,\CC)\times SL(2,\CC)$ defines a transformation $\Lambda(A,B)$ of $\CC^4$ via 
\begin{align}
\utilde{ \Lambda(A,B) z} &= A\utilde{z} B^t. \label{eqLorc}  
\end{align} 
By virtue of \eqref{eqdetz}, $\Lambda(A,B)$ is in $\calL_+^c$. 

We first discuss the case $m>0$. Then $\bar p=(m,0,0,0)$ and $\utilde{\bar p}=m\unity$. Hence $\Lambda(A,B)$ leaves $\bar p$ invariant iff $B^t=A^{-1}$. Then $\utilde{\Lambda(A,B)z}=A\utilde{z}A^{-1}$. 
Let now $e$ and $\hat e$ be in $\Tub$ such that $\bar p\cdot e=\bar p\cdot \hat e$, i.e.\ $e_0=\hat{e}_0$. We have to show that $\utilde{e}$ and $\utilde{\hat{e}}$ are related by a similarity transformation (similar). By eqs.~\eqref{eqdetz} and \eqref{eqtracez}, $\utilde{e}$ and $\utilde{\hat{e}}$ have the same determinant, $-1$, and trace, $2e_0$.  Hence they have the same characteristic polynomial, namely $x^2-2e_0x-1$, and the same eigenvalues, $\lambda_\pm=e_0\pm\sqrt{1+e_0^2}$. If $e_0\neq \pm i$, these eigenvalues are different, and therefore $\utilde{e}$ and $\utilde{\hat{e}}$ are both similar to diag$(\lambda_+,\lambda_-)$, hence related by a similarity transformation. If $e_0=i$ or $-i$, then $e_0$ is a two-fold root of the characteristic polynomial. Such matrix is either equal to $e_0\unity$ or similar to the elementary Jordan matrix with diagonal $(e_0,e_0)$. 
But the first case has been excluded, for $\utilde{e}=\pm i\unity$ iff $e=(\pm i,0,0,0)$. (These points have been excluded because they are orbits by their own.) Hence $e$ and $\hat e$ are both similar to the same Jordan matrix, and therefore similar. 
This proves the claim for $m>0$. 

In the case $m=0$, $\bar p=\half\,(1,0,0,1)$ and $\utilde{\bar p}=\text{diag}(1,0)$. One checks that $\Lambda(A,B)$ leaves $\bar p$ invariant iff 
\begin{align} \label{eqAB0}
A=\left(\begin{matrix}c&a\\0&c^{-1}\end{matrix}\right)\quad \text{ and} \quad  B=\left(\begin{matrix}c^{-1}&b\\0&c\end{matrix}\right),\quad a,b,c\in\CC, c\neq0. 
\end{align}
Let now $e$ and $\hat e$ be in $\Spdc$ such that $\bar p\cdot e=\bar p\cdot \hat e$. We have to show that there is a $\Lambda$ of the above form which maps $e$ to $\hat e$. But $\bar p\cdot e= e_0-e_3=(\utilde{e})_{2,2}$. Hence we have to show that for any two matrices $\utilde{e},\utilde{\hat{e}}$ with the same determinant and $2,2$ component, there are $A,B$ as in \eqref{eqAB0} such that $\utilde{\hat{e}}=A\utilde{e}B^t$. One checks that the choice  
$$ 
a:=\frac{(\utilde{\hat e})_{1,2}\,c^{-1}-(\utilde{e})_{1,2}\, c}{(\utilde{e})_{2,2}},\quad 
b:= \frac{(\utilde{\hat e})_{2,1}\, c-(\utilde{e})_{2,1}\, c^{-1}}{(\utilde{e})_{2,2}}, 
$$
$c\neq0$ arbitrary, does it. This proves the claim for $m=0$. 

The three-dimensional case follows along similar lines. 

Ad $ii)$ Since $G$--related points have conjugate stability groups, it suffices to
consider one point in each $G$-orbit of $\Spd$. 
In the case $m>0$, $G$ is isomorphic to $SO(3)$ which acts transitively on the spheres $e_0=$constant, and we  consider for each $e_0\in\RR$ the point $(e_0,0,0,\sqrt{1+e_0^2})$. Clearly, the stability subgroup, in $SO(3)$, of these points are the rotations around the $3$--axis, which are represented by 
$D(R_3(\omega))_{\sz\sz'}=e^{i\sz\omega}\delta_{\sz,\sz'}$. 
Hence $v\in\calh=\Bc^{2s+1}$ is invariant if and only if
$v_\sz=c\delta_{\sz,0}$. This proves the claim for $m>0$. 

In order to conveniently discuss the case $m=0$, we give an explicit
formula for the action of the little
group $G\cong E(2)$ on $\Bc^{4}$. To this end, we use  coordinates $z_\pm:= z_0\pm
z_{3}$, and identify points $z$ in complexified Minkowski space $\Bc^{4}$ with
tuples $(z_+,z_-,\underline{z})$ with $z_\pm\in\Bc$ and 
$\underline{z}\in\Bc^{2}$, the metric being written as $z\cdot
z=z_+z_--\underline{z}\cdot\underline{z}$. 
In these coordinates, the action of $G\cong E(d-2)$ reads 
\begin{align} \label{eqE(2)Action} 
\lor(\underline{c},R_\phi)\,(z_+,z_-,\underline{z})=(z_++2\underline{c}\cdot
R_\phi\underline{z}+|\underline{c}|^2z_- \,,\, z_- \,,\, R_\phi
\underline{z}+z_- \underline{c} ).   
\end{align}
(This follows from the identification of $G$ with $E(d-2)$ acting in 
$\Gamma$=$\{z: z_-=1\}\cap\HypNull$ by linear extension.) 
For $t\in\RR$, consider now the sub-manifold $\Spd_t$ of $e\in\Spd$ with 
$e_-=t$. For $t\neq0$, it is isomophic to $\RR^2$ via $e\mapsto (e_1,e_2)$, a
nd the action of $G$ on $\Spd_t$ can be identified, by virtue of 
Eq.~\eqref{eqE(2)Action}, with the natural action of the Euclidean group. 
It is therefore transitive. Hence 
every $e\in\Spd$ with $e_-\neq0$ is $G$-related to some $e(t):=(-1/t,t,\underline{0})$. 
Equation~\eqref{eqE(2)Action} shows that the stability subgroup, in $G$, of $e(t)$ are the rotations
$\lor(0,R_\phi)$. These are represented in $\calh= L^2(\Bb^2,d \nu_\kappa)$ as
$\big(D(\lor(0,R_\phi))v\big)(k)=v(R_\phi^{-1} k)$. Hence $v\in\calh$ 
is invariant if  and only if it is the constant function, proving the claim for $t\neq0$. For $t=0$, $\Spd_{t=0}$ is isomorphic to $\RR\times S^1$, and from Eq.~\eqref{eqE(2)Action}, one sees that $G$ acts  transitively. Hence every $e\in\Spd_{t=0}$ is $G$-related to $e(0):=(0,0,\underline{e}_0)$ with $\underline{e}_0=(1,0)\in S^1$.   
Eq.~\eqref{eqE(2)Action} shows that the stability subgroup, in $G$, of $e(0)$ are ``translations'' of the form $\Lambda(\underline{c},1)$, with $\underline{c}=(0,c_2)\perp\underline{e}_0$. It follows that $v\in\calh$ is invariant iff for all $k\in\RR^2$ with $|k|=\kappa$ there holds $e^{ic_2k_2}v(k)=v(k)$. Such $v$ must vanish except at the points $k=(\pm\kappa,0)$, hence almost everywhere. Thus, in the case $e_-=0$ the invariant subspace is trivial.  

Ad $iii)$ As we have seen in the proof of $ii)$, the stability group of $e$ are the rotations $\lor(0,R_\phi)$ if $\bar p\cdot e\neq0$. But the representation \eqref{eqDHel} does not contain any invariant vector. 
This proves the claim.  
\end{Proof}
\begin{Lem} \label{StabPhot}
Let $G_e$ be the stability group, in $G$, of a fixed point $e\in \Spd$ satisfying $e_0\neq e_3$. 
Then there is precisely one vector in $\CC^4$, up to a constant, satisfying the eigenvalue condition 
\begin{align} 
\Lambda(c,R_\phi) v&= e^{i\lambda \phi} v, \quad \Lambda(c,R_\phi)\in G_e, \label{eqEV}
\end{align} 
where $\lambda\in\{1,-1\}$. 
\end{Lem}
\begin{Proof}
As in the proof of Lemma~\ref{OrbitsStab}, we use coordinates $(z_+,z_-,z_1,z_2)$ in which the action of $G\cong E(2)$ is given by Eq.~\eqref{eqE(2)Action}. 
Again, it suffices to consider one point $e$ in each $G$-orbit, the latter being characterized by the value of $e_-$. 
We consider the point $e=(e_+,e_-,0,0)$, with $e_+e_-=-1$. 
We know that then $G_e$ consists of the rotations $\Lambda(0,R_\phi)$. 
Then the eigenvalue equation~\eqref{eqEV} reads 
$$ (v_+,v_-,\cos(\phi) v_1+\sin (\phi) v_2, \cos(\phi) v_2-\sin(\phi))=e^{\pm i\phi}(v_+,v_-,v_1,v_2) 
$$
and implies that $v_+=0=v_-$ and $v_2=\pm i v_1$, hence $v\sim (0,0,1,\pm i)=:\hat e_\pm$. 
\end{Proof}
We finally prove a fact about the complexified boosts which we have frequently used. 
\begin{Lem} \label{LWte}
i) Every point in the complexified $\Spdc$ is of the form 
$z=\Boo{W}{i\theta}e$, 
where $W$ is some wedge, $e\in\Spd$ and  $\theta\in[0,\pi)$. 

ii) Every point in $\Tub$ is of the same form, but with $e\in W$ and $\theta \in(0,\pi)$.  
\end{Lem}
\begin{Proof}
\newcommand{\xNull}{\bar{x}}
\newcommand{\yNull}{\bar{y}}  
Let us first recall that $z=x+iy\in\Spdc$ if and only if $x^2-y^2=-1$ and $x\cdot y=0$. Note that, by the latter condition, $y^2>0$ implies that $x$ is space-like or zero and hence $y^2\leq1$. 

Ad $i)$. Clearly, $z=\Boo{W}{i\theta}\hat e$ iff $\Lambda z=\BooNull(i\theta) e$, where $\Lambda\in\Lor$ is such that $\lor W=W_0$, the standard wedge~\eqref{eqW0}, and where $e=\Lambda \hat e$. One calculates 
\begin{equation} \label{eqLWte} 
\BooNull(i\theta) e \equiv \big(\cos(\theta)e_0,e_1,e_2,\cos(\theta)e_3\big)+i\sin(\theta)\,(e_3,0,0,e_0).
\end{equation} 
We have to show that for every $z\in\Spdc$ there are $\lor\in\Lor$, $e\in \Spd$ and $\theta\in[0,\pi)$ such that $\lor z$ coincides with the above vector. 
We denote this vector by $\xNull+i\yNull$. We first claim that for our given $z=x+iy\in\Spdc$ on can choose $\theta\in[0,\pi)$ and $e$ so that $\xNull$ is in the same $\Lor$-orbit as $x$, and $\yNull$ is in the same $\Lor$-orbit as $y$. 

This can be achieved as follows. 
Case 1: $y^2>0$, $y_0\gtrless 0$.  Then $0<y^2\leq1$ (see above), hence $y^2=\sin^2\theta$  and $x^2=-\cos^2\theta$ for some $\theta\in(0,\pi)$. (Note that $y^2=1$ implies $x\equiv0$.) 
Putting $e:=(0,0,0,\pm 1)$ yields $\xNull=\cos(\theta)(0,0,0,\pm1)$ and $\yNull=\sin(\theta)(\pm1,0,0,0)$, hence  does the job. 
Case 2: $y^2<0$. Then $y^2=-\sinh^2\chi$ and $x^2=-\cosh^2\chi$ for some $\chi\in\RR$. 
Putting $e:=(\sin\chi,\cosh\chi,0,0)$ and $\theta:=\pi/2$ yields $\xNull=(0,\cosh\chi,0,0)$ and $\yNull=(0,0,0,\sinh\chi)$, hence  does the job. 
Case 3: $y^2=0$, $y_0\gtrless0$  and $x^2=-1$. 
Putting $e:=(1,1,0,\pm1)$ and $\theta:=\pi/2$ yields $\xNull=(0,1,0,0)$ and $\yNull=(\pm1,0,0,1)$, hence  does the job. In the remaining case $y\equiv 0$ nothing has to be shown. 

With this choice of $\theta$ and $e$ there is, in particular, some $\lor_1$ such that $\lor_1x=\xNull$. Suppose we can find some $\lor_2$ which leaves $\xNull$ invariant and maps $\lor_1 y$ to $\yNull$. Then $\lor:=\lor_2\lor_1$ satisfies $\lor z=\xNull+i\yNull\equiv \BooNull(i\theta) e$, as claimed.  It remains to prove the existence of such $\lor_2$.  
Since $y$ is orthogonal to $x$, $\lor_1 y$ is orthogonal to $\lor_1 x\equiv \xNull$. 
Suppose first that $\xNull^2<0$. Then its orthogonal complement $\xNull^\perp$ is a three-dimensional Minkowski space, and the stability group, in $\Lor$, of $\xNull$ is the corresponding Lorentz group. It acts transitively on the intersection of $\xNull^\perp$ with the $\Lor$-orbit of (any given) $\yNull$. Hence there is a $\lor_2$ with the mentioned properties. 
The only other case is $x=\xNull=0$ (see above), which is trivial. 

Ad $ii)$. Note that $\hat e$ from above Eq.~\eqref{eqLWte} is in $W$ iff $e\in W_0$. Thus we only have to show that in the above argument we can choose $e\in W_0$ and $\theta\in(0,\pi)$. But this has been achieved above, cf.\ case 1. 
This completes the proof. 
\end{Proof}

\section{Principal Series Representations of the Lorentz Group}
We recall the principal series representation of the Lorentz group
as outlined by Bros and Moschella~\cite{BrosMos} (cf.\ also Vilenkin).
Consider the space $C^\degree(\HypNull)$ of continuous complex valued
functions on the mass zero hyperboloid which are homogeneous of degree
$\degree$,  i.e.\ $\tilde{h}(\lambda \lc)=\lambda^\alpha \tilde{h}(\lc)$ for $\lambda>0$. 
On this space the Lorentz transforms are represented naturally according to 
\begin{equation} 
(D(\Lambda) \tilde{h})(\lc)\doteq \tilde{h}(\Lambda^{-1}\lc)\,. 
\end{equation}
Let $\Gamma$ be any two--dimensional cycle 
which encloses the origin,\footnote{Note that a homogeneous function is
determined by its restriction to $\Gamma$.} and let
$\d\nu_\Gamma$ be the restriction to $\Gamma$ of the Lorentz invariant 
measure $\d\nu$ on $H_0^+$ . Then define
\begin{align} 
 \Lsp \tilde h, \tilde f \Rsp \doteq \int_\Gamma
\overline{\tilde h(\lc)}\tilde f(\lc)\d\nu_\Gamma(\lc).  
\end{align}
As Bros and Moschella point out~\cite{BrosMos}, if $\Re \degree=-(d-2)/2$ 
(where $d$ is the dimension of ambient Minkowski space) then this pairing is a  well--defined 
({\em i.e.\ }independent of $\Gamma$) scalar product on
$C^\degree(\HypNull)$, with respect to which the representation $D$ of the
Lorentz group is unitary. The resulting unitary representation on the
Hilbert space completion of $C^\degree(\HypNull)$ is the irreducible principal series representation
corresponding to the value $\degree(\degree+d-2)=-|\degree|^2$ of the Casimir operator.

In their article~\cite{BrosMos}, Bros and Moschella consider a particular class of cycles 
$\Gamma_e$ of $\HypNull$, indexed by vectors $e$ in Minkowski
space. Namely, if $e$ is time-like and future--pointing, then
$\Gamma_e$ is defined as the set of vectors $\lc$ in $\Lc$ which satisfy
$\lc\cdot e=1$. If $e$ is space-like, then $\Gamma_e$ is the disjoint
union of the sets of vectors $\lc$ in $\HypNull$ which satisfy 
$\lc\cdot e=1$ or $-1$. They show that all of these cycles are
homologous within a suitable homology group. We observe that this
fact extends to a third class of cycles, namely 
\begin{align} \label{eqGammaq}
\Gamma_e\doteq\{\lc\in\HypNull:\;e \cdot \lc=1\}, 
\end{align} 
where $e$ is a  {\em light-like} and future--pointing vector. 

\paragraph{Acknowledgments.} 
JM gratefully acknowledges financial support by FAPESP, and thanks D.~Buchholz for pointing out ineq.~\eqref{eqVacFluc} to him. 
B.S. thanks the ESI, Vienna, and J.Y. the MPI for Physics, Munich and
the Science Institute of the University of Iceland for hospitality
during the completion of this paper. JY's research is partially supported by a grant P17176-N02 of the Austrian Science Fund (FWF) and  
the Network HPRN-CT-2002-00277 of the European Union.

\providecommand{\bysame}{\leavevmode\hbox to3em{\hrulefill}\thinspace}
\providecommand{\bysame}{\leavevmode\hbox to3em{\hrulefill}\thinspace}


\begin{thebibliography}{10}

\bibitem{Abbott} 
L.~F.~Abbott,  \emph{Massless particles with continuous spin indices}, 
Phys.~Rev. D~\textbf{13} (1976), 2291--2294.

\bibitem{Babujian}H.\  Babujian, A.\  Foerster, M.\ Karowski, 
\emph{Exact form factors in integrable quantum field theories: the 
scaling $Z(N)$-Ising model},
ArXiv: hep-th/0510062.


\bibitem{Bahns} D.\ Bahns,  \emph{The Invariant Charges of the Nambu-Goto String and Canonical
Quantization}, J.Math.Phys. \textbf{45} (2004), 4640-4660. 
 
\bibitem{BiWi} 
{J.~J.~Bisognano, E.~H.~Wichmann}, 
\emph{On the duality condition for a {Hermitian} scalar field},
J.~Math.~Phys.\ \textbf{16} (1975),  985--1007. 

\bibitem{BBS}
H-J.~Borchers, D.~Buchholz and B.~Schroer,  
\emph{Polarization-Free Generators and the {S}-Matrix},
Commun.~Math.~Phys. \textbf{219} (2001), 125--140. 

\bibitem{BrBu94}
J.~Bros and D.~Buchholz, \emph{Towards a relativistic {KMS}-condition}, Nucl.
  Phys. B \textbf{429} (1994), 291--318.

\bibitem{BrosMos}
J.~Bros and U.~Moschella, \emph{Two-point functions and quantum fields in de
  {S}itter universe}, Rev.\ Math.\ Phys.\ \textbf{8} (1996), 327--391.

\bibitem{BGL}
R.~Brunetti, D.~Guido, and R.~Longo, \emph{Modular localization and {W}igner
  particles}, Rev.\ Math.\ Phs. \textbf{14} (2002), 759--786.

  
  \bibitem{BDF} D. Buchholz, C. D'Antoni, K. Fredenhagen, \emph{The universal 
  structure of local algebras}, Commun. Math. Phys. 
{\bf 111} (1987), 123--135.
  
  
\bibitem{BF} 
D.~Buchholz and K.~Fredenhagen, \emph{Locality and the structure of particle
  states}, Commun. Math. Phys \textbf{84} (1982), 1--54.

\bibitem{BuchholzSummers2004} D.~Buchholz and S.~J.~Summers, \emph{Quantum
Statistics and Locality}, Phys. Lett. A {\bf 337} (2005),  17--21.

\bibitem{Bu-Yng_nucl}D.~Buchholz, J.~Yngvason, 
\emph{ Generalized Nuclearity Conditions and the Split 
Property in Quantum Field theory.} Lett.\ Math.\ Phys.\ {\bf 23}  (1991), 159--167.


\bibitem{Bu-Yng} D.~Buchholz and J.~Yngvason,
Phys.\ Rev\ .Lett. \textbf{73} (1994),  613--616. 

\bibitem{BW} D.~Buchholz, E.H.~Wichmann, 
\emph{Causal independence and the energy density of states in 
local quantum field theory},   Commun. Math. Phys. {\bf 106} (1986), 
321--344

\bibitem{Chang}S.-J.~Chang, \emph{Lagrange Formulation for Systems with Higher Spin
}, Phys. Rev. \textbf{161} (1967), 1308--1315.


\bibitem{Dimock} J.~Dimock, \emph{Locality in free string field 
theory}, J. Math. Phys. \textbf{41} (2000), 40--61.

  
\bibitem{DHRII}
S.~Doplicher, R.~Haag, and J.E.~Roberts, \emph{Fields, observables and gauge
  transformations {II}}, Commun. Math. Phys. \textbf{15} (1969), 173--200.




\bibitem{DL} S.~Doplicher and R.~Longo, \emph{Standard and split inclusions of von Neumann 
algebras}, Invent. Math. {\bf 75}, 493--536 (1984)

\bibitem{DS} M.~Duetsch and B.~Schroer, \emph{Massive vector mesons and gauge theory}, 
J.~Phys.~A: Math.~Gen.~{\bf 30}, 4317 (2000), and previous work by G. Scharf cited therein.

\bibitem{Ep-Gl}H.~Epstein and V.~Glaser, \emph{The role of locality in perturbation theory}, 
Ann. Inst. H. Poincar\'{e} A
\textbf{19}, (1973) 211--295.


\bibitem{Erler-Gross} D.~G.~ Erler and D.~J.~Gross, \emph{Locality ,
Causality and an Initial Value Formulation of Open Bosonic String
Field Theory}, arXiv:hep-th/0409179. 

\bibitem{FS02} 
{L.~Fassarella, B.~Schroer},
\emph{Wigner particle theory and local quantum physics},
{J.~Phys.~A} \textbf{35} ({2002}), {9123-9164}. 

\bibitem{Fermi} E. Fermi, \emph{Quantum Theory of Radiation}, {\em Rev. 
Mod. Phys.\/} {\bf 4}  (1932),  87--132.

\bibitem{FGR} 
K.~Fredenhagen, M.~Gaberdiel, and S.M. R\"{u}ger, \emph{Scattering states of
  plektons (particles with braid group statistics) in 2+1 dimensional field
  theory}, Commun. Math. Phys. \textbf{175} (1996), 319--355.

\bibitem{FRSII} 
K.~Fredenhagen, K.-H.~Rehren, and B.~Schroer, \emph{Superselection sectors with
  braid group statistics and exchange algebras {II}: Geometric aspects and
  conformal covariance}, Rev.~Math.~Phys.\ \textbf{SI1} (1992), 113--157.

\bibitem{Froehlich}J.~Fr\"{o}hlich, \emph{New super-selection sectors (``soliton-states'') in two dimensional Bose quantum field models},
 Commun. Math. Phys. \textbf{47}
(1976), 269--310. 

\bibitem{FG} 
J.~Fr\"{o}hlich, F.~Gabbiani,
\emph{Braid Statistics in Local  Quantum Field Theory},
{Rev.~Math.~Phys.}, \textbf{2} (1990), 251-353. 

\bibitem{H96}
R.~Haag, \emph{Local quantum physics}, second ed., Texts and Monographs in
  Physics, Springer, Berlin, Heidelberg, 1996.
  
  \bibitem{Hegerfeldt} G.C. Hegerfeldt, \emph{Causality 
Problems in Fermi's Two Atom System}, {\em Phys. Rev. Lett.\/}, {\bf 
72}  (1994), 
596-599.
  
  \bibitem{Hirata} K.\ Hirata, \emph{Quantization of Massless Fields with 
  Continuous Spin}, Prog.\ Theor.\ Phys.\ {\bf 58} (1977), 652--666 

\bibitem{IM}
G.J. Iverson and G.~Mack, \emph{Quantum fields and interactions of massless
  particles: The continuous spin case}, Ann. Phys. \textbf{64} (1971),
  211--253.
 
\bibitem{Joos}H. Joos, \emph{Zur Darstellungstheorie der inhomogenen
Lorentzgruppe als Grundlage quantenmechanischer Kinematik}, 
Fortschritte der Physik \textbf{10}, (1962) 65--146

\bibitem{Jordan1} 
P.~Jordan, \emph{Zur Quantenelektrodynamik, I. Eichinvariante
Operatoren},
Zeitschrift f\"ur Physik \textbf{95}, (1935) 202. 


\bibitem{Jordan2}\bysame, 
\emph{Beitr\"{a}ge zur Neutrinotheorie des Lichts}, 
 Zeitschr. f\"{u}r Physik \textbf{114}, (1937) 229.  

\bibitem{Klaiber} B.~Klaiber, \emph{The Thirring model}, in
A. O. Barut and W. E. Brittin (eds.) Lectures in theoretical physics,
Vol 10A, pp. 141-176, Gordon and Breach, New York 1968. 

\bibitem{Lechner}G.\ Lechner, {\it Towards the construction of quantum field theories from a factorizing S-matrix}, ArXiv: hep-th/0502184. 

\bibitem{LM} J.~M.~Leinaas and J.~Myrheim,
 \emph{On the Theory of Identical   Particles},
  {Il Nuovo Cimento}  \textbf{37 b} (1977), 1-23.  

\bibitem{LRT}
P.~Leyland, J.~Roberts and D.~Testard, 
\emph{Duality for Quantum Free Fields},
{unpublished notes}, {CNRS Marseille}, (1978). 


\bibitem{Licht} A.~L.~Licht, \emph{Local States},
Journ. Math. Phys. \textbf{7} (1966), 1656. 

\bibitem{Malament} D.\ 
Malament, \emph{In defence of dogma: Why there cannot be a relativistic 
quantum mechanics of (localizable) particles},  in R. K. Clifton (Ed.) 
Perspectives of quantum reality, Dortrecht Kluwer, 1996. 



\bibitem{Mandelstam}
S.~Mandelstam, \emph{Quantum electrodynamics without potentials},
Ann.\ Phys. \textbf{19} (1962), 1--24.  



\bibitem{Mourad} J.~Mourad, \emph{Continuous spin and tensionless 
strings}, ArXiv: hep-th/0410009

\bibitem{M} 
J.~Mund, \emph{No-go theorem for `free' relativistic anyons in $d=2+1$}, Lett.
  Math. Phys. \textbf{43} (1998), 319--328.


\bibitem{M01a}
\bysame, \emph{The {B}isognano-{W}ichmann theorem for massive theories}, Ann.
  H. Poinc. \textbf{2} (2001), 907--926.

\bibitem{M02a} 
\bysame, \emph{Modular localization of
massive particles with ``any'' spin in d=2+1}, J.\ Math.\
Phys. \textbf{44} (2003), 2037--2057.

\bibitem{MundSaclay} 
\bysame, \emph{String-Localized Covariant Quantum Fields}, ArXiv:hep-th/0502014. 

\bibitem{MSY1}
J.~Mund, B.~Schroer, and J.~Yngvason, \emph{String--localized quantum fields
  from {W}igner representations}, Phys.~Lett.~B {\bf 596} (2004), 156-162. 


\bibitem{NW} 
T.D.~Newton, E.~P.~Wigner,
\emph{Localized States for Elementary Systems},
{Rev.~Mod.~Phys.}\textbf{21} ({1949}), {400--406}. 

\bibitem{PerezWilde}
J.~F.~Perez, I.~F.~Wilde, 
\emph{Localization and causality in relativistic quantum mechanics},  
Phys.~Rev.Ð \textbf{16} (1977), 315--317. 
 


\bibitem{Polchinski} J.~Polchinski, \emph{String Theory Vol. I and II}, 
Cambridge Univ. Press, 1998

\bibitem{ReSi}
M.~Reed and B.~Simon, \emph{Methods of modern mathematical physics {II}},
  Academic Press, New York, 1975.
  
  \bibitem{Reeh-Schlieder} H.~Reeh, S.~Schlieder, \emph{Bemerkungen zur 
  unit\"ar\"aquivalenz von
Lorentzinvarianten Feldern},  Nuovo Cimento {\bf 22} (1961), 1051­1068.



\bibitem{RvD} 
M.~A.~Rieffel, A.~Van Daele,
\emph{A bounded operator approach to {T}omita-{T}akesaki theory}, 
Pacific Journ. Math.\ \textbf{1} (1977),  187-221. 

\bibitem{Savvidy} G.~Savvidy, \emph{Tensionless strings, correspondence 
with SO(D,D) sigma model}, Phys.~Lett.~  B {\bf 615} (2005), 285-290

\bibitem{SchroerAOP} B.~Schroer, \emph{Modular Wedge Localization and the 
d=1+1 Formfactor Program}, Ann.\ Phys. \textbf{295} (1999), 190--223. 


\bibitem{crossing} B.~Schroer, \emph{Constructive proposals based on
the crossing property and the lightfront holography} Ann.\ Phys. {\bf 319} (2005) 48



\bibitem{St}
O.~Steinmann, \emph{A {Jost-Schroer} theorem for string fields}, Commun. Math.
  Phys. \textbf{87} (1982), 259--264.

\bibitem{SW}
R.F.~Streater and A.S.~Wightman, \emph{{PCT}, spin and statistics, and all
  that}, W. A. Benjamin Inc., New York, 1964.

\bibitem{Streater-Wild} R.~F.~Streater and I.~F.~Wilde, \emph{Fermion
states of a Bose field}, Nucl. Phys.\ \textbf{ B 24} (1970), 561. 

\bibitem{Strocchi} F.~Strocchi, \emph{} Phys.~Rev.\ \textbf{166} (1969), 1302-1307. 

\bibitem{Summers} S.~Summers, \emph{On the independence of local algebras 
in 
quantum field theory}, {\em Rev. Math. Phys.\/} {\bf 2} (1990), 201--247.


\bibitem{Weinberg} S.~Weinberg, \emph{What is quantum field theory,
and what did we think it is?}, ArXiv:hep-th/9702027. 

\bibitem{Wei-book}S.~Weinberg, \textit{The Quantum Theory of Fields I},
Cambridge University Press 1995.

\bibitem{Wei-article}S.~Weinberg, \emph{Feynman Rules For Any Spin}, 
Phys. Rev. \textbf{133}, (1964) B1318--30.

\bibitem{Wern} R. Werner, \emph{Local preparability of States and the 
Split Property in Quantum Field Theory}, {\em Lett. Math. Phys.\/} 
{\bf 13} (1987),  325--329.

\bibitem{Wig48} 
E.P. Wigner, \emph{Relativistische {W}ellengleichungen}, Z.~Physik \textbf{124}
  (1948), 665--684.

\bibitem{Wilczek} F.~Wilczek, \emph{Quantum Mechanics of 
   Fractional-Spin Particles}, 
 Phys.~Rev.~Lett.~{\bf 49} (1982), 957-1149. 

\bibitem{Wilson} K.~G.~Wilson, {\it Confinement of Quarks}, Phys. Rev. 
D \textbf{10} (1974), 2445-2459. 

\bibitem{Yng70} 
J.~Yngvason, \emph{Zero-mass infinite spin representations of the {P}oincar\'e
  group and quantum field theory}, Commun. Math. Phys. \textbf{18} (1970),
  195--203.
  
  \bibitem{TypeIII} J.~Yngvason, \emph{The Role of Type III Factors in 
  Quantum Field Theory}, Rep.\ 
Math.\ Phys. {\bf 55}, 135--147, (2005).  


\end{thebibliography}
\end{document}